\documentclass[preprint,12pt]{elsarticle}
\usepackage[top=1in, bottom=1in, left=1in, right=1in]{geometry}
\usepackage{bm}
\usepackage{amsmath}
\usepackage{amssymb}
\usepackage{setspace}
\usepackage{bigints}
\usepackage{array}
\usepackage{multirow}
\usepackage{colortbl}
\usepackage{graphicx}
\usepackage{subfigure}
\usepackage[table,xcdraw]{xcolor}
\usepackage[outdir=./]{epstopdf}
\usepackage{booktabs}
\usepackage{tabu}
\usepackage{bbm}
\usepackage{float}
\usepackage[singlelinecheck=false]{caption}
\usepackage[utf8]{inputenc}
\usepackage[english]{babel}
\usepackage{natbib}
\usepackage{bookmark}
\usepackage{mathrsfs}
\usepackage{color}
\usepackage{mathtools} 
\newcommand{\comment}[1]{}
\usepackage{amssymb}
\usepackage{xfrac}
\usepackage{algorithm}
\usepackage{algorithmic}
\usepackage{lscape}
\usepackage{pdflscape}

\journal{Elsevier}
\begin{document}
\begin{spacing}{1.2}
\begin{frontmatter}

%\title{The effect of prior knowledge on the uncertainty quantification and efficient propagation from limited data}
\title{Probability measure changes in Monte Carlo simulation}
\author{ Jiaxin Zhang, Michael D. Shields$^{*}$ \footnote{*Corresponding author.\\ Email: jiaxin.zhang@jhu.edu (Jiaxin Zhang), michael.shields@jhu.edu (Michael D. Shields)} }
\address{Department of Civil Engineering, Johns Hopkins University
	Baltimore, MD 21218, USA.}

\begin{abstract}

The objective of Bayesian inference is often to infer, from data, a probability measure for a random variable that can be used as input for Monte Carlo simulation. When datasets for Bayesian inference are small, a principle challenge is that, as additional data are collected, the probability measure inferred from Bayesian inference may change significantly. That is, the original probability density inferred from Bayesian inference may differ considerably from the updated probability density both in its model form and parameters. In such cases, expensive Monte Carlo simulations may have already been performed using the original distribution and it is infeasible to start again and perform a new Monte Carlo analysis using the updated density due to the large added computational cost. In this work, we discuss four strategies for updating Mote Carlo simulations for such a change in probability measure: 1. Importance sampling reweighting; 2. A sample augmenting strategy; 3. A sample filtering strategy; and 4. A mixed augmenting-filtering strategy. The efficiency of each strategy is compared and the ultimate aim is to achieve the change in distribution with a minimal number of added computational simulations. The comparison results show that when the change in measure is small importance sampling reweighting can be very effective. Otherwise, a proposed novel mixed augmenting-filtering algorithm can robustly and efficiently accommodate a measure change in Monte Carlo simulation that minimizes the impact on the sample set and saves a large amount of additional computational cost. The strategy is then applied for uncertainty quantification in the buckling strength of a simple plate given ongoing data collection to estimate uncertainty in the yield stress.
\end{abstract}

\end{frontmatter}
%%%%%%%%%%%%%%%%%%%
%             INTRODUCATION            %
%%%%%%%%%%%%%%%%%%%
\section{Introduction}

Uncertainty quantification (UQ) is playing an increasingly important role in computational analysis of physical and engineering systems, particularly performance prediction, risk analysis and decision making. Generally speaking, there are two types of problems in UQ. One is the so-called \emph{Forward UQ}, also named uncertainty propagation, where the many sources of uncertainty in system input and/or parameters are propagated through the computational model to analyze and predict the overall uncertainty in the system response. On the other hand, \emph{Inverse UQ}, aims to infer uncertainty in a model or its parameters from measured data or simulations.  

The general application of UQ includes aspects of both forward and inverse UQ. Often, this starts from the Bayesian inference of uncertainty measured data and proceeds to the propagation of those uncertainties through a computational model using Monte Carlo simulation or some other approach. Finally, the response outputs are systematically analyzed and assessed. One challenge that may arise in this process occurs when data collection and simulation occur concurrently. In such cases, forward UQ must be performed initially from limited data and then must be updated as additional data are collected. More specifically, Bayesian inference is used to define initial probability distributions for system parameters/inputs from limited data. These distributions are propagated through the model using Monte Carlo simulation. When new data are collected, Bayesian inference is once more conducted and the distributions updated. But, when the distributions are updated the corresponding Monte Carlo simulations must also be updated. Given the computational expense of Monte Carlo simulation, it is undesirable (and potentially infeasible) to conduct an entirely new study from the updated probabilities. 

% Zhang and Shields [ref] propose a Bayesian multimodel UQ methodology to quantify the uncertainty and employs the optimal importance sampling to efficiently propagate the uncertainties resulting from small datasets.  A potential challenge is that the optimal sampling density loss the optimality when there is a significant probability measure change resulting from additional collected data, and consequently it leads to a bias estimate with high variance. On the contrary, if the optimality is retained, that means the optimal sampling density needs to be updated sequentially as data collection. At the same time, all the random samples drawn from the optimal sampling density will be regenerated. In other words, the efficiency of the algorithm will significantly decrease due to the large computational simulations required for uncertainty propagation. Therefore, a balanced method is necessary to take into account the accuracy and efficiency in uncertainty propagation simultaneously. 

In this work, we discuss four strategies to update Monte Carlo simulations to accommodate a measure change in the input distributions. The objective is to minimize the number of additional calculations/simulations required while maintaining a high level of statistical confidence in the Monte Carlo results. We specifically compare the widely used importance sampling reweighting strategy \cite{ONeill2009, fetz2016imprecise, pmlr-v62-troffaes17a,zhang2018mssp} with three new approaches based on augmenting the original sample set, filtering the original sample set, and a combined augmenting and filtering approach. The four approaches are systematically compared for various types of measure changes including cases where the support of the distribution changes. These examples serve to show that the importance sampling reweighting is effective when the change in distribution is modest and the support does not increase. The proposed combined augmenting and filtering approach, on the other hand, is robust and efficient in the sense that it minimizes the number of additional simulations needed even when the change in distribution is significant. However, augmenting or filtering alone are highly inefficient and are not recommended in practice. The updating strategy is then put into practice for a plate buckling analysis given ongoing data collection efforts to quantify the distribution of the material yield stress.  

% develop a set of resampling strategies that efficiently accommodate a measure change in Monte Carlo simulation that minimizes the impact on the sample set. In other words, it retains as many as samples as possible from the original Monte Carlo set drawn from density and adds a minimal number of samples from a ``correction" density such that the combined set follows the desired new density while density while keeping the sample size constant.  

% The paper is structured as follows. Section 2 provide the basic theory for Bayesian method for model-form and parameter uncertainty. Additionally, the probability measure changes are discussed as changes of the datasets. Section 3 presents the proposed methods for changing measure in Monte Carlo simulation. Four resampling strategies are formulated and compared in terms of their advantages and limitations.  Numerical illustration,  including distribution with common support and changing support are systematically shown for the efficiency and degeneracy in Section 4.  A plate buckling strength problem as the engineering example are illustrated to demonstrate the effectiveness of the proposed methodology in Section 5. Finally, some concluding remarks are provided in Section 6. 

\section{Bayesian inference and probability model selection}

Let $\bm{X}:\Omega\to \mathbb{R}^n$ be a random variable defined on a complete probability space $(\Omega,\mathcal{F},\mathcal{P})$ where $\Omega$ is the sample space of events, $\mathcal{F}$ is the sigma-algebra, and $\mathcal{P}$ is a probability measure. The objective of Bayesian inverse UQ is to infer a probability measure $\mathcal{P}$ from a given dataset $\bm{d}$. The measure $\mathcal{P}$ is assumed to follow a parametric form defined through a probability model $M$ having parameters $\boldsymbol{\theta}$. 

Here, Bayesian inference occurs in two stages. First, the model form is inferred. Given a collection of $m$ candidate models $\mathcal{M}=\{M_j\}$ with parameters $\boldsymbol{\theta}_j, j=1,\dots,m$ and associated prior probabilities $\pi_j=p(M_j)$ with $\sum_{j=1}^m\pi_j=1$, Bayes' rule is applied to determine posterior model probabilities given the data $\bm{d}$ as:
\begin{equation}
\hat{\pi}_j = p(M_j| \bm{d}) = \frac{p(\bm{d} | M_j)p(M_j )}{\sum_{j=1}^{m} p(\bm{d} | M_j)p(M_j )},  \quad j=1,\dots, m \label{eq:Bayes_model}
\end{equation}
having $\sum_{j=1}^m\hat{\pi}_j=1$ and where 
\begin{equation}
p(\bm{d}|M_j)=\int_{\boldsymbol{\theta}_j}p(\bm{d}|\boldsymbol{\theta}_j,M_j)p(\boldsymbol{\theta}_j|M_j)d\boldsymbol{\theta}_j, \quad j=1,\dots,m
\label{eq:Model_evidence}
\end{equation}
is the marginal likelihood or evidence of model $M_j$. 

One challenge in determining $\hat{\pi}_j$ is that Eq.\ \eqref{eq:Model_evidence} can be difficult to evaluate. Several methods have been proposed to estimate the evidence such as Laplace's method \cite{konishi2008} and information theoretic approximations \cite{akaike1974new, schwarz1978, hurvich1989regression}. A detailed discussion of the evidence calculation can be found in \cite{ZHANG2018cmame2}. For simplicity, in this work we employ a Monte Carlo estimator given by
\begin{equation}
\hat{p}(\bm{d}|M_j)=\dfrac{1}{N_k}\sum_{k=1}^{M_k}p(\bm{d}|\boldsymbol{\theta}_j^k,M_j), \quad \boldsymbol{\theta}_j^k\sim p(\boldsymbol{\theta}_j|M_j), \quad j=1,\dots,m
\end{equation}
in which samples $\boldsymbol{\theta}_j^k$ are drawn from the parameter prior distribution and $N_k$ is the number of samples.

The model with maximum a posteriori probability (MAP) according to Eq.\ \eqref{eq:Bayes_model} is selected and the second stage of the inference is to identify the parameters of this model. Again, Bayes' rule is applied such that the posterior parameter probabilities $p(\bm{\theta} | \bm{d}, M)$ given the model $M$ and data $\bm{d}$ can be computed from the prior probabilities as
\begin{equation}
p(\bm{\theta} | \bm{d}, M) = \frac{p(\bm{d} | \bm{\theta}, M) p(\bm{\theta}, M)}{p(\bm{d}, M)} \propto p(\bm{d} | \bm{\theta}, M) p(\bm{\theta}, M) \label{eq: BI}
\end{equation}
Here, the model evidence $p(\bm{d},M)$ does not need to be evaluated explicitly as $p(\boldsymbol{\theta}|\bm{d},M)$ can be estimated implicitly from samples drawn using the Markov Chain Monte Carlo method.

From the posterior parameter density $p(\bm{\theta} | \bm{d}, M)$, the precise parameters are estimated using an MAP estimate, which is closely related to the method of maximum likelihood (ML) estimation but employs an augmented optimization objective that incorporates a prior distribution over the quantity one wants to estimate. MAP estimates $\bm{\theta}$ as the mode of the posterior probability measure $p(\bm{\theta} | \bm{d}, M)$ \cite{gauvain1994maximum, Bassett2018}
\begin{equation}
\hat{\bm{\theta}}_{\textup{MAP}}(\bm{d}, M)  = \underset {\bm{\theta}  }{ \arg \max } \ p(\bm{\theta} | \bm{d}, M) = \underset {\bm{\theta}  }{ \arg \max } \ {p(\bm{d} | \bm{\theta}, M) p(\bm{\theta}, M)}  
\end{equation}
Note that the MAP estimate of $\bm{\theta}$ coincides with the maximum likelihood estimate when the prior probability measure is uniform (that is, a constant function).

In an effort where data collection and simulation occur concurrently, as motivated here, it is straightforward to apply this two-stage Bayesian inference approach to update both the probability model $M$ and the associated parameters $\boldsymbol{\theta}$ as new data are collected.

\section{Methods for changing measure in Monte Carlo simulation}
The Monte Carlo method is used to estimate the probabilistic response of the system $Y=g(\bm{X})$ from the deterministic response of the system evaluated at independent statistical samples of $\bm{X}$. Specifically, given $N$ independent samples of $\bm{X}$ drawn independently from $p(\bm{x})$, the expected value of $Y$, $E[Y]$ can be estimated using Monte Carlo simulation by
\begin{equation}
E_p[Y]=\int_{\mathcal{S}_p} g(\bm{x})p(\bm{x})d\bm{x}\approx \mu_p = \dfrac{1}{N}\sum_{i=1}^Ng(\bm{x}_i)
\label{eqn:MCS_p}
\end{equation}
where the subscript $p$ denotes expectation with respect to $p(\bm{x})$ and $\mathcal{S}_p=\{\bm{x}\in\mathbb{R}^n:p(\bm{x})>0\}$ is the support of $p(\bm{x})$.

The challenge addressed here is that, as additional data are collected, the probability measure inferred from Bayesian updating may change considerably. That is, if the probability density inferred from the initial Bayesian inference is $p(\bm{x})$, the updated probability density may be given by $q(\bm{x})$ where $p(\bm{x})\ne q(\bm{x})$. Moreover, if $\mathcal{S}_p=\{\bm{x}\in\mathbb{R}^n:p(\bm{x})>0\}$ is the support of $p(\bm{x})$ and 
$\mathcal{S}_p=\{\bm{x}\in\mathbb{R}^n:q(\bm{x})>0\}$ is the support of $q(\bm{x})$, then in general $\mathcal{S}_p\nsubseteq\mathcal{S}_q$ and $\mathcal{S}_p\nsupseteq\mathcal{S}_q$. That is, $p(\bm{x})$ and $q(\bm{x})$ do not necessarily have the same support.

What is to be done then, when Monte Carlo simulation has already been performed using the density $p(\bm{x})$ but  the density of $\bm{X}$ has been updated (or changed) to $q(\bm{x})$? Certainly, it is undesirable to start again and perform a new Monte Carlo analysis using density $q(\bm{x})$. How can the results of the original Monte Carlo analysis be leveraged for Monte Carlo analysis on the updated distribution? In this section, we discuss four different strategies for updating Monte Carlo simulations for such a change in measure. Formally stated, consider Monte Carlo analysis has been used to estimate $E_p[Y]$ using \eqref{eqn:MCS_p}. Using the results from Eq.\ \eqref{eqn:MCS_p}, and perhaps some additional samples, we show four strategies to obtain a Monte Carlo estimator for $E_q[Y]$. The efficiency of each strategy is compared with the prospect of a new Monte Carlo simulation on the updated density and the ultimate aim is to achieve the change in distribution with a minimal number of added simulations.

\subsection{Importance sampling reweighting}
\label{Sec:IS_Reweight}
Perhaps the most attractive option is to simply reweight the existing samples drawn from $p(\bm{x})$ to convert an evenly weighted Monte Carlo estimator on $p(\bm{x})$ into an unevenly weighted estimator on $q(\bm{x})$. This can be achieved using the principals of importance sampling as follows. Consider the desired expected value with respect to $q(\bm{x})$, $E_q[\cdot]$, which can be written as:
\begin{equation}
E_q[g(\bm{X})] = \int_{\mathcal{S}_q} {g(\bm{x})q(\bm{\bm{x}})d\bm{\bm{x}}} = \int_{\mathcal{S}_p}g(\bm{x})\frac{q(\bm{x})}{p(\bm{x})}p(\bm{x})d\bm{x} =  {E}_{p}\left[g(\bm{X})\frac{q(\bm{X})}{p(\bm{X})}\right]
\end{equation}
where, again, $E_p[\cdot]$ is the expectation with respect to $p(\bm{x})$. Defining weights $w(\bm{x})={q(\bm{x})}/{p(\bm{x})}$ (referred to as the importance weights), the importance sampling estimate of $E_q[g(\bm{X})]$ can be expressed as 
\begin{equation}
E_q[g(\bm{X})] \approx \mu_q = \frac{1}{N} \sum_{i=1}^{N} g(\bm{x}_i) \frac{q(\bm{x}_i)}{p(\bm{x}_i)} = \frac{1}{N} \sum_{i=1}^{N} g(\bm{x}_i) w(\bm{x}_i)
\label{eq:E_ISs} 
\end{equation}
Thus, the new estimate is obtained simply from the original estimate by applying sample weights $w(\bm{x})$.

This option is attractive because it does not require any additional simulations. But, depending on the change in measure, it can have a strong adverse effect on the estimator. That is, if $p(\bm{x})$ and $q(\bm{x})$ have large discrepancy, the quality of $\mu_q$ may be much less than the quality of $\mu_p$. To formalize this, note that the variance of the importance sampling estimator is given by:
\begin{equation}
\begin{aligned}
\text{Var}_p(\mu_q) &= \dfrac{1}{N}\text{Var}_p(g(\bm{X})w(\bm{X}))\\
& = \dfrac{1}{N}\int_{\mathcal{S}_p}\left(g(\bm{x})w(\bm{x})-E_p[g(\bm{X})w(\bm{X})]\right)^2p(\bm{x})d\bm{x}\\
& = \dfrac{1}{N}\left\{\int_{\mathcal{S}_p}g^2(\bm{x})w^2(\bm{x})p(\bm{x})d\bm{x}-E_p[g(\bm{X})w(\bm{X})]^2\right\}\\
& = \dfrac{1}{N}\left\{\int_{\mathcal{S}_q}g^2(\bm{x})w(\bm{x})q(\bm{x})d\bm{x}-E_q[g(\bm{X})]^2\right\}\\
& = \dfrac{1}{N}\left\{E_q[g^2(\bm{x})w(\bm{x})]-E_q[g(\bm{X})]^2\right\}
\end{aligned}
\label{eqn:Var_IS}
\end{equation}
From Eq.\ \eqref{eqn:Var_IS}, it is apparent that we must have $\mathcal{S}_q \subseteq \mathcal{S}_p$. This is a well-known requirement of importance sampling as $w(\bm{x})\to\infty$ when $p(\bm{x})=0$ and $q(\bm{x})\ne 0$. In such cases importance sampling cannot be used. 

Another challenge arises when $p(\bm{x})\to 0$ faster than $q(\bm{x})\to 0$, usually in the tails of the distribution. When this occurs, $w(\bm{x})\to \infty$ as $p(\bm{x})\to 0$ and the importance sampling estimator $\text{Var}_p(\mu_q)\to\infty$ yielding a poor IS estimator. In such cases, the importance sampling density $p(\bm{x})$ is said to be degenerate.

A useful measure for the quality of an importance sampling estimator is the so-called Effective Sample Size ($N_{ESS}$). Formally defined as the ratio of the estimator variances times the number of samples as
\begin{equation}
N_{ESS} = \dfrac{\text{Var}_p(\mu_q)}{\text{Var}_q(\mu_q)}N,
\end{equation}
where $\text{Var}_p(\mu_q)$ is the variance of the importance sampling estimator of $E_q[g(\bm{X})]$ with samples drawn from $p(\bm{x})$ given by Eq.\ \eqref{eq:E_ISs} and $\text{Var}_p(\mu_q)$ is the variance of the classical Monte Carlo estimator of $E_q[g(\bm{X})]$ with samples drawn from $q(\bm{x})$ itself. It is rarely feasible to calculate $N_{ESS}$ directly, but several approximations have been proposed as discussed in \cite{martino2017effective}. The most common approximation is given by 
\begin{equation}
\hat{N}_{ESS} = \frac{\left(\sum_{i=1}^{N}w(\bm{x}_i)\right)^2}{\sum_{i=1}^{N}w(\bm{x}_i)^2}, 
\label{eqn:ESS_approx}
\end{equation}
which, as shown by \cite{martino2017effective}, is related to the $L_2$-norm between the importance sampling weights $w(\bm{x})$ and the uniform Monte Carlo weights $\bar{w}(\bm{x})=\frac{1}{N},~\forall \bm{x}$. It follows from Eq.\ \eqref{eqn:ESS_approx} that the effective sample size decreases as the weights become degenerate (some very large weights and some very small weights) and $N_{ESS}\to N$ as $w(\bm{x})\to \bar{w}(\bm{x})$. The effective sample size in Eq.\ \eqref{eqn:ESS_approx} will be the primary metric of assessing importance sampling quality throughout.  

\subsection{Augmenting sample sets}
\label{Sec:Augment}
When the conditions for importance sampling reweighting are not ideal (i.e. effective sample size is small) or are simply not met (i.e. a change in support), it may be necessary to augment the existing sample set with additional samples and perform further simulations. In this section, a simple method is introduced for augmenting a sample set $\mathbf{x}=\{\bm{x}_1,\bm{x}_2,\dots,\bm{x}_N\}\sim p(\bm{x})$ with $N_a$ new samples such that the augmented sample set $\mathbf{x}_a=\{\bm{x}_1,\bm{x}_2,\dots,\bm{x}_N\,\dots, \bm{x}_{N+N_a}\}\sim q(\bm{x})$.

Consider the updated probability density $q(\bm{x})$ as a mixture distribution (convex combination) of the original distribution $p(\bm{x})$ and a new, as yet unknown, distribution $f(\bm{x})$ as:
\begin{equation}
q(\bm{x})=\dfrac{Bf(\bm{x})+p(\bm{x})}{A}
\label{eqn:augment_mixture}
\end{equation}
where $A$ and $B$ are constants subject to constraints discussed below. Solving for $f(\bm{x})$ yields:
\begin{equation}
f(\bm{x})=\dfrac{Aq(\bm{x})-p(\bm{x})}{B}.
\end{equation}
To be a valid density, $f(\bm{x})$ must satisfy $f(\bm{x})\ge 0~\forall \bm{x}$ and $\int f(\bm{x)}d\bm{x} = 1$. The first condition implies:
\begin{equation}
Aq(\bm{x})-p(\bm{x})\ge 0\Rightarrow A \ge \dfrac{p(\bm{x})}{q(\bm{x})} \quad \forall \bm{x}
\end{equation}
while the second condition states:
\begin{equation}
B=\int_\Omega Aq(\bm{x})-p(\bm{x})d\bm{x}
\end{equation}
Hence, the constants $A$ and $B$ are not unique, which implies that several distributions $f(\bm{x})$ can be selected that satisfy Eq.\ \eqref{eqn:augment_mixture}. For reasons that will become clear shortly, we select the constants such that $A$ is minimized.  This yields
\begin{equation}
\begin{aligned}
A &= \max{\left\{\dfrac{p(\bm{x})}{q(\bm{x})}\right\}}\\
B & = \int_\Omega \max{\left\{\dfrac{p(\bm{x})}{q(\bm{x})}\right\}}q(\bm{x})-p(\bm{x})d\bm{x}
\label{eqn:AB_vals}
\end{aligned}
\end{equation}
yielding the ``augmentation density'' or ``correction density''
\begin{equation}
q_c(\bm{x})=f(\bm{x})=\dfrac{\max{\left\{\dfrac{p(\bm{x})}{q(\bm{x})}\right\}}q(\bm{x})-p(\bm{x})}{\bigintss_\Omega \max{\left\{\dfrac{p(\bm{x})}{q(\bm{x})}\right\}}q(\bm{x})-p(\bm{x})d\bm{x}}
\end{equation}
Hence, when sampled in correct proportion, samples from $q_c(\bm{x})$ and $p(\bm{x})$ will combine to form a sample set that follows $q(\bm{x})$. 

Returning to our initial problem, we have $N$ samples drawn from $p(\bm{x})$ and we wish to draw an additional $N_a$ samples from $q_c(\bm{x})$ such that the total set of $N^* = N+N_a$ samples follows $q(\bm{x})$. How many samples, $N_a$, are required to match $q(\bm{x})$? According to the composition method for generation of random variables \cite{law2015simulation}, the distributions $q_c(\bm{x})$ and $p(\bm{x})$ should be sampled in proportions $\frac{B}{A}$ and $\frac{1}{A}$ respectively in the general case given in Eq.\ \eqref{eqn:augment_mixture}. Applying the values in Eq.\ \eqref{eqn:AB_vals} implies
\begin{subequations}
\begin{equation}
\dfrac{N}{N+N_a}=\dfrac{1}{A}=\dfrac{1}{\max{\left\{\dfrac{p(\bm{x})}{q(\bm{x})}\right\}}}
\label{eqn:prop1}
\end{equation}
\begin{equation}
\dfrac{N_a}{N+N_a}=\dfrac{B}{A}=\dfrac{\bigintss_\Omega \max{\left\{\dfrac{p(\bm{x})}
{q(\bm{x})}\right\}}q(\bm{x})-p(\bm{x})d\bm{x}}{\max{\left\{\dfrac{p(\bm{x})}{q(\bm{x})}\right\}}}
\end{equation}
\end{subequations}
Solving Eq.\ \eqref{eqn:prop1} for $N_a$ yields
\begin{equation}
N_a = \left(\max{\left\{\dfrac{p(\bm{x})}{q(\bm{x})}\right\}}-1\right)N
\label{eqn:N_a}
\end{equation}
From Eq.\ \eqref{eqn:N_a}, it is clear that defining $q_c(\bm{x})$ with any value of $A>\max{\left\{\dfrac{p(\bm{x})}{q(\bm{x})}\right\}}$ will require a larger number of augmented samples.

Similar to importance sampling reweighting, this approach has notable shortcomings. Most notably, Eq.\ \eqref{eqn:AB_vals} requires that $\mathcal{S}_p\subseteq\mathcal{S}_q$. That is, the support of $q(\bm{x})$ must cover the full support of $p(\bm{x})$. When this condition is satisfied, the primary drawback of this approach is that $N_a$ is can become very large. As expected, when $\max{\left\{\dfrac{p(\bm{x})}{q(\bm{x})}\right\}}=1$ (i.e. $p(\bm{x})=q(\bm{x})$), $N_a\to 0$. However, when $q(\bm{x})\to 0$ faster than $p(\bm{x})\to 0$, $N_a\to \infty$ and the approach is not viable. 

\subsection{Filtering sample sets}
\label{Sec:Filter}
Another alternative that does not require additional simulations is to ``filter'' the existing $N$ samples drawn from $p(\bm{x})$ according to the well-known acceptance/rejection (A/R) method. To apply A/R, let us scale the original density $p(\bm{x})$ to serve as a majorizing function for $q(\bm{x})$. Recall that the majorizing function, $t(\bm{x})$, by definition must envelope $q(\bm{x})$ such that $t(\bm{x})\ge q(\bm{x})~\forall \bm{x}$ and, in general, $t(\bm{x})$ is not a probability density. Letting $t(\bm{x})=cp(\bm{x})$, we have $\int t(\bm{x})d\bm{x} = c$. To satisfy the conditions on $t(\bm{x})$ requires that
\begin{equation}
c\ge \dfrac{q(\bm{x})}{p(\bm{x})} \quad \forall \bm{x}
\end{equation}
which can be satisfied by setting
\begin{equation}
c = \max\left\{\dfrac{q(\bm{x})}{p(\bm{x})}\right\}
\end{equation}
Applying the A/R method, we then accept samples with probability 
\begin{equation}
P_{a}(\bm{x})=\dfrac{q(\bm{x})}{t(\bm{x})}= \dfrac{\dfrac{q(\bm{x})}{p(\bm{x})}}{\max\left\{\dfrac{q(\bm{x})}{p(\bm{x})}\right\}}
\end{equation}
Notice that when $c>\max\left\{\dfrac{q(\bm{x})}{p(\bm{x})}\right\}$, the acceptance rate will decline and a larger proportion of the original $N$ samples will be rejected. 

Applying A/R, the overall proportion of samples that will be accepted, $N_f$, to the $N$ original samples drawn from $p(\bm{x})$ is given by
\begin{equation}
\dfrac{N_f}{N} = \dfrac{\bigintss_{\mathcal{S}_q} q(\bm{x})d\bm{x}}{\bigintss_{\mathcal{S}_p} \max\left\{\dfrac{q(\bm{x})}{p(\bm{x})}\right\}p(\bm{x})d\bm{x}}=\dfrac{1}{\max\left\{\dfrac{q(\bm{x})}{p(\bm{x})}\right\}}
\end{equation}

As in the previous cases, we note that the A/R method has the following limitations. Most notably, the method requires $\mathcal{S}_q\subseteq \mathcal{S}_p$. That is, the approach is not viable if the support of the updated density is larger than the support of the original density. Notice also that, again, when $\max{\left\{\dfrac{q(\bm{x})}{p(\bm{x})}\right\}}=1$ (i.e. $p(\bm{x})=q(\bm{x})$), $N_f/N\to 1$, which means that no samples will be rejected. However, when $p(\bm{x})\to 0$ faster than $q(\bm{x})\to 0$, $N_f\to 0$ and the approach is not viable.

\subsection{Augmenting and filtering sample sets}
\label{Sec:AandF}
\begin{figure}[!ht]	
	\centering
	\includegraphics[height=2.5in]{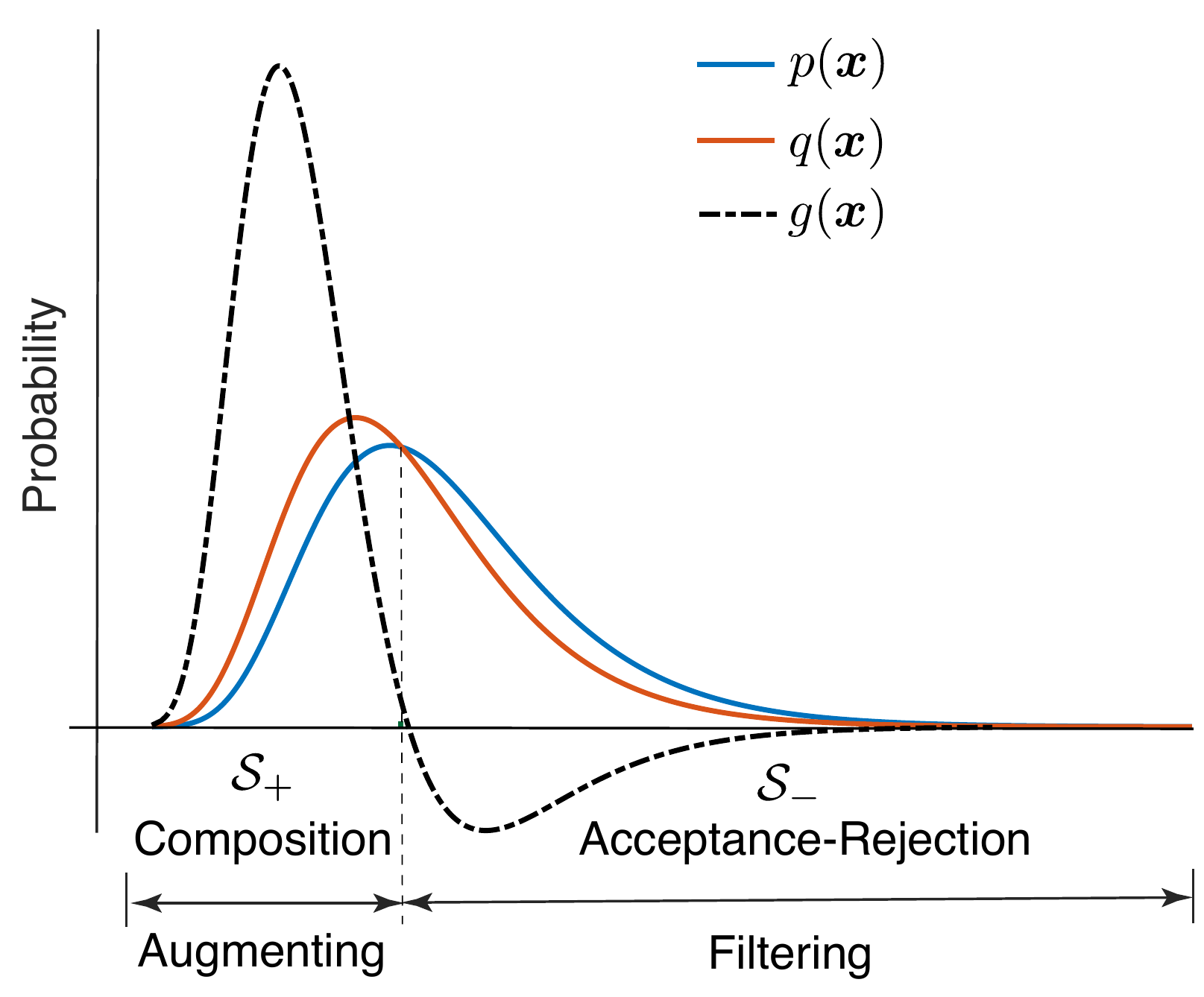}
	\caption[]{Corrected sampling density}  \label{fig:correct_pdf}
\end{figure}

The final approach combines the benefits of the augmenting and filtering approaches while also addressing their limitations.Consider the original density $p(\bm{x})$ and the updated density $q(\bm{x})$ such as those shown in Figure \ref{fig:correct_pdf}. For illustration, it is convenient to define the function $g(\bm{x})=\dfrac{p(\bm{x})-q(\bm{x})}{p(\bm{x})}$ having the property $g(\bm{x})>0$ when $p(\bm{x})>q(\bm{x})$ and $g(\bm{x})<0$ when $p(\bm{x})<q(\bm{x})$ as also illustrated in Figure \ref{fig:correct_pdf}. 
It is convenient to partition the total support $\mathcal{S}=\mathcal{S}_p\cup\mathcal{S}_q$ according to the support $\mathcal{S}_+=\{\bm{x}:q(\bm{x})\ge p(\bm{x})\}$ and $\mathcal{S}_-=\{\bm{x}:q(\bm{x})<p(\bm{x})\}$ with $\mathcal{S}=\mathcal{S}_+\cup\mathcal{S}_-$ and $\mathcal{S}_+\cap\mathcal{S}_1=\emptyset$. The point(s) $\bm{x}^*$ satisfying $p(\bm{x}^*)=q(\bm{x}^*)$ ($g(\bm{x}^*)=0$) represents the point(s) at which the support is divided into $\mathcal{S}_+$ and $\mathcal{S}_-$. The region, $\mathcal{S}_-$ where $g(\bm{x})<0$ is referred to as the filtered region while the region $\mathcal{S}_+$ where $g(\bm{x})>0$ is called the augmented region.Next, define
\begin{subequations}
\begin{equation}
\pi_{p+}=\int_{\mathcal{S}_+}p(\bm{x})d\bm{x}
\label{eqn:pi1}
\end{equation}
\begin{equation}
\pi_{q+}=\int_{\mathcal{S}_+}q(\bm{x})d\bm{x}
\end{equation}
\begin{equation}
\pi_{p-}=\int_{\mathcal{S}_-}p(\bm{x})d\bm{x}
\end{equation}
\begin{equation}
\pi_{q-}=\int_{\mathcal{S}_-}q(\bm{x})d\bm{x}.
\label{eqn:pi4}
\end{equation}
\end{subequations}

According to this partitioning, the distributions $p(\bm{x})$ and $q(\bm{x})$ can be expressed as mixture distributions:
\begin{subequations}
\begin{equation}
p(\bm{x})=\pi_{p+}p_+(\bm{x})+\pi_{p-}p_-(\bm{x})
\end{equation}
\begin{equation}
q(\bm{x})=\pi_{q+}q_+(\bm{x})+\pi_{q-}q_-(\bm{x})
\label{eqn:mix_updated}
\end{equation}
\label{eqn:mix_total}
\end{subequations}
where 
\begin{subequations}
\begin{equation}
p_+(\bm{x})=\dfrac{p(\bm{x})}{\pi_{p+}},\hspace{3pt}\bm{x}\in\mathcal{S}_+
\end{equation}
\begin{equation}
p_-(\bm{x})=\dfrac{p(\bm{x})}{\pi_{p-}},\hspace{3pt}\bm{x}\in\mathcal{S}_-
\end{equation}
\begin{equation}
q_+(\bm{x})=\dfrac{q(\bm{x})}{\pi_{q+}},\hspace{3pt}\bm{x}\in\mathcal{S}_+
\end{equation}
\begin{equation}
q_-(\bm{x})=\dfrac{q(\bm{x})}{\pi_{q-}},\hspace{3pt}\bm{x}\in\mathcal{S}_-
\end{equation}
\end{subequations}

The general idea of the proposed approach is to augment the samples in the region $\mathcal{S}_+$ with new samples and filter the samples in region $\mathcal{S}_-$. According to the mixture distributions in Eq.\ \eqref{eqn:mix_total} and given that $\mathcal{S}_+$ and $\mathcal{S}_-$ are disjoint, we can sample independently over the support partitions $\mathcal{S}_+$ and $\mathcal{S}_-$. 

Let us focus on $\mathcal{S}_+$ first. The objective is to obtain a set of samples that follows $q_+(\bm{x})$ from Eq.\ \eqref{eqn:mix_updated}. Let us treat $q_+(\bm{x})$ as a mixture distribution defined by:
\begin{equation}
q_+(\bm{x})=\dfrac{\pi_{p+}}{\pi_{q+}}p_+(\bm{x})+\dfrac{\pi_{q+}-\pi_{p+}}{\pi_{q+}}q_c(\bm{x})
\label{eqn:mix1}
\end{equation}
where the ``correction'' distribution $q_c(\bm{x})$ is defined by:
\begin{equation}
q_c(\bm{x})=\dfrac{\pi_{q+}}{\pi_{q+}-\pi_{p+}}q_+(\bm{x})-\dfrac{\pi_{p+}}{\pi_{q+}-\pi_{p+}}p_+(\bm{x})
\label{eqn:correction}
\end{equation}
It is straightforward to show that Eqs.\ \eqref{eqn:mix1} and \eqref{eqn:correction} are consistent and that $q_c(\bm{x})$ is a valid pdf with $\int_{\mathcal{S}_+}q_c(\bm{x})d\bm{x}=1$.

Given a set of $N$ samples following $p(\bm{x})$, we have $N_+$ samples following $p_+(\bm{x})$ and $N_-$ samples following $p_-(\bm{x})$ with $N_++N_-=N$, $\dfrac{N_+}{N}\approx \pi_+$, and $\dfrac{N_-}{N}\approx \pi_-$. Over $\mathcal{S}_+$, we propose to add $N_{a+}$ samples from $q_c(\bm{x})$ such that the combined distribution of the $N_++N_{a+}$ samples follows $q_+(\bm{x})$. According to the mixture distribution in Eq.\ \eqref{eqn:mix1}, this requires on average $\dfrac{\pi_{p+}}{\pi_{q+}}(N_++N_{a+})$ samples from $p(\bm{x})$
and $\dfrac{\pi_{q+}-\pi_{p+}}{\pi_{q+}}(N_++N_{a+})$ from $q_c(\bm{x})$. Therefore, the required number of added samples is given by
\begin{equation}
N_{a+}=\dfrac{\pi_{q+}}{\pi_{p+}}N_+-N_+=\left(\dfrac{\pi_{q+}}{\pi_{p+}}-1\right)N_+=(\pi_{q+}-\pi_{p+})N
\label{eqn:num_add}
\end{equation}
It follows directly from the proof of the composition method that the augmented samples set follows $q_+(\bm{x})$. Moreover, it is interesting to note that the number of new samples required to correct the distribution scales as half the total variation distance between the original and the updated densities given by:
\begin{equation}
d_1=\int|p(\bm{x})-q(\bm{x})|d\bm{x}=2(\pi_{q+}-\pi_{p+}).
\label{eqn:TVD}
\end{equation}
That is, $N_{a+}=\frac{d_1}{2}N_+$. Proof of Eq.\ \eqref{eqn:TVD} can be found in Appendix A.

Next, consider the range of support $\mathcal{S}_-$ for $p_-(\bm{x})$ (i.e. where $p(\bm{x})>q(\bm{x})$). The objective is to obtain a set of samples that follows $q_-(\bm{x})$ from the mixture model in Eq.\ \eqref{eqn:mix_updated}. Given a set of $N$ samples following $p(\bm{x})$, again we have $N_+$ samples following $p_+(\bm{x})$ and $N_-$ samples following $p_-(\bm{x})$ with $N_++N_-=N$, $\dfrac{N_+}{N}\approx \pi_+$, and $\dfrac{N_-}{N}\approx \pi_-$. 

Let us define $t_-(\bm{x})=\dfrac{\pi_{p-}}{\pi_{q-}}p_-(\bm{x})$ as a majorizing function for $q_-(\bm{x})$ and sample according to the A/R method. Note that $p(\bm{x})>q(\bm{x}),\hspace{3pt}\forall \bm{x}\in\mathcal{S}_-\Rightarrow \dfrac{\pi_{p-}}{\pi_{q-}}p_-(\bm{x})>q_-(\bm{x})$ -- proof of which is straightforward. The A/R method proceeds as follows. Each of the $N_-$ samples from $p_-(\bm{x})$ is accepted with probability $\dfrac{q_-(\bm{x})}{t_-(\bm{x})}$. On average, this will yield a set of $N_{f-}=\dfrac{\pi_{q-}}{\pi_{p-}}N_-=\pi_{q-}N$ samples. Note that the number of rejected samples is given by
\begin{equation}
N_{\text{reject}}=N_--N_{f-}=(\pi_{p-}-\pi_{q-})N=(\pi_{q+}-\pi_{q-})N=N_{a+}. 
\label{eqn:num_reject}
\end{equation}
That is, the same number of samples are rejected in region $\mathcal{S}_-$ as are added in $\mathcal{S}_+$ (with that number being proportional to the total variation distance), which keeps the total sample size constant on average.

The proposed method has been proven to modify a sample set to follow the $q_+(\bm{x})$ over the range $\mathcal{S}_+$ by adding samples according to a carefully defined ``correction'' (see Eq.\ \eqref{eqn:correction}) and to follow $q_-(\bm{x})$ over the range $\mathcal{S}_-$ by filtering the existing sample set according to the A/R method with a carefully defined majorizing function. Moreover, the samples from $q_+(\bm{x})$ and $q_-(\bm{x})$ are proven to satisfy the correct proportions so as to follow the mixture distribution given by Eq.\ \eqref{eqn:mix_updated}. It therefore follows that the combined samples set follows the updated distribution $q(\bm{x})$. 

\subsection{Discussion of efficiency and degeneracy}

The method proposed in Section \ref{Sec:AandF} has considerable advantages over the methods in Sections \ref{Sec:IS_Reweight} - \ref{Sec:Filter}. It eliminates the degeneracy problems present in all of the other methods, dramatically reduces the number of additional simulations required when compared to the purely augmented sample sets presented in Section \ref{Sec:Augment}, and dramatically reduces the number of rejected samples when compared to the purely filtered samples sets presented in Section \ref{Sec:Filter}. 

Consider first the degeneracy issue. Using the method proposed in Section \ref{Sec:AandF}, there are no restrictions on the support of $p(\bm{x})$ and $q(\bm{x})$ as there are in the other methods. In the augmented region $\mathcal{S}_+$, where $q(\bm{x})\ge p(\bm{x})$, the correction distribution is simply equal to a scaled version of $q_+(\bm{x})$ when $p(\bm{x})=0$ and, because $q(\bm{x})\ge p(\bm{x})$, $q_c(\bm{x})=0$ when $q_+(\bm{x})=0$. Moreover, the number of added samples in $\mathcal{S}_+$ is shown in Eq.\ \eqref{eqn:num_add} is stable (proportional to the total variation distance) and cannot cause the sample size to more than double. That is, in the most extreme case where $p(\bm{x})$ and $q(\bm{x})$ have disjoint support, $N_{a+}=N$. This is in contrast to the purely augmented case where the number of added samples depends on the ratio $\max{\left\{\dfrac{p(\bm{x})}{q(\bm{x})}\right\}}$, which goes to infinity under certain conditions (discussed in Section \ref{Sec:Augment}). 

In the filtered region, $\mathcal{S}_-$ where $q(\bm{x})< p(\bm{x})$, again the method proposed in Section \ref{Sec:AandF} places no restrictions on the support of $p(\bm{x})$ and $q(\bm{x})$. When $p_-(\bm{x})=0$, $q_-(\bm{x})$ must also be zero and, when $q_-(\bm{x})=0$, the acceptance probability $\dfrac{q_-(\bm{x})}{t_-(\bm{x})}=0$. Also, the number of rejected samples in $\mathcal{S}_-$ given in Eq.\ \eqref{eqn:num_reject} is stable (again proportional to the total variation distance) and equal to the number of samples added in the augmented region. In contrast, the acceptance rate for the purely filtered method in Section \ref{Sec:Filter} converges to zero under certain conditions. 

Finally, as will be shown in the following Section, the method proposed in Section \ref{Sec:Augment} proves far more efficient than the other methods in that it requires the addition of fewer samples than the purely augmented case and rejects fewer samples than the purely filtered case. This is directly related to the nature of the scaling. The number of added/rejected samples in both the purely augmented and purely filtered scale with the maximum ratio of the densities, which can be very large. The method proposed in Section \ref{Sec:Augment}, however, scales with half the total variation distance between the densities which is bounded on $[0,1]$. Therfore, as previously stated, the method will never add more than $N$ samples and it keeps the same size constant.

Of course, if the support of the distributions allows it and the effective sample size is satisfactorily large, the Importance Sampling reweighting option affords the benefit that no new simulations are necessary. However, if the support of the distributions are different or if the penalty in effective sample size is not acceptable, the proposed method provides an efficient alternative that often requires only a small number of correction samples. 

These approaches are explored in further detail in the following section.

%%%%%%%%%%%%%%%%%%%
%             Numerical illustration         %
%%%%%%%%%%%%%%%%%%%

\section{Numerical illustrations}

In this section, we illustrate the performance of these four strategies for two support relationships between the original distribution and the updated distribution. The first section considers distributions with common support and the second considers distributions with changing support. The benefits and limitations of each proposed strategy are compared and discussed for representative numerical examples. 
 
\subsection{Distributions with common support}

\begin{figure}[!ht]	
	\centering
	\includegraphics[height=2in]{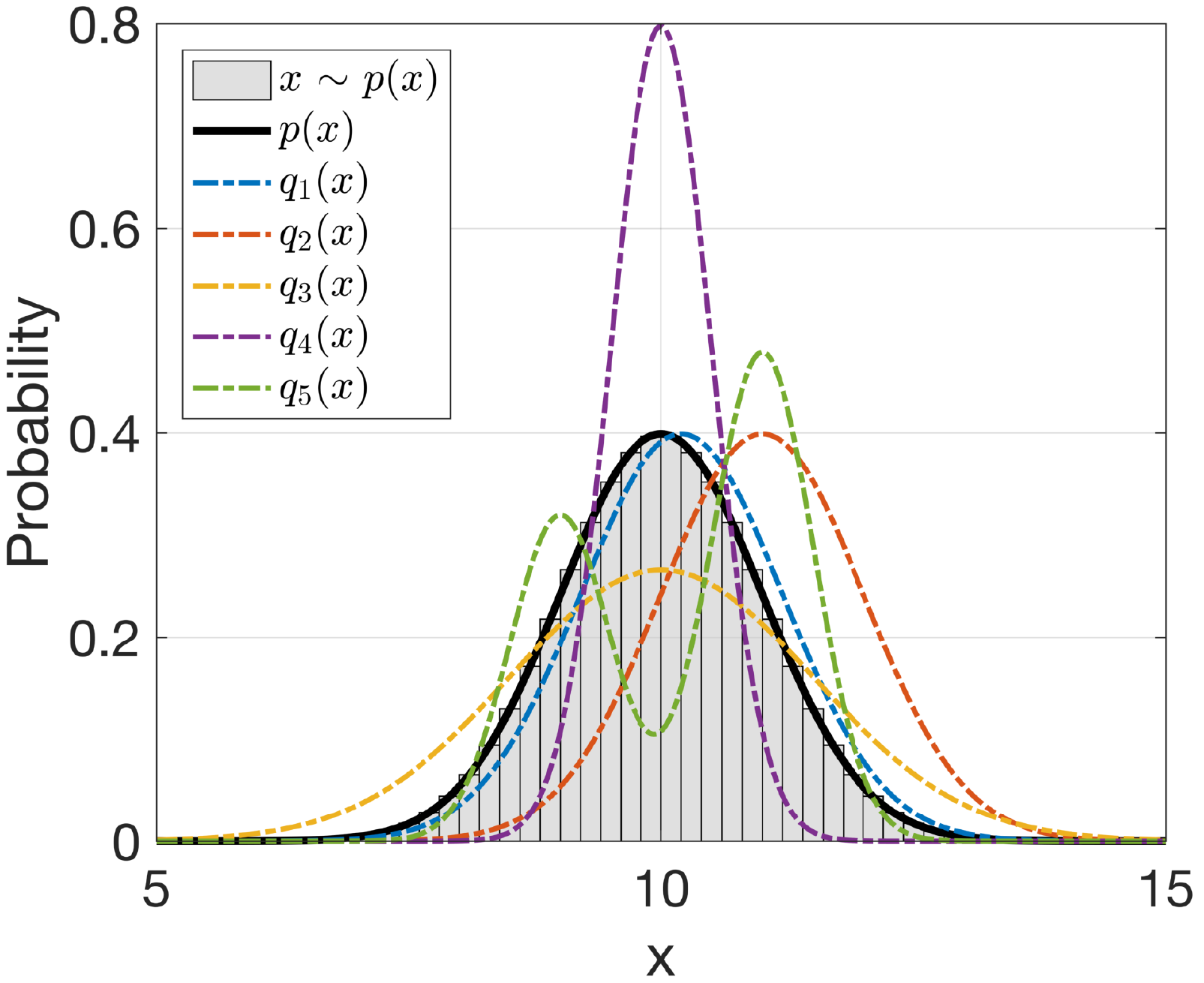}
	\caption[]{Five updated probability densities with common support}  \label{fig:total_five}
\end{figure}

Consider $10,000$ random samples $\bm{x}$ drawn from an original probability density $p(\bm{x})\sim N(10,1)$. Here, we explore the performance of the various sample adjustment strategies for five updated probability densities with different shapes and locations but identical support $\mathcal{S}=(-\infty,\infty)$, as follows and illustrated in Fig.\ref{fig:total_five}. The objective, in each case, is to maintain a Monte Carlo sample size as close as possible to the original 10,000 samples with minimal additional samples.

%Normal distribution N(10,1) is assumed as the original sampling density $p(\bm{x})$, and we draw $10,000$ random samples $\bm{x}$ from $p(\bm{x})$. Various updated sampling densities are considered with different shapes and locations as follows and illustrated in Fig.\ref{fig:common}. 

\begin{itemize}
\item Case 1: Normal distribution with a small shift -- $q_1(\bm{x}) = N(10.2, 1)$
\item Case 2: Normal distribution with a larger shift -- $q_2(\bm{x}) = N(11, 1)$
\item Case 3: Wider normal distribution -- $q_3(\bm{x}) = N(10, 1.5)$
\item Case 4: Narrower normal distribution -- $q_4(\bm{x})  = N(10, 0.5)$
\item Case 5: Multimodal mixture distribution -- $q_5(\bm{x})  = 0.4 \times N(9,0.5) + 0.6 \times N(11,0.5) $
\end{itemize} 

The results for each sample strategy applied to all five cases are summarized in Table \ref{tab:summary1} and a discussion of each case follows.
\begin{table}[!ht] \footnotesize
\centering
\caption{Comparison of four strategies for five cases of different updated probability densities.}
\label{tab:summary1}
\resizebox{\textwidth}{!}{
\begin{tabular}{cccccccccc}
\hline
     &          & IS   & \multicolumn{2}{c}{Augmenting}            & \multicolumn{2}{c}{Filtering} & \multicolumn{2}{c}{Mixed} \\ \hline
Case &          & ESS  & $N_a$               & $N^*$               & $N_{reject}$         & $N^*$         & $N_{a+}$     & $N^*$     \\
1    & $q_1(x)$ & 9608 & 12214               & 22214               & 5322          & 4678          & 796           & 10000     \\
2    & $q_2(x)$ & 3717 & 796882              & 806882              & 9660          & 340           & 3812          & 10000     \\
3    & $q_3(x)$ & 4626 & 5000                & 15000               & 9784          & 216           & 1843          & 10000     \\
4    & $q_4(x)$ & 6614 & 3.63E+13 ($\infty$) & 3.63E+13 ($\infty$) & 5001          & 4999          & 3227          & 10000     \\
5    & $q_5(x)$ & 7115 & 1.17E+8 ($\infty$)  & 1.17E+8 ($\infty$)  & 5731          & 4269          & 2735          & 10000     \\ \hline
\end{tabular}}
\end{table}

\begin{figure}[!ht]	
 	\centering
 	\subfigure[]{\includegraphics[height=1.6in]{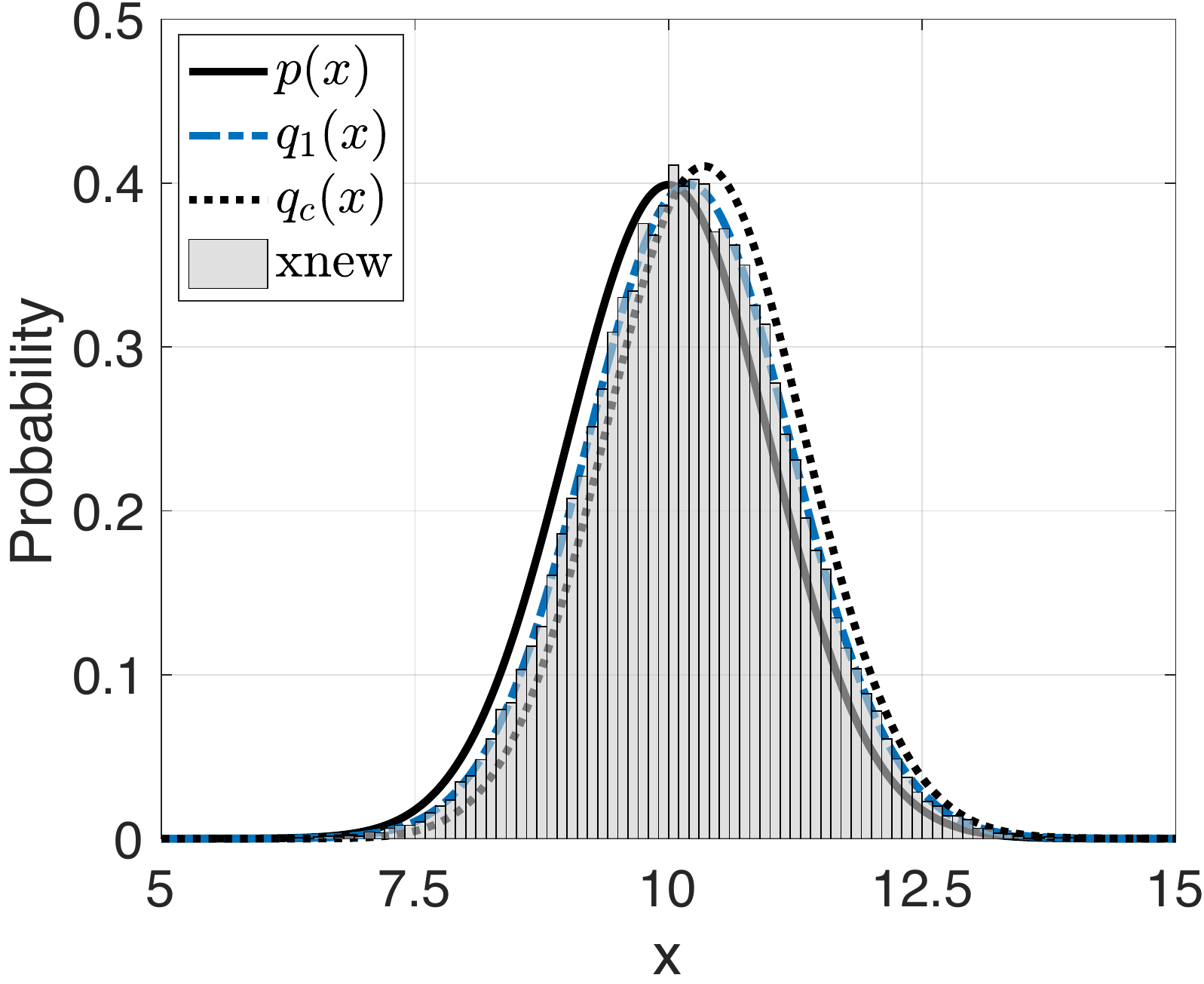}}
 	\subfigure[]{\includegraphics[height=1.6in]{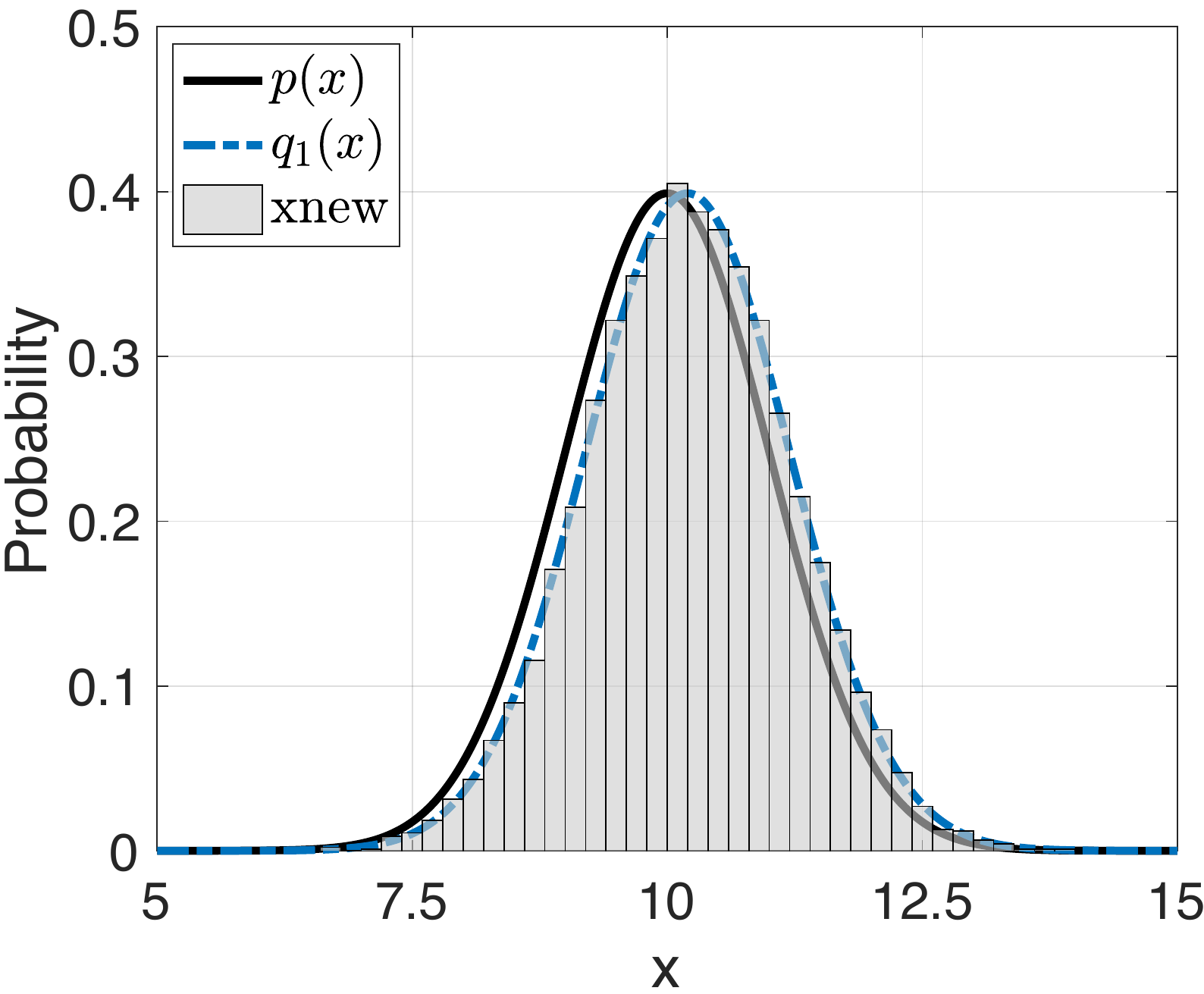}}
 	\subfigure[]{\includegraphics[height=1.6in]{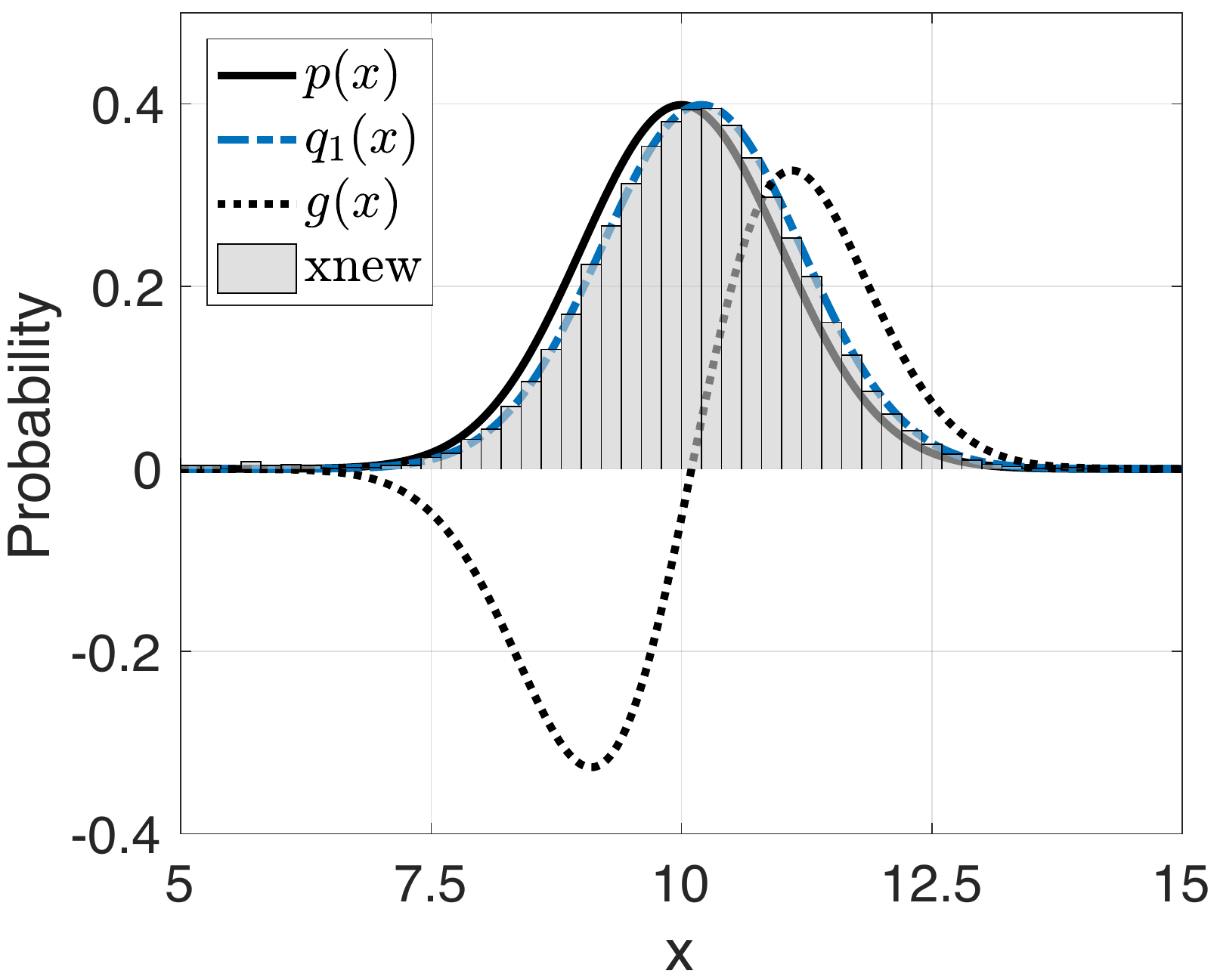}}
 	\caption[]{Case 1 - Resampling strategy for an updated distribution with a small shift: (a) augmenting (b) filtering and (c) mixed augmenting and filtering.}  \label{fig:case1} 
 \end{figure}

\noindent{\bf Case 1: Normal distribution with a small shift} -- In this case, the difference between the original and updated distributions is very small. Consequently, importance sampling is effective for reweighting the original samples $\bm{x}$ generated from $p(\bm{x})$. The corresponding effective sample size $\hat{N}_{ESS} = 9608$ remains quite large. A purely augmenting strategy requires a large number of additional samples, $N_{a} = 12214$, to be drawn from the correction distribution $q_c(\bm{x})$ (black dashed line in Figure \ref{fig:case1}a). As a result, the total number of samples $N^*$ is more than double the number of original samples. This is not considered a viable option. A purely filtering strategy, on the other hand, does not come at any additional computational cost, but this strategy eliminates $N_{reject} = 5322$ samples, or approximately half the original samples. Although the reduced sample set fits the updated probability density well, as shown in Figure \ref{fig:case1}b, the smaller sample size reduces the accuracy of statistical estimates from Monte Carlo simulation appreciably. Figure  \ref{fig:case1}c presents the mixed augmenting and filtering strategy where again $g(\bm{x})$ illustrates the separation between the augmented and filtered regions. In this case, only an additional $N_{a+} = 796$ samples are required to maintain a sample size of $N^*=10,000$ and the samples match the updated distribution accurately. 

\noindent{\it Best Option: Importance Sampling Reweighting} -- Although augmenting and filtering requires only a relatively small number of new samples ($N_{a+}=796$), importance sampling reweighting requires no new calculations and maintains a large effective sample size ($>9600$ samples). 
 
\begin{figure}[!ht]	
 	\centering
 	\subfigure[]{\includegraphics[height=1.6in]{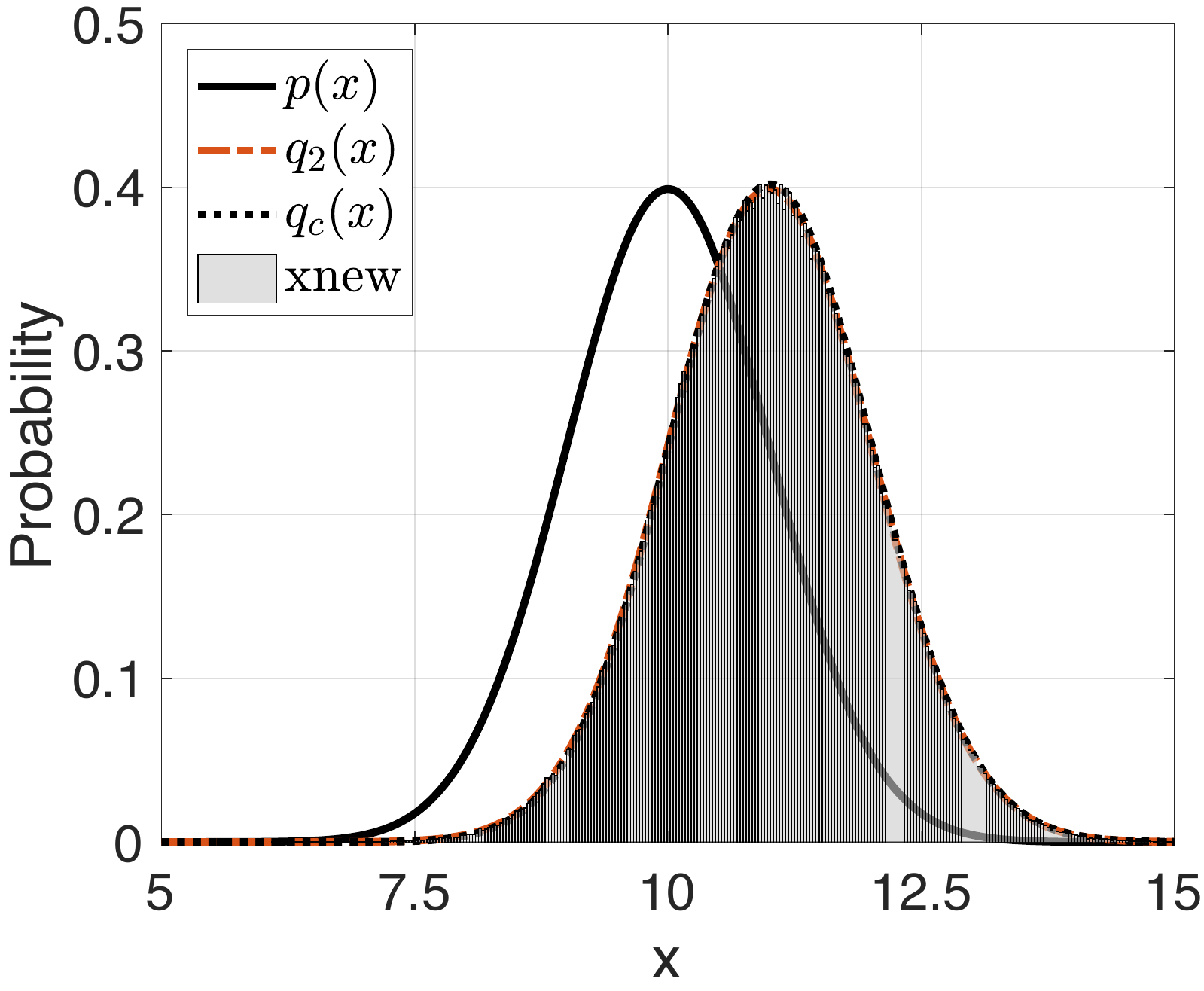}}
 	\subfigure[]{\includegraphics[height=1.6in]{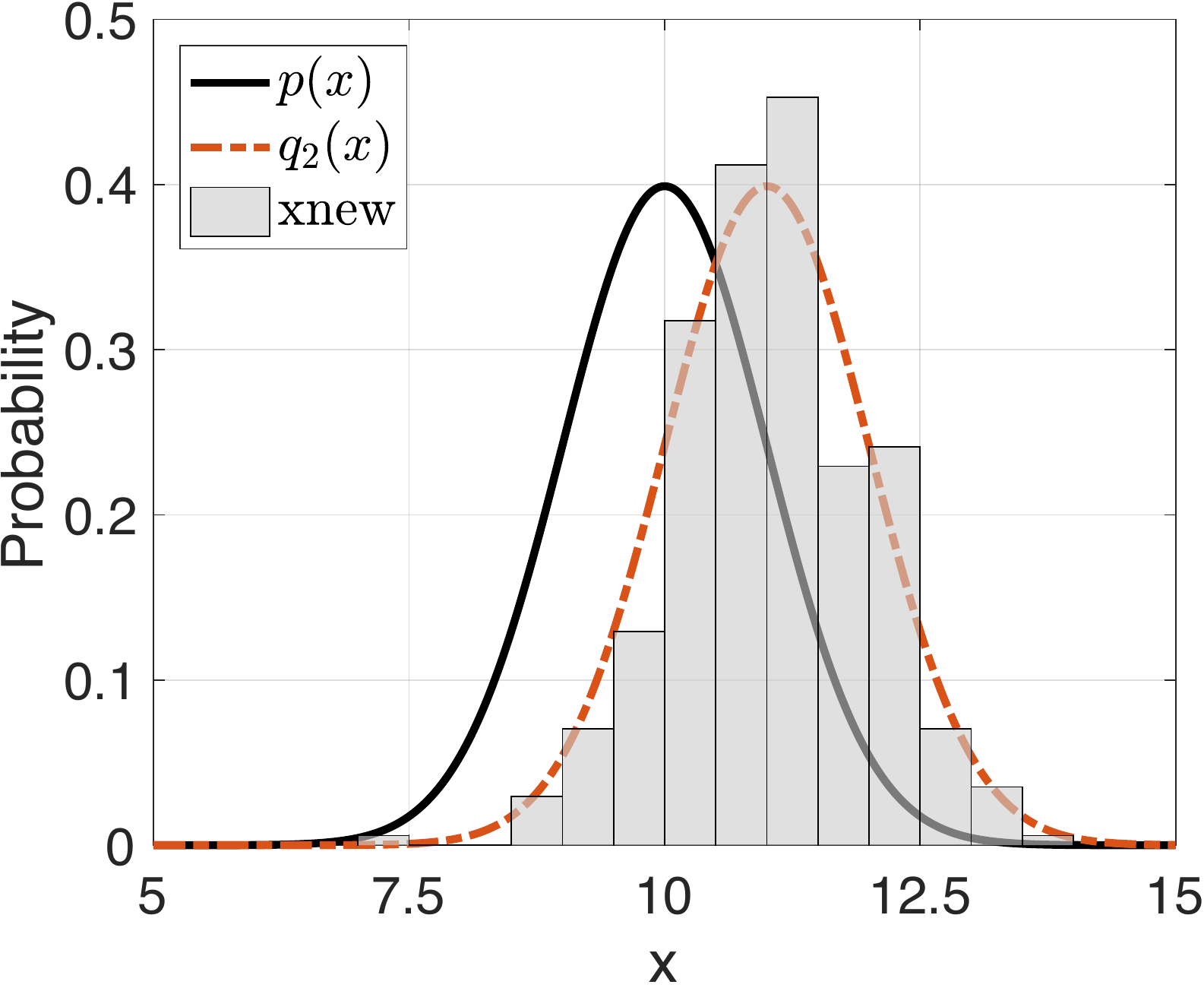}}
 	\subfigure[]{\includegraphics[height=1.6in]{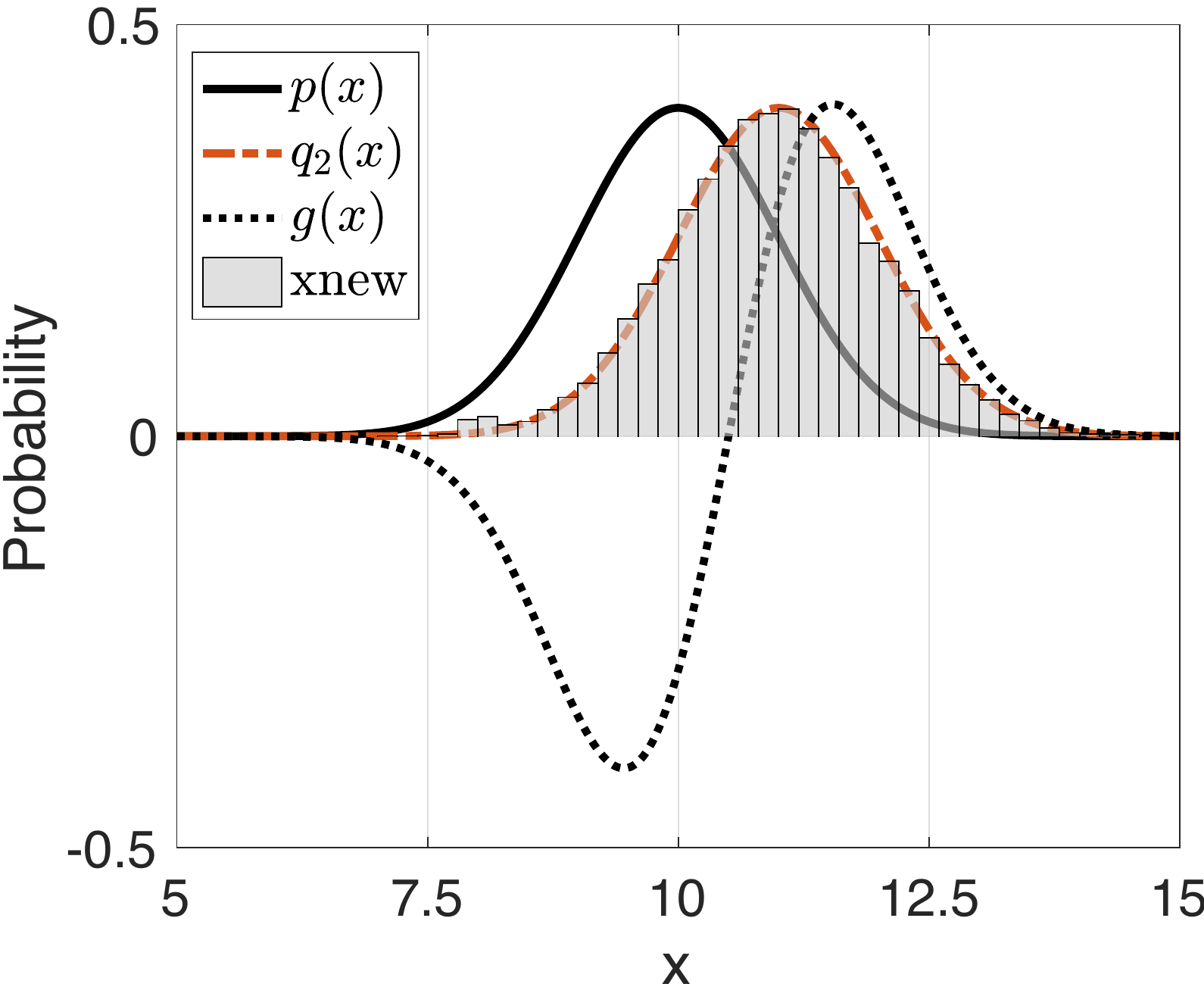}}
 	\caption[]{Case 2 - Resampling strategy for an updated distribution with a large shift: (a) augmenting (b) filtering and (c) mixed augmenting and filtering.} \label{fig:case2}
 \end{figure}

\noindent{\bf Case 2: Normal distribution with a large shift} -- A large shift between $p(\bm{x})$ and $q_2(\bm{x})$ leads to large variation in importance weights which yields a poor importance sampling estimator with effective sample size only $\hat{N}_{ESS} = 3713$ (Table \ref{tab:summary1}). The performance of the other three strategies are shown in Figure \ref{fig:case2}. The number of samples generated from the augmenting strategy (Figure \ref{fig:case2}a), $N_a=796,882$ (Table \ref{tab:summary1}) is unacceptably large. The filtering strategy, meanwhile, removes more than 96$\%$ of the original samples and retains only 340 samples, as shown in Figure. \ref{fig:case2}b, which isn't nearly enough for Monte Carlo simulation. The mixed augmenting and filtering strategy, as shown in Figure \ref{fig:case2}c requires $N_{a+}=3812$ to maintain the full 10,000 sample size, and while it is not ideal to perform an additional 3800 calculations, it is preferable to the other options.

\noindent {\it Best Option: Mixed Augmenting and Filtering}
 
\begin{figure}[!ht]	
 	\centering
 	\subfigure[]{\includegraphics[height=1.6in]{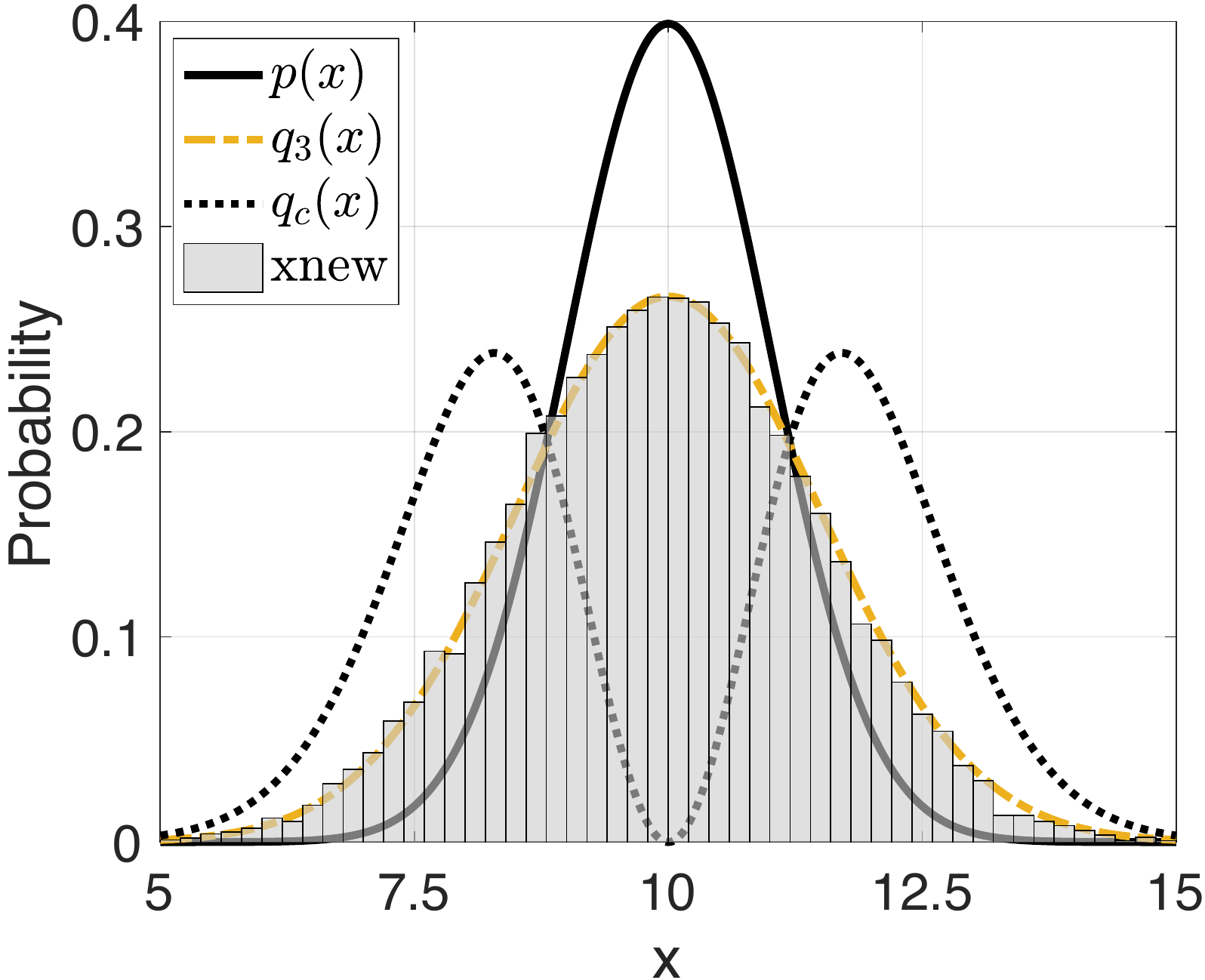}}
 	\subfigure[]{\includegraphics[height=1.6in]{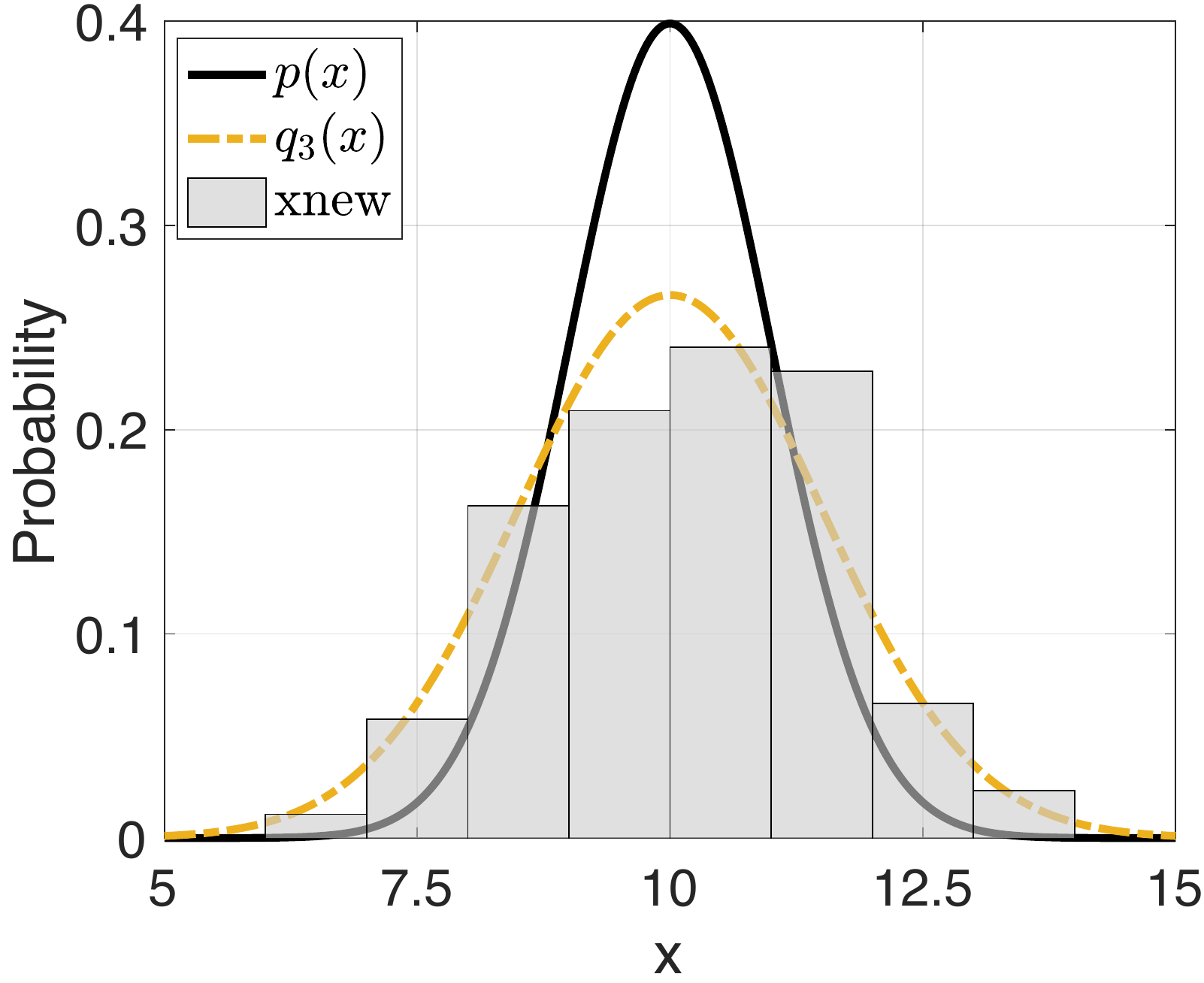}}
 	\subfigure[]{\includegraphics[height=1.6in]{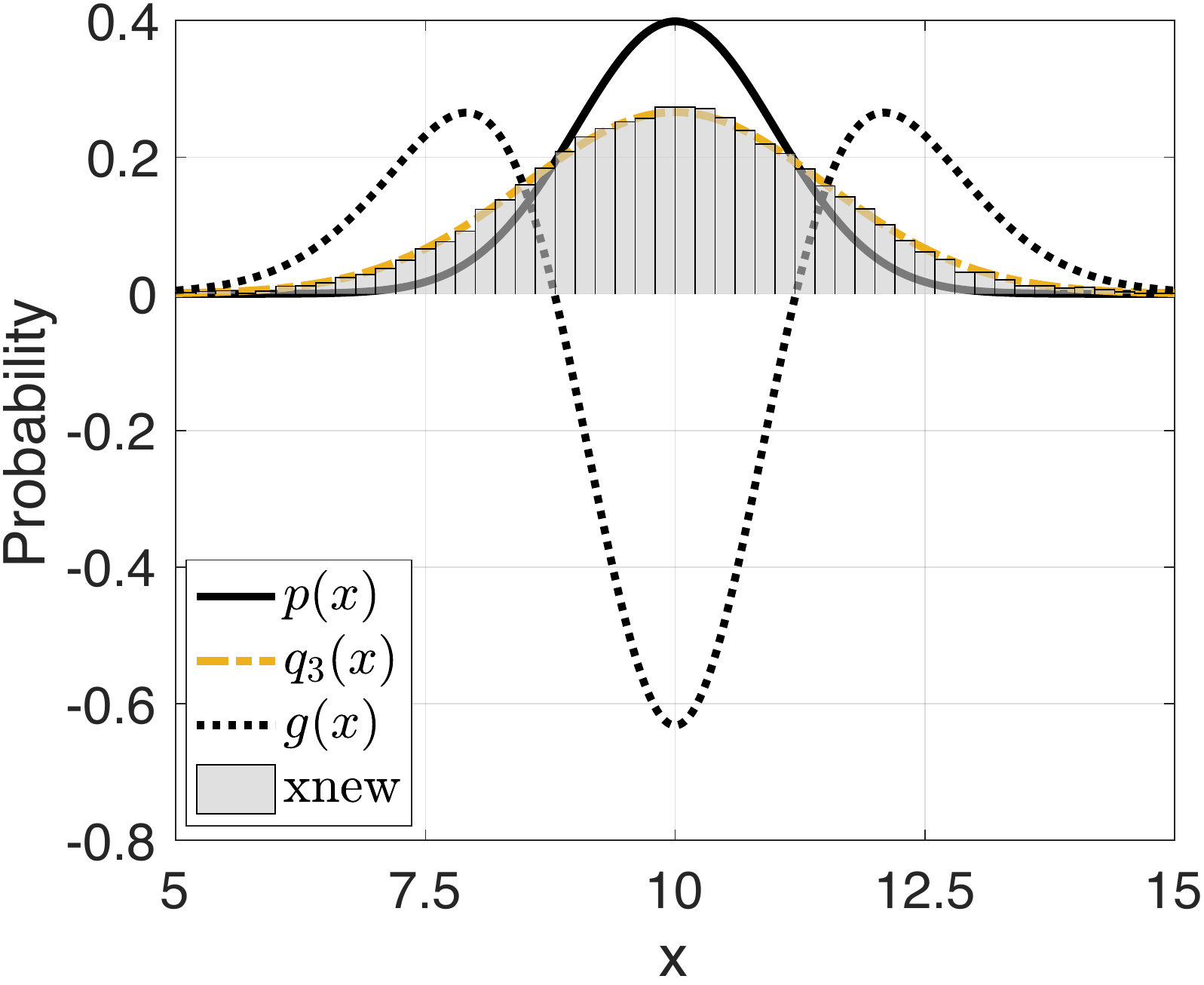}}
 	\caption[]{Case 3 - Resampling strategy for a wider updated distribution: (a) augmenting (b) filtering and (c) mixed augmenting and filtering.}  \label{fig:case3}
 \end{figure}

\noindent{\bf Case 3: Wider distribution} -- Significantly widening the distribution results in undersampling the tails of the updated distribution. This will cause importance sampling to place large weight on samples in the tails of the original distribution. Consequently, the importance sampling weights have high variance, which yields a poor estimator with small effective sample size, $\hat{N}_{ESS} = 4626$ (Table \ref{tab:summary1}). Regarding the augmenting strategy, many original samples can be effectively retained such that new samples only need to be added in the tails, as the correction distribution $q_c(\bm{x})$ in Figure \ref{fig:case3}a shows. Nonetheless, $N_a=5000$, which is still quite large. The filtering strategy, meanwhile, retains only 2$\%$ of the original samples leaving only $N^*=216$ samples as illustrated in Figure \ref{fig:case3}b. The mixed strategy is much more efficient as it adds/filters $N_{a+}=1843$ samples (Figure \ref{fig:case3}c and Table \ref{tab:summary1}), which is less than 20\% of the original number, in order to maintain a sample size of 10,000. 

\noindent {\it Best Option: Mixed Augmenting and Filtering}
 
\begin{figure}[!ht]	
 	\centering
 	%\subfigure[]{\includegraphics[height=1.6in]{narrower_augmenting.pdf}}
 	\subfigure[]{\includegraphics[height=1.6in]{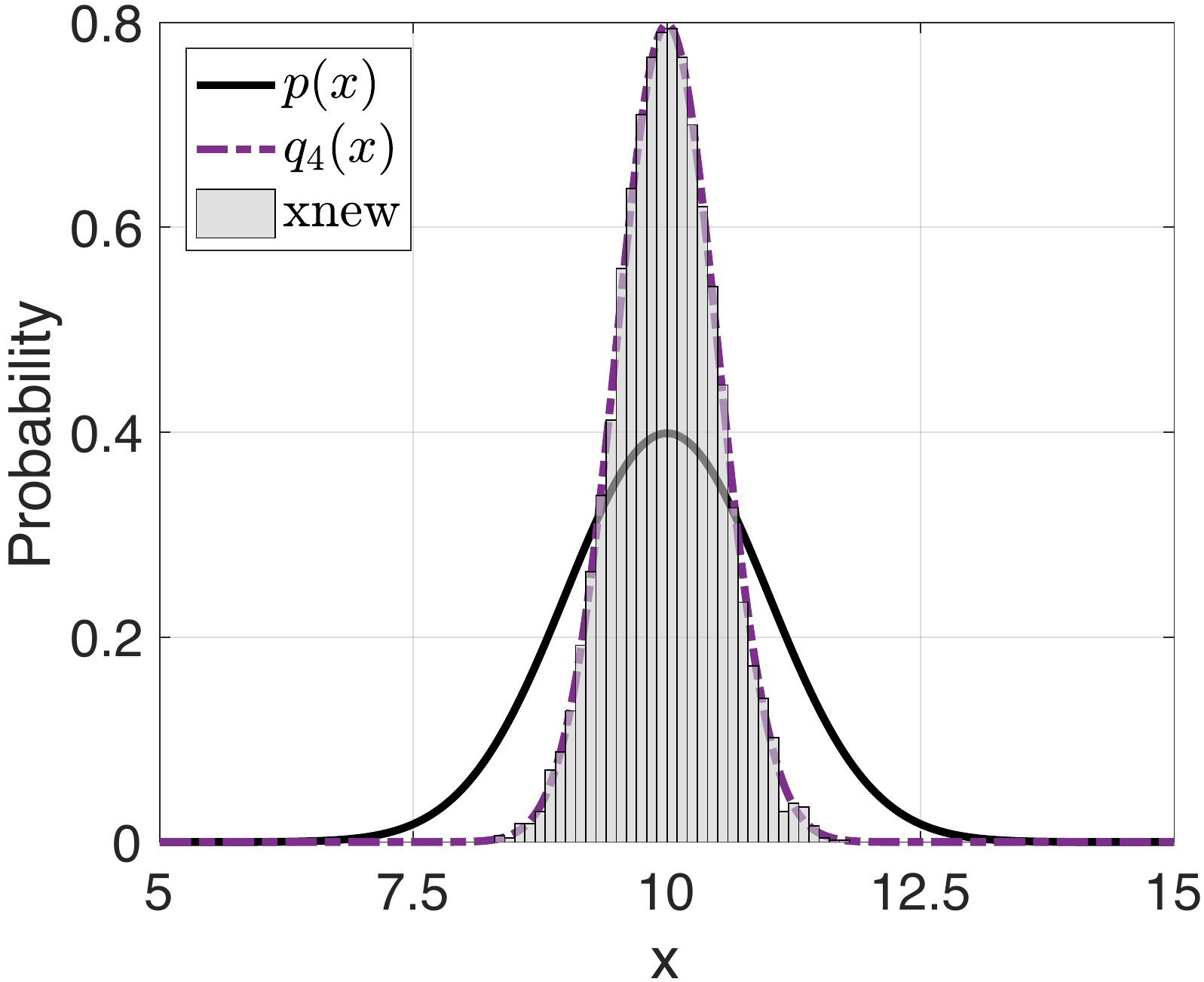}}
 	\subfigure[]{\includegraphics[height=1.6in]{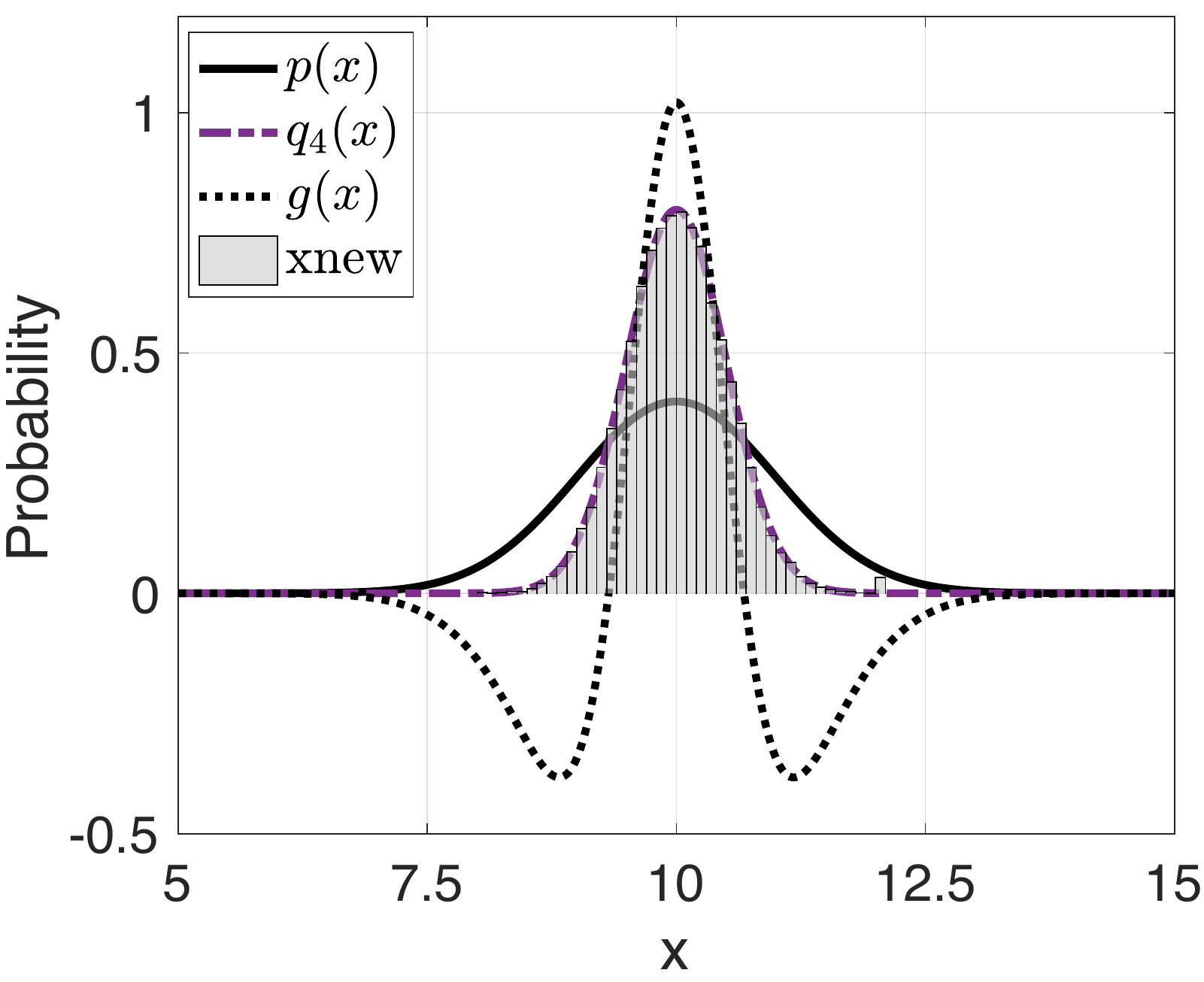}}
 	\caption[]{Case 4 - Resampling strategy for a narrower updated distribution: (a) filtering and (b) mixed augmenting and filtering.}  \label{fig:case4}
 \end{figure}

\noindent{\bf Case 4: Narrower distribution} -- Narrowing the distribution results in the tails of the updated distribution being oversampled. For importance sampling, this is better than undersampling the tails; it more evenly distributes the sample weights resulting in an effective sample size $\hat{N}_{ESS} = 6614$. While larger than Cases 2 and 3, it is still is only $\sim 2/3$ of the total number of samples and will result in a fair, but not great, estimator. It is easy to show that the augmenting strategy requires infinite additional samples since $\lim\limits_{\bm{x}\to\pm\infty}\dfrac{p(\bm{x})}{q(\bm{x})}=\infty$. This is reflected in Table \ref{tab:summary1} as $N_+=3.63E+13$, which is finite only due to numerical discretization of $p(\bm{x})$ and $q(\bm{x})$ in implementation. It is therefore, not a viable option. The filtering strategy, as shown in Figure \ref{fig:case4}a remains viable as $\max\left\{\dfrac{q(\bm{x})}{p(\bm{x})}\right\}$ is bounded and retains approximately half of the original samples, thus performing slightly better than Cases 2 and 3.  The mixed strategy, by comparison as shown in Fig \ref{fig:case4}b, requires 3227 new samples to maintain the sample size $N^*=10000$. 

\noindent {\it Best Option: Mixed augmenting and filtering OR Importance sampling reweighting} -- If $\sim3000$ additional simulations are affordable, the mixed strategy is preferred. If these additional simulations are not affordable, importance sampling reweighting is appropriate.

\begin{figure}[!ht]	
 	\centering
 	%\subfigure[]{\includegraphics[height=1.6in]{narrower_augmenting.pdf}}
 	\subfigure[]{\includegraphics[height=1.6in]{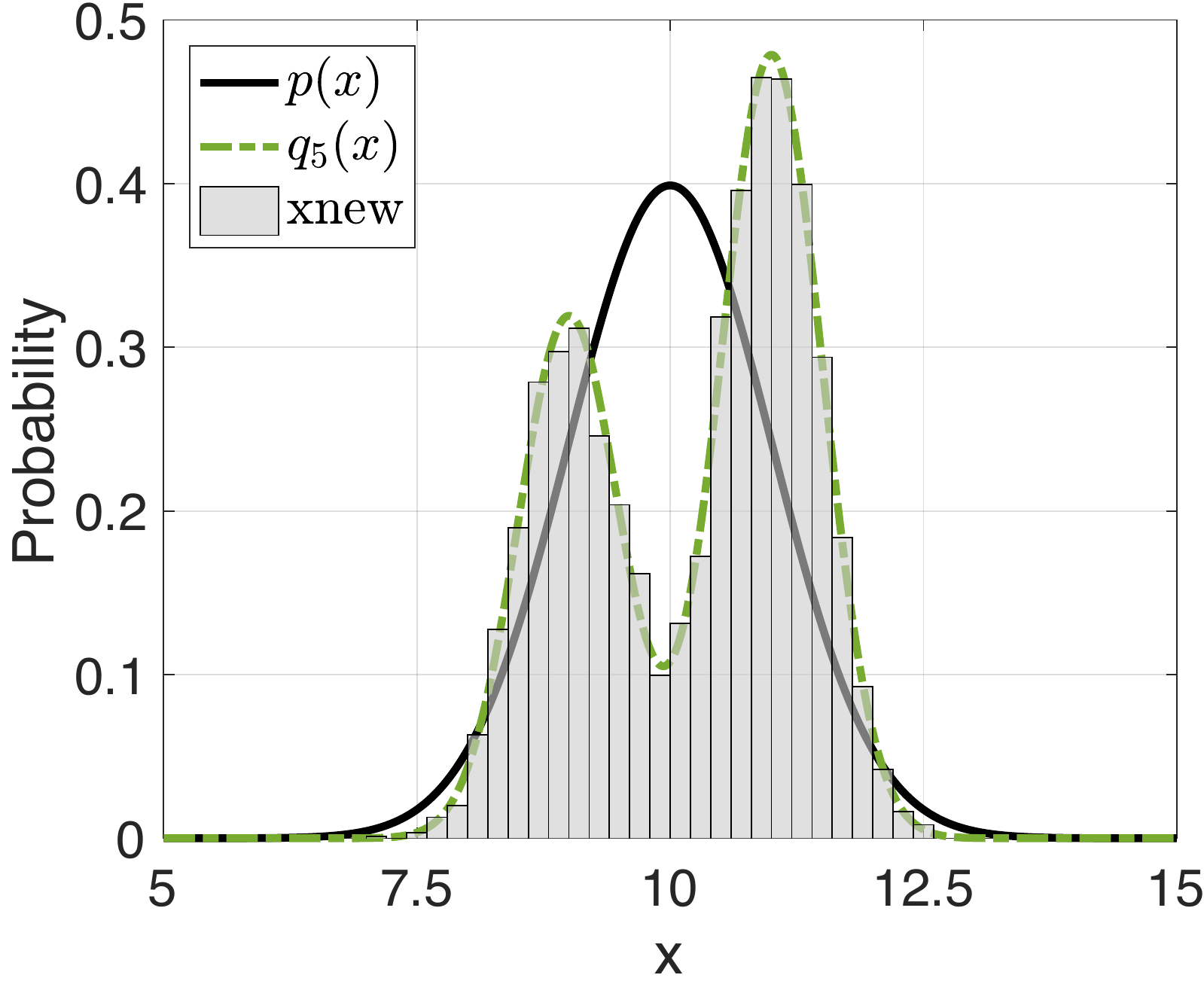}}
 	\subfigure[]{\includegraphics[height=1.6in]{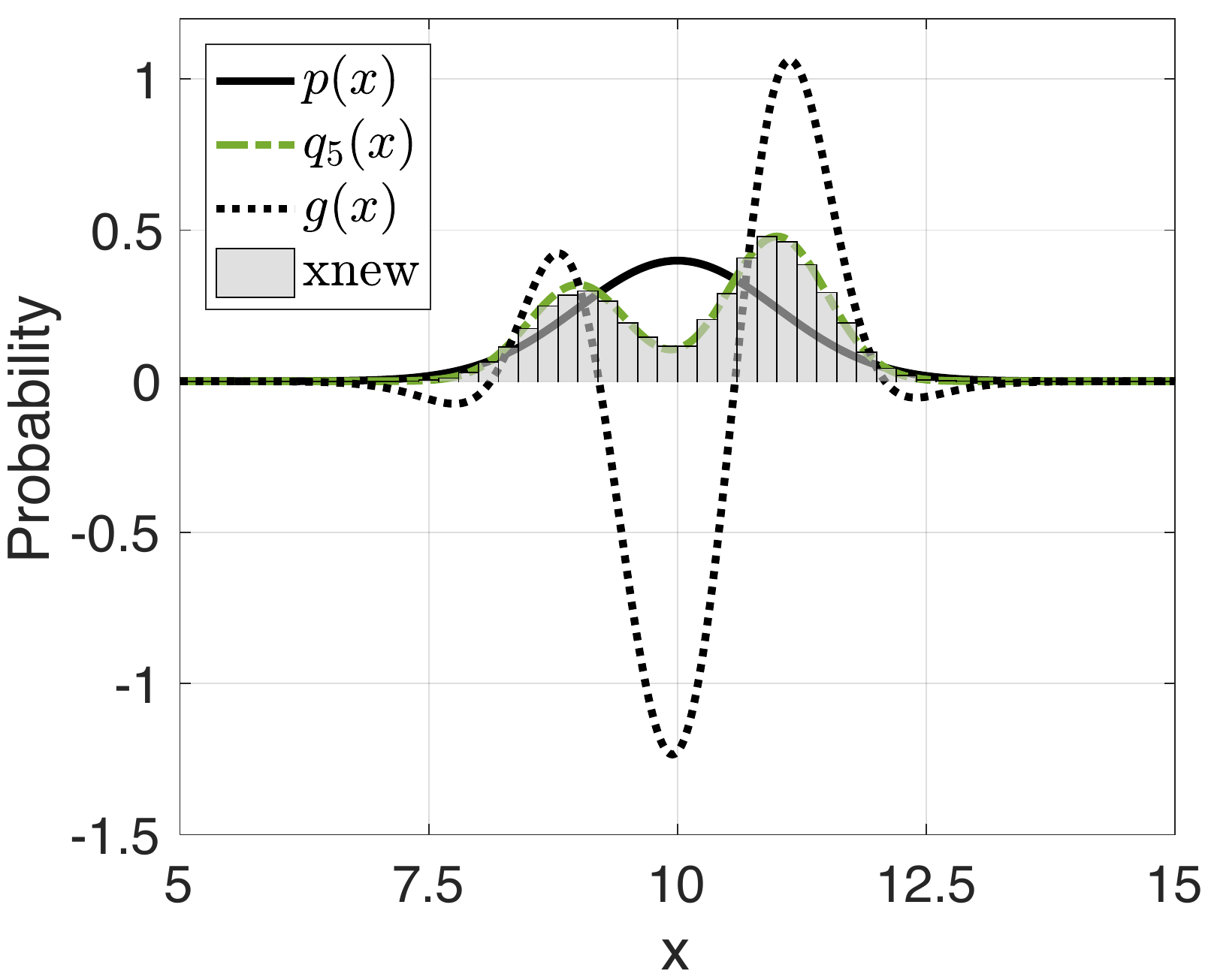}}
 	\caption[]{Case 5 - Resampling strategy for a multimodel updated distribution: (a) filtering and (b) mixed augmenting and filtering.}  \label{fig:case5}
 \end{figure}

\noindent{\bf Case 5: Unimodal to multimodal distribution} --  This case represents an extreme where the distribution radically changes form. In this specific example (Figure \ref{fig:case5}), the multimodal updated distribution has tails that are slightly oversampled by the original distribution making it similar to Case 4. But also the central region of the distribution is oversampled by the original distribution similar to Case 3. These two facts govern the importance sampling reweighting yielding an effective sample size $N_{ESS}=7115$. Because it is similar in the tails to Case 4, again it is straightforward to show that $\lim\limits_{\bm{x}\to\pm\infty}\dfrac{p(\bm{x})}{q(\bm{x})}=\infty$, thus requiring infinite additional samples to match the updated distribution according to the augmenting strategy. This is reflected in Table \ref{tab:summary1} by $N_a=1.17E8$ samples which, again, results from numerical discretization of the pdf. The filtering strategy, as shown in Figure \ref{fig:case5}a, retains only $\sim40\%$ of the original samples. For the mixed strategy, shown in Figure \ref{fig:case5}b, this example illustrates that the regions $\mathcal{S}_+$ and $\mathcal{S}_-$ may be complex. Here, there are filtering regions ($\mathcal{S}_-$ with $g(\bm{x}<0$)) in both tails and the center of the distribution separated by augmenting regions ($\mathcal{S}_+$ with $g(\bm{x})>0$). The resulting number of additional calculations $N_{a+}=2735$ is reasonable in comparison to the other strategies.

\noindent {\it Best Option: Mixed augmenting and filtering OR Importance sampling reweighting} -- If $\sim3000$ additional simulations are affordable, the mixed strategy is preferred. If these additional simulations are not affordable, importance sampling reweighting is appropriate.

%%% another example
\subsection{Distributions with changing support}

When the support changes from the original distribution to the updated distribution the resampling process may be more complicated. Here, we explore several such examples based on the following three general classes of support bounds: 
\begin{itemize}
\item {\it Infinite Support}: $\mathcal{S}_1=(-\infty,\infty)$, e.g. Normal distribution. 
\item {\it Bounded support}: $\mathcal{S}_2=[a,b]$, e.g. Beta distribution. 
\item {\it Semi-infinite Support}: $\mathcal{S}_3=[a,\infty)$ or $\mathcal{S}_3=(-\infty,b]$, e.g. Lognormal distribution.
\end{itemize}

\begin{figure}[!ht]	
	\centering
	\includegraphics[height=2in]{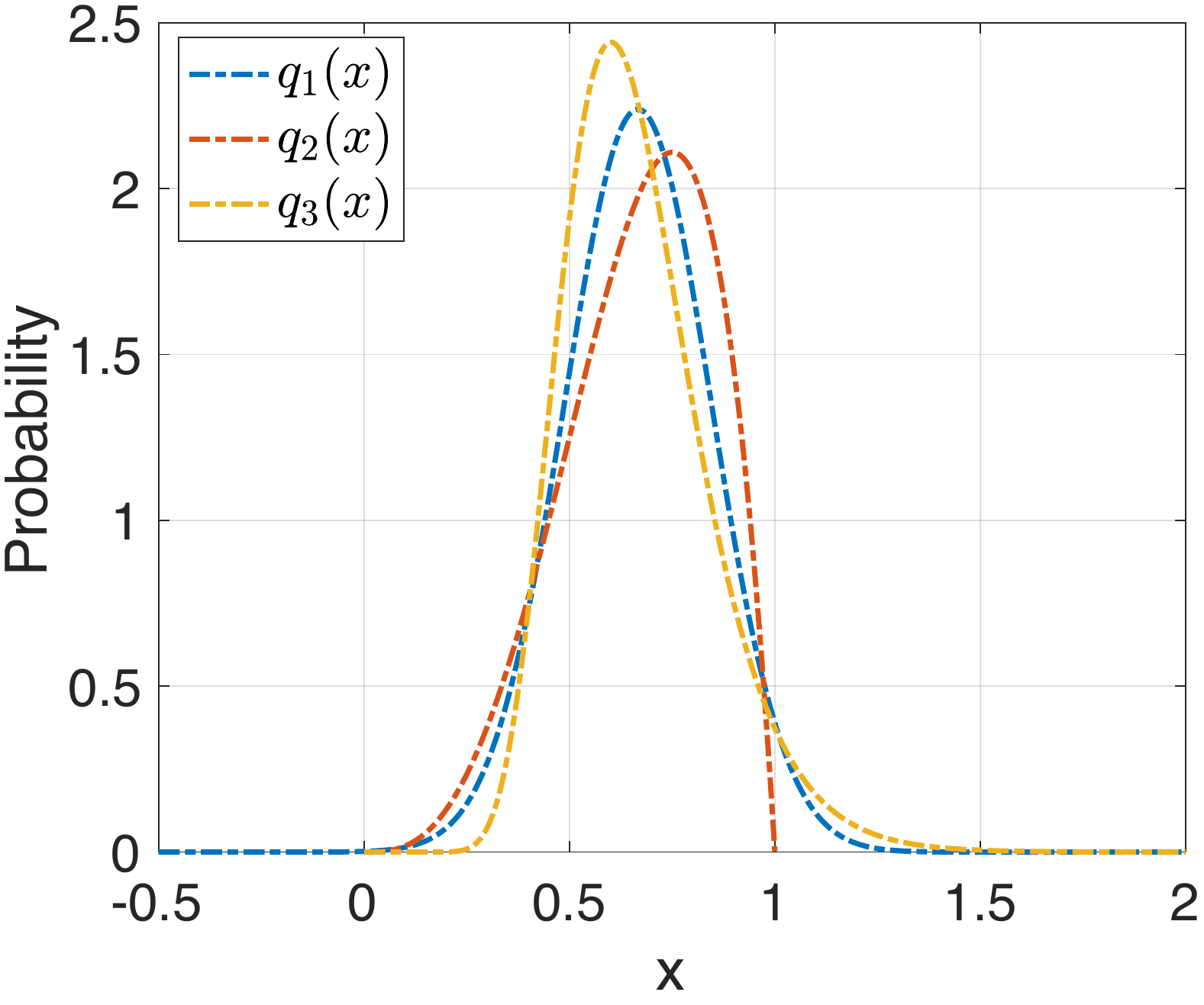}
	\caption[]{Three representative distributions with different support bounds}  \label{fig:change_support}
\end{figure}

We specifically select a representative distribution for each support condition with $q_1(\bm{x}) \sim \text{Normal}(0.667, 0.0317)\in \mathcal{S}_1$,  $q_2(\bm{x}) \sim \text{Beta}(4,2)\in \mathcal{S}_2=[0,1]$, and $q_3(\bm{x}) \sim \text{LogNormal}(-0.44, 0.2627)\in S_3=[0,\infty)$ as shown in Figure  \ref{fig:change_support}.  
We further select these three distributions to have identical mean $\mu = 0.667$ and standard deviation $\sigma = 0.0317$. A total of six cases are considered such that each permutation of original and updated distribution is studied. The results for each resampling strategy are given in Table \ref{tab:change_support} and discussed below.\\
\begin{table}[!ht] \footnotesize
\centering
\caption{Comparison of four strategies for six cases of updated probability densities with changing support.}
\label{tab:change_support}
\resizebox{\textwidth}{!}{
\begin{tabular}{ccccccccccc}
\hline
     &          &          &                                          & IS   & \multicolumn{2}{c}{Augmenting} & \multicolumn{2}{c}{Filtering} & \multicolumn{2}{c}{Mixed} \\ \hline
Case & Original $p(\bm{x})$ & Updated $q(\bm{x})$  & Support                                  & ESS  & $N_a$             & $N^*$            & $N_{reject}$            & $N^*$            & $N_{a+}$       & $N^*$          \\
1    & $q_1(x)$ & $q_2(x)$ & $\mathcal{S}_p \supseteq  \mathcal{S}_q$ & 9218 & N/A            & N/A           & 4266          & 5734          & 1046        & 10000       \\
2    & $q_1(x)$ & $q_3(x)$ & $\mathcal{S}_p \supseteq  \mathcal{S}_q$ & 9012 & N/A            & N/A           & 9406          & 594           & 989         & 10000       \\
3    & $q_2(x)$ & $q_1(x)$ & $\mathcal{S}_p \subseteq  \mathcal{S}_q$ & N/A  & 7811           & 17811         & N/A	         & N/A          & 728         & 10000       \\
4    & $q_2(x)$ & $q_3(x)$ & $\mathcal{S}_p \subseteq  \mathcal{S}_q$ & N/A  & 5.19E+22 ($\infty$)      & 5.19E+22 ($\infty$)     & N/A          & N/A          & 1522        & 10000       \\
5    & $q_3(x)$ & $q_1(x)$ & $\mathcal{S}_p \supseteq  \mathcal{S}_q$ & 6125 & N/A            & N/A           & 9718          & 282           & 1875        & 10000       \\
6    & $q_3(x)$ & $q_2(x)$ & $\mathcal{S}_p \subseteq  \mathcal{S}_q$ & N/A  & N/A       & N/A      & N/A          & N/A           & 931         & 10000       \\ \hline
\end{tabular}}
\end{table}

\begin{figure}[!ht]
 	\centering
 	\subfigure[]{\includegraphics[height=1.6in]{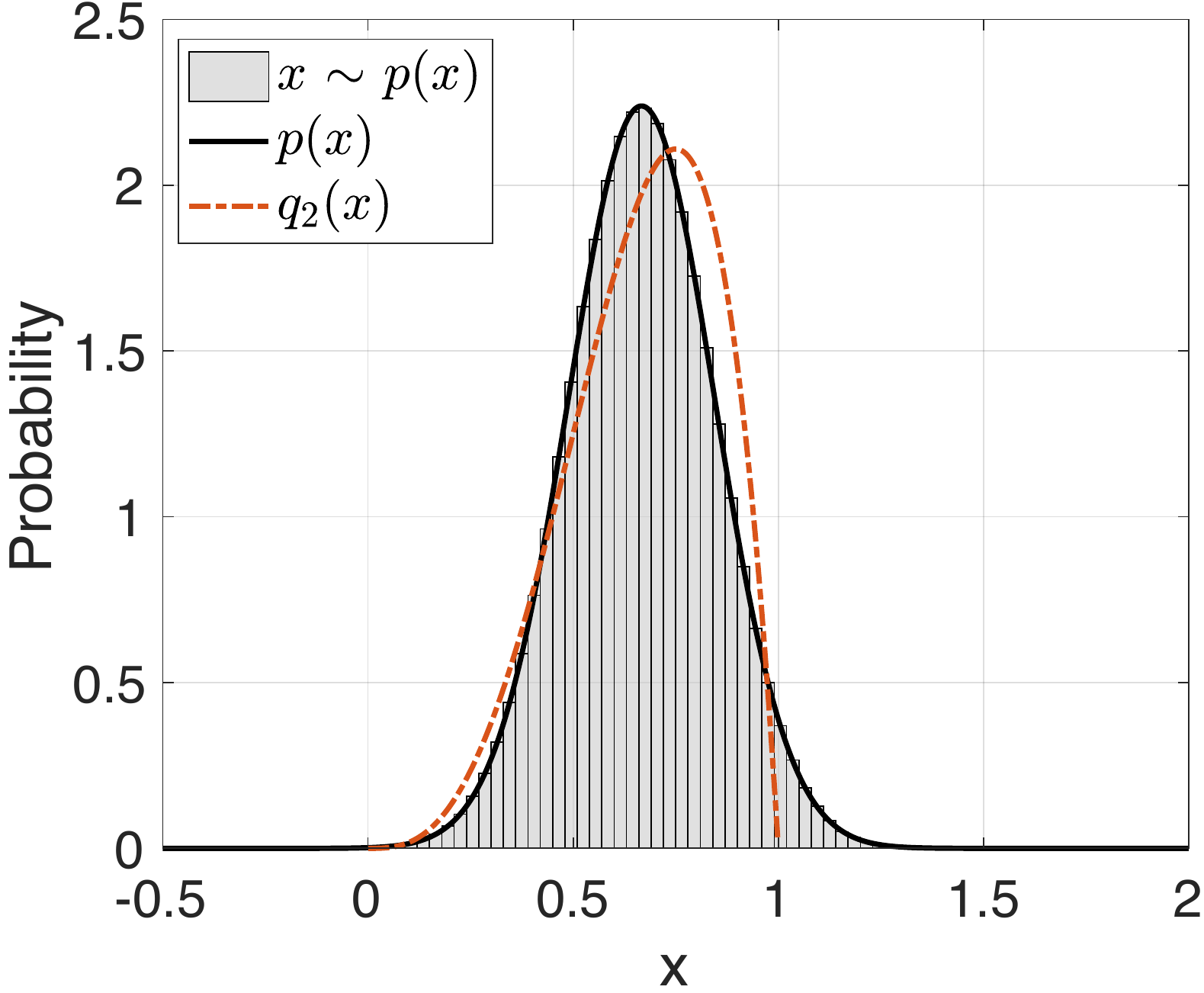}}
 	\subfigure[]{\includegraphics[height=1.6in]{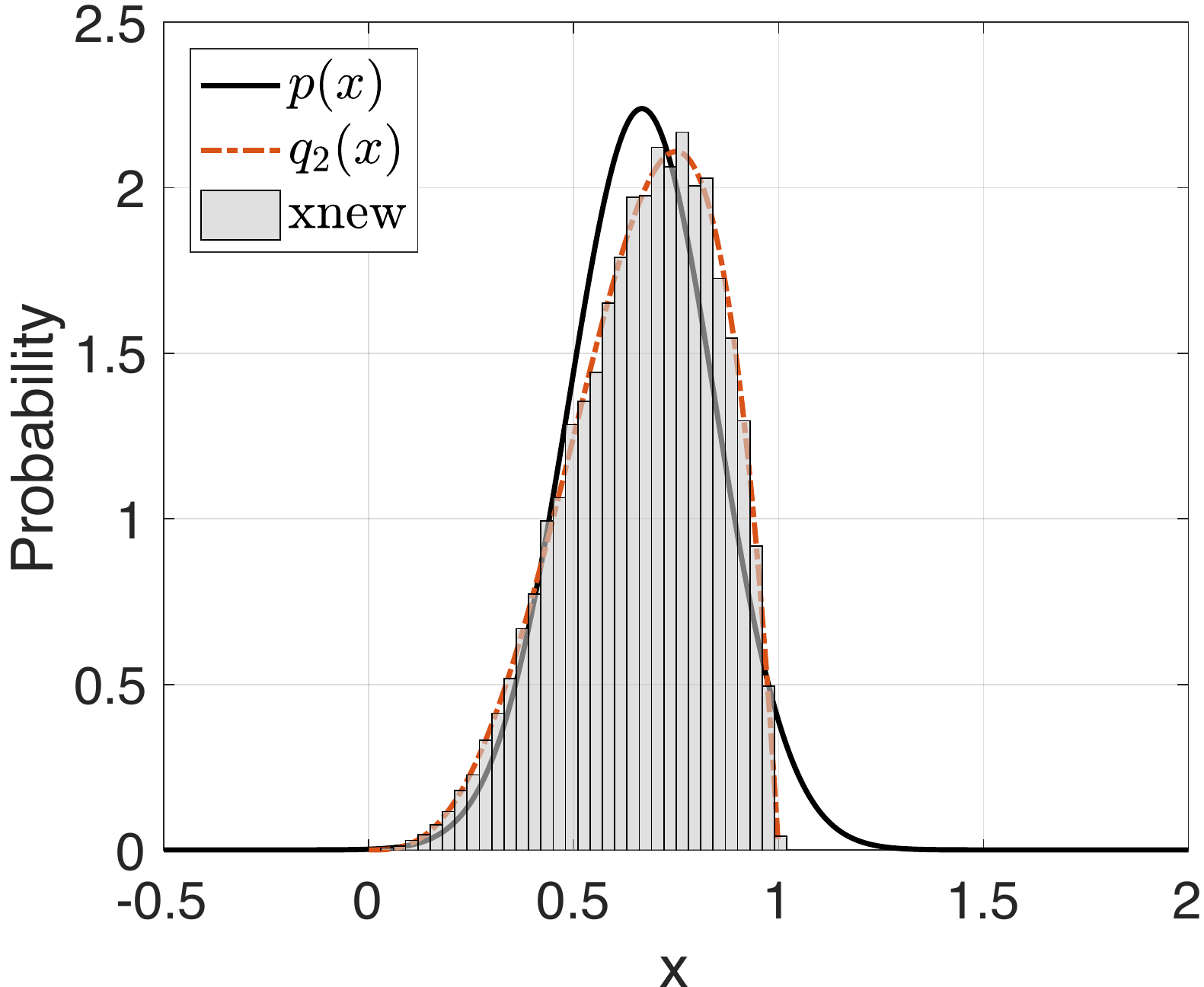}}
 	\subfigure[]{\includegraphics[height=1.6in]{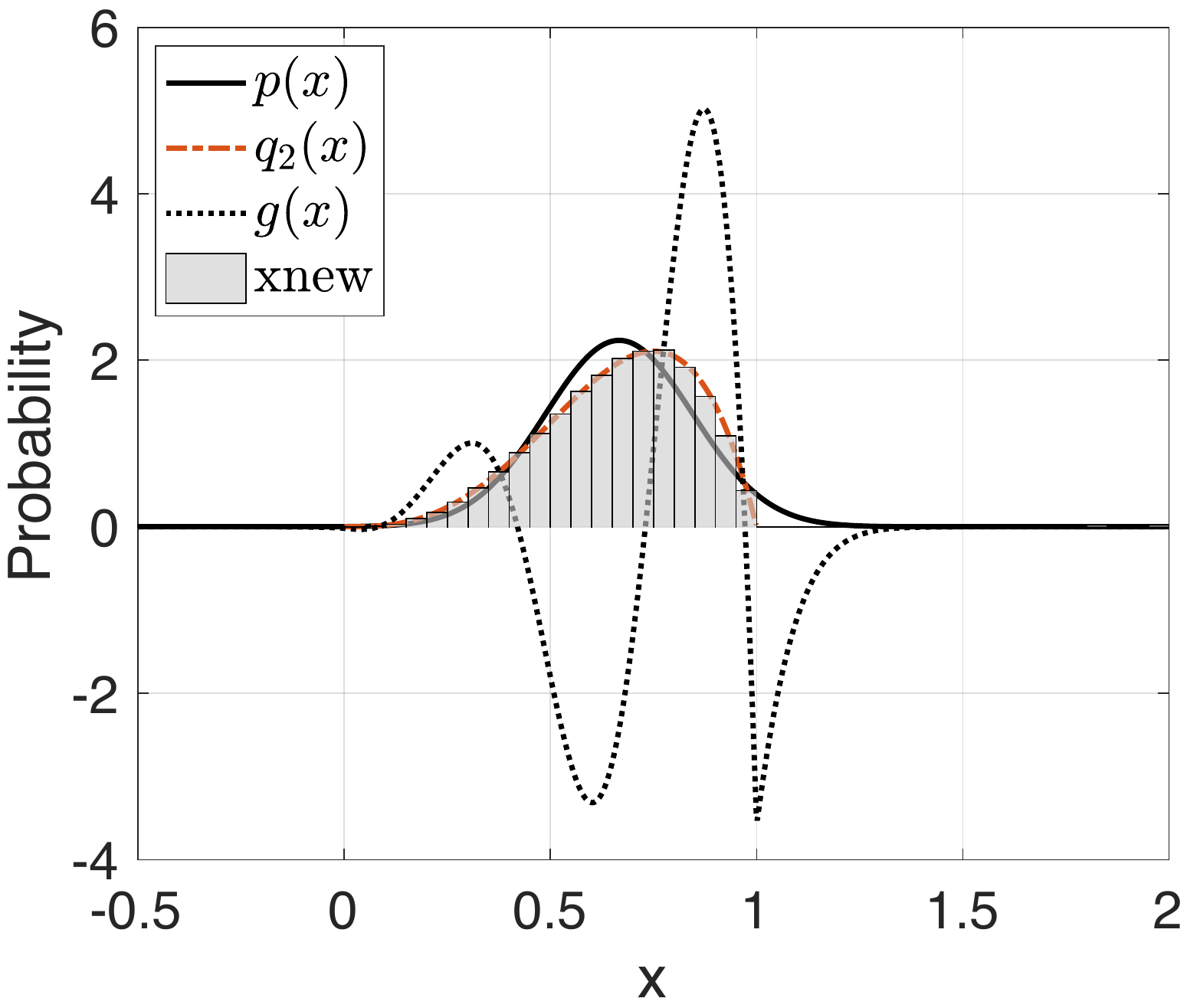}}
 	\caption[]{Case1 -- Resampling strategy for original distribution with infinite support and updated distribution with bounded support: (a) original samples, original distribution, and updated distribution, (b) filtering, and (c) mixed augmenting and filtering.}  \label{fig: change_case1}
\end{figure}

\noindent{\bf Case 1: Infinite to bounded support} -- If the original distribution $p(\bm{x})=q_1(\bm{x})$ has infinite support while and the updated distribution $q_2(\bm{x})$ is bounded as shown in Figure \ref{fig: change_case1}, i.e.\ $\mathcal{S}_p \supseteq  \mathcal{S}_q$, the importance sampling reweighting is effective with large effective sample size $\hat{N}_{ESS} = 9218$. It is capable of assigning zero weight to all samples outside $\mathcal{S}_q$ and requires only minor non-uniformity for the remaining samples. The augmenting strategy, however, does not work given this support relationship because $\dfrac{p(\bm{x})}{q(\bm{x})}=\infty,\forall \bm{x}\notin \mathcal{S}_q$. The filtering strategy can be used here but nearly half of the total original samples are removed with $N_{reject}=4266$. The mixed strategy is capable of adding samples where needed and removing all samples outside $\mathcal{S}_q$. It is quite efficient and only requires $N_{a+}= 1046$ additional samples to maintain the 10,000 sample size.

\noindent{\it Best Option: Importance Sampling Reweighting}
 
\noindent{\bf Case 2: Infinite to semi-infinite support} -- For importance sampling reweighting, this case exhibits similar performance to Case 1 with slighly smaller effective sample size $\hat{N}_{ESS} = 9012$. Again it can assign zero weight to those samples not in $\mathcal{S}_q$ and requires only small changes in the weights for other samples. Also, the augmenting strategy cannot be applied because $\dfrac{p(\bm{x})}{q(\bm{x})}=\infty,\forall x\notin \mathcal{S}_q$. The filtering strategy is not a viable option because it removes $N_{reject}=9406$ samples - leaving the Monte Carlo sample size very small. The mixed augmenting and filtering strategy is very effective here requiring only $N_{a+}=989$ samples to maintain the 10,000 sample size.

\noindent{\it Best Option: Importance Sampling Reweighting}
\begin{figure}[H]	
 	\centering
 	\subfigure[]{\includegraphics[height=1.6in]{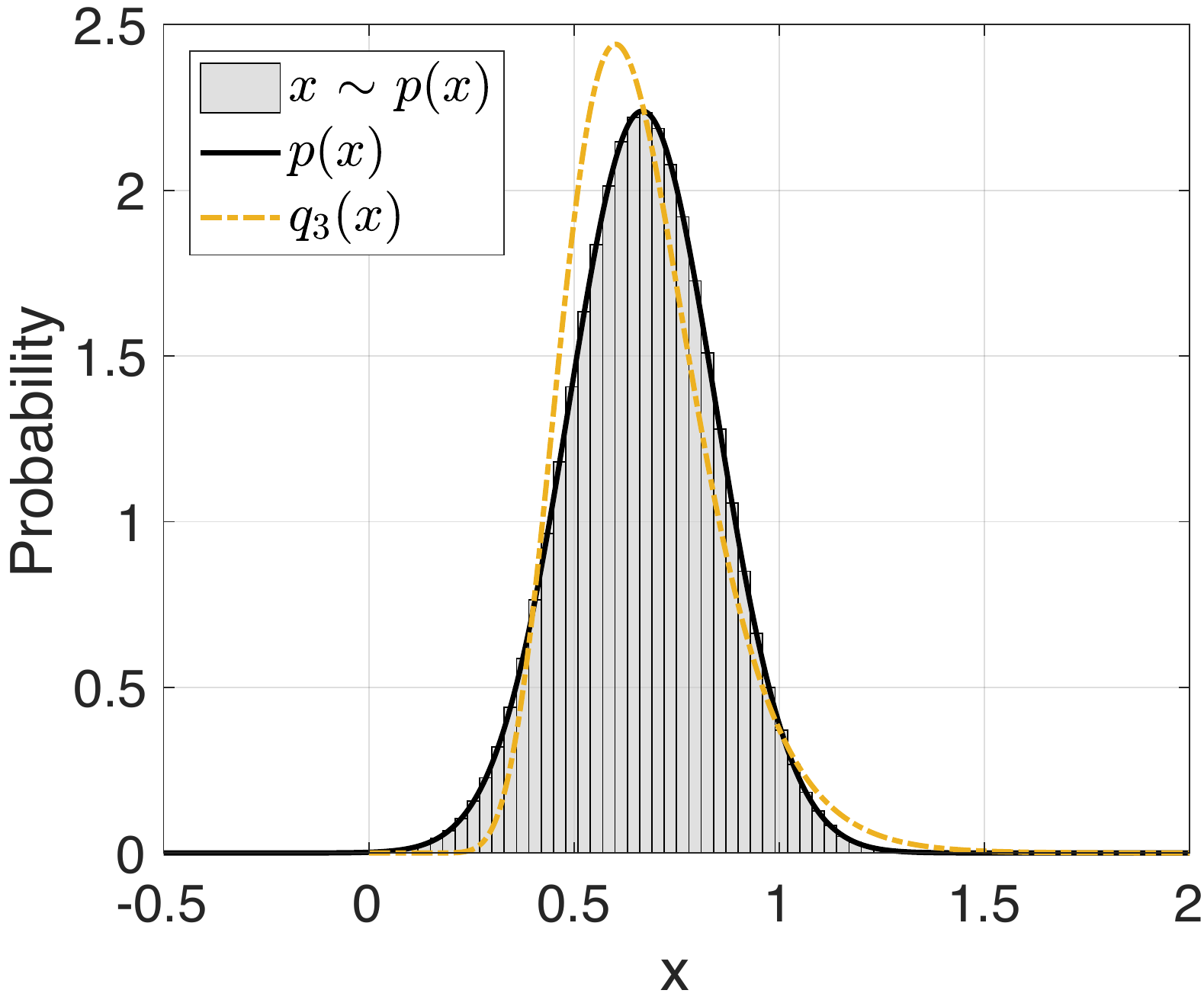}}
 	\subfigure[]{\includegraphics[height=1.6in]{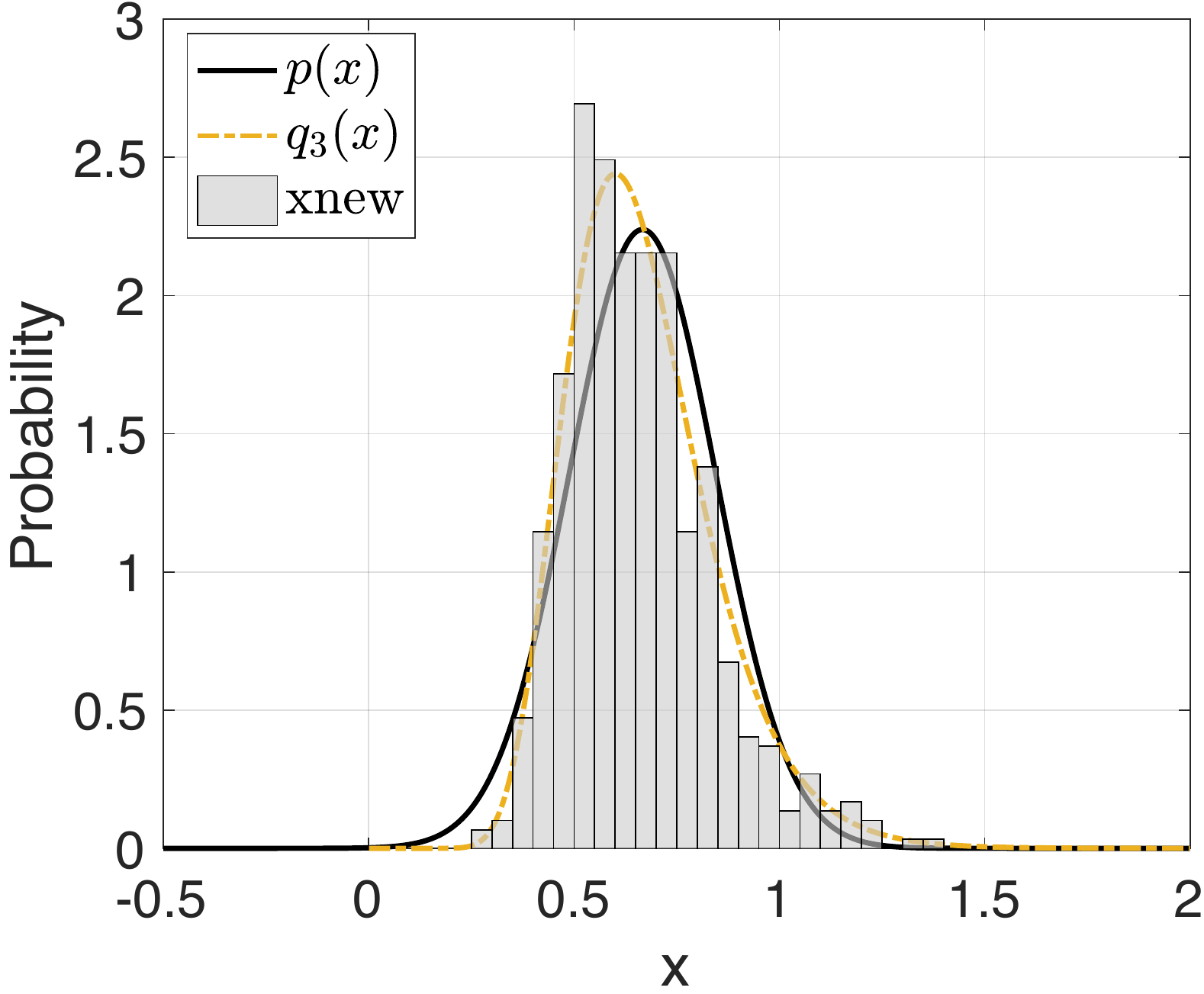}}
 	\subfigure[]{\includegraphics[height=1.6in]{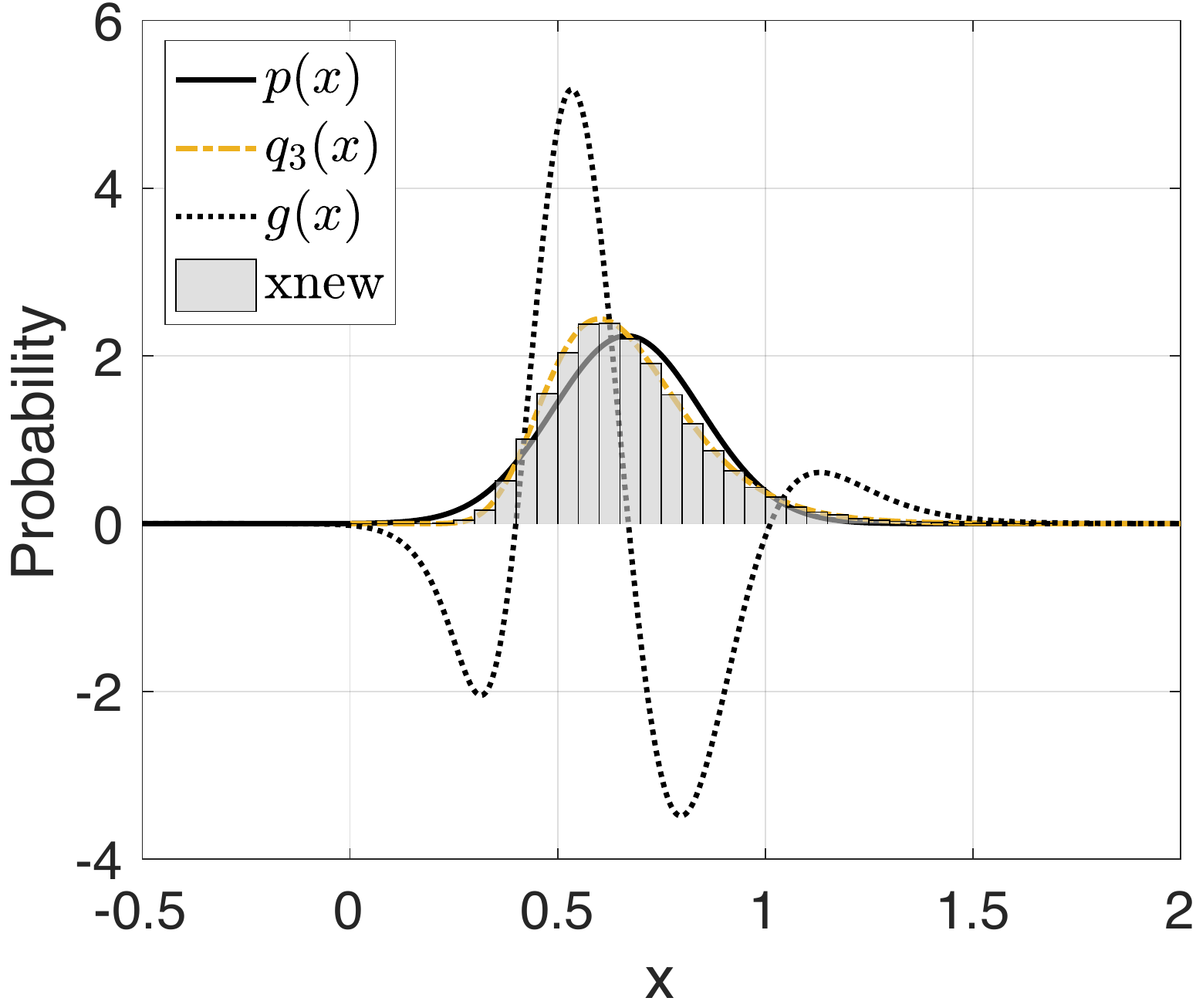}}
 	\caption[]{Case 2 - Resampling strategy for original distribution with infinite support and updated distribution with semi-infinite support: (a) original samples, original distribution, and updated distribution, (b) filtering, and (c) mixed augmenting and filtering.}  \label{fig: change_case2}
\end{figure} 
 
\begin{figure}[!ht]
 	\centering
 	\subfigure[]{\includegraphics[height=1.6in]{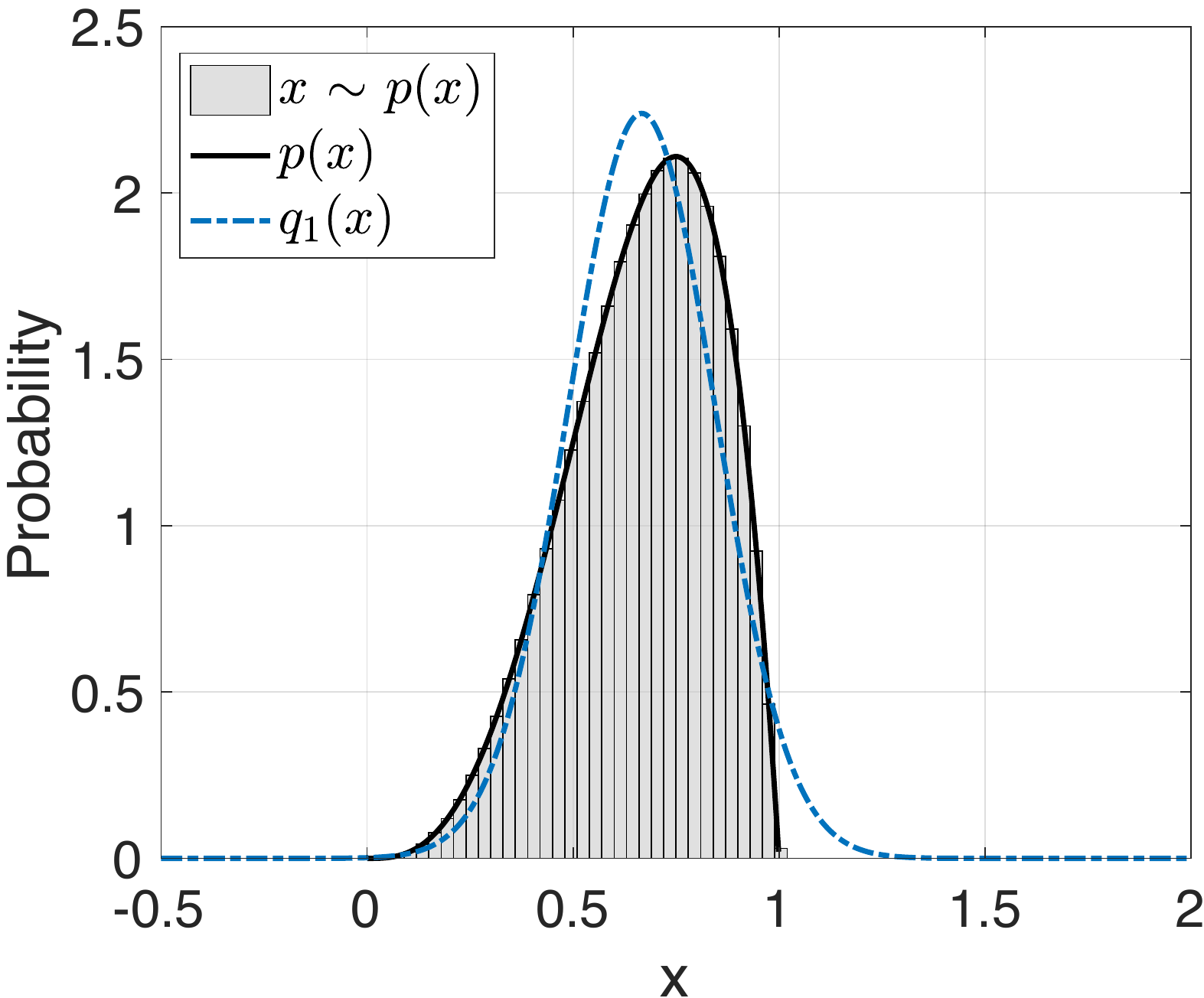}}
 	\subfigure[]{\includegraphics[height=1.6in]{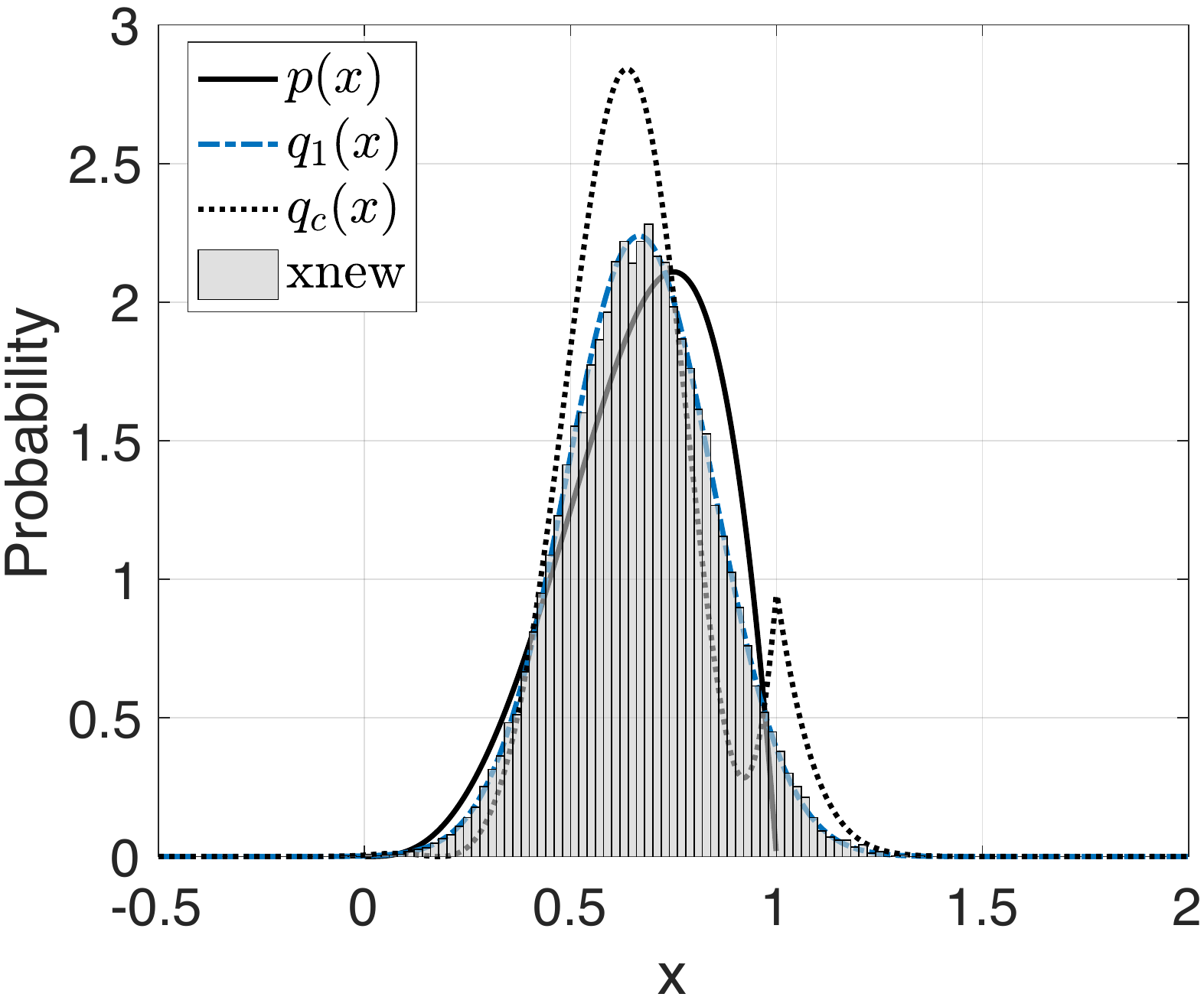}} 
%  	\subfigure[]{\includegraphics[height=1.6in]{bound_unbound_filtering.pdf}}
	\subfigure[]{\includegraphics[height=1.6in]{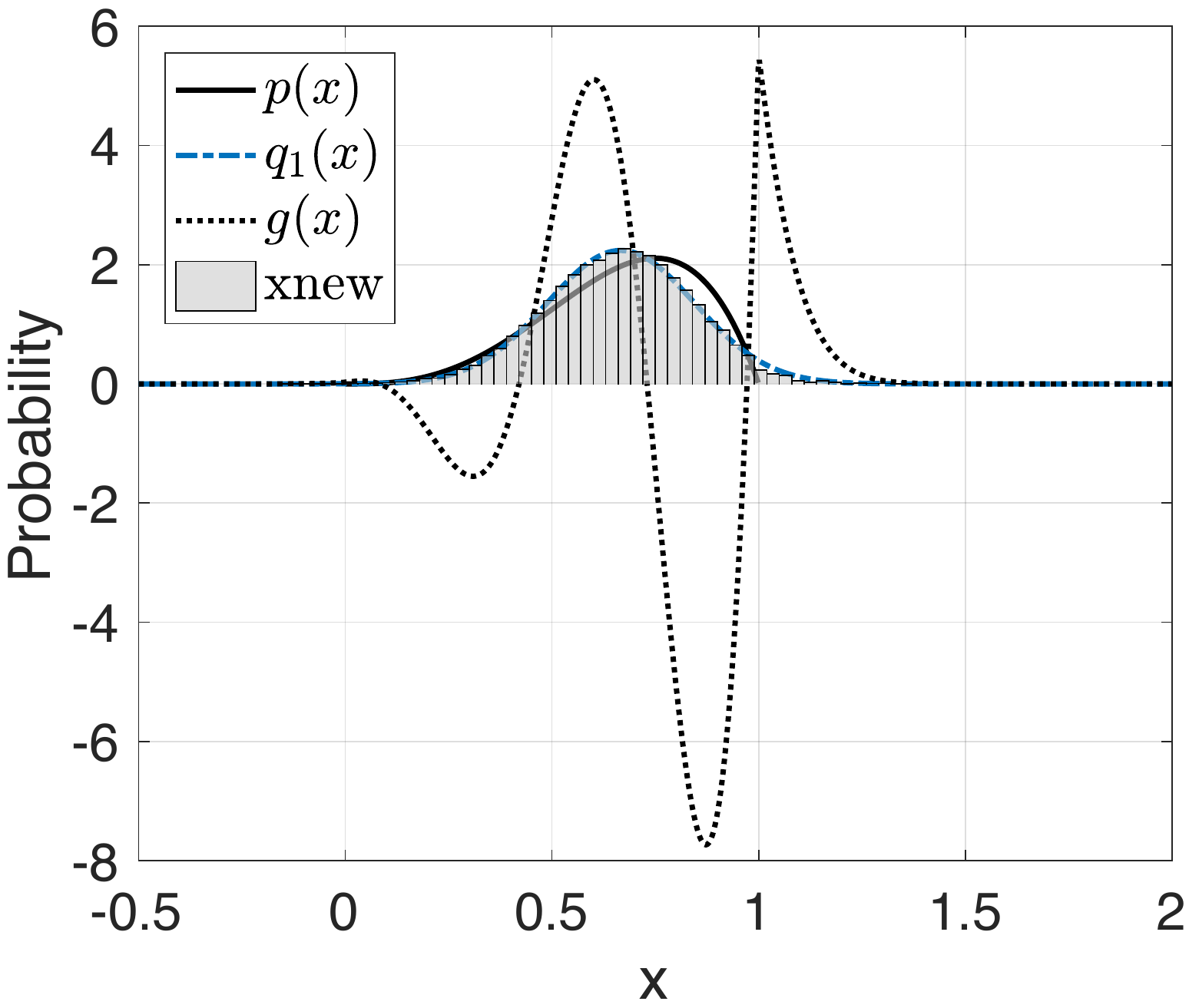}}
 	\caption[]{Case 3 - Resampling strategy for original distribution with bounded support and updated distribution with infinite support: (a) original samples, original distribution, and updated distribution, (b) augmenting, and (c) mixed augmenting and filtering.} \label{fig: change_case3}
 \end{figure}

\noindent{\bf Case 3: Bounded to infinite support} -- When $\mathcal{S}_p \subseteq \mathcal{S}_q$, importance sampling reweighting and the filtering approach cannot be applied. The original samples do not span $\mathcal{S}_p$ and therefore must be supplemented. In this case, the purely augmenting strategy requires $N_{a} = 7811$ additional samples as shown in Figure \ref{fig: change_case3}b and Table \ref{tab:change_support}. The mixed augmenting and filtering, by contrast, requires only $N_{a+}=728$ additional samples -- see Figure \ref{fig: change_case3}c. This is clearly the preferred approach.

\noindent{\it Best Option: Mixed Augmenting and Filtering}\\

\begin{figure}[!ht]
 	\centering
 	\subfigure[]{\includegraphics[height=1.6in]{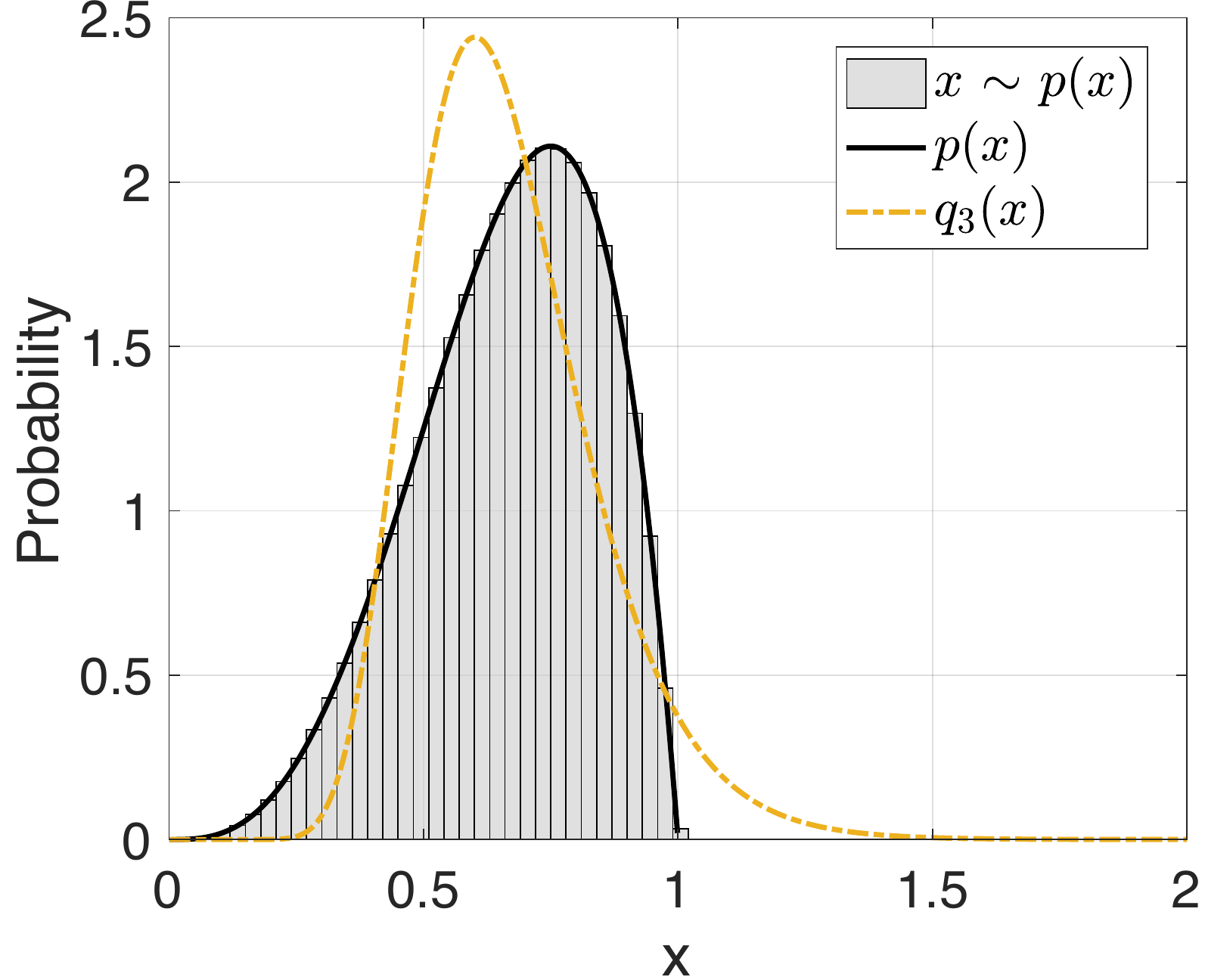}}
%  	\subfigure[]{\includegraphics[height=1.6in]{bound_one_bound_filtering.pdf}} 
 	\subfigure[]{\includegraphics[height=1.6in]{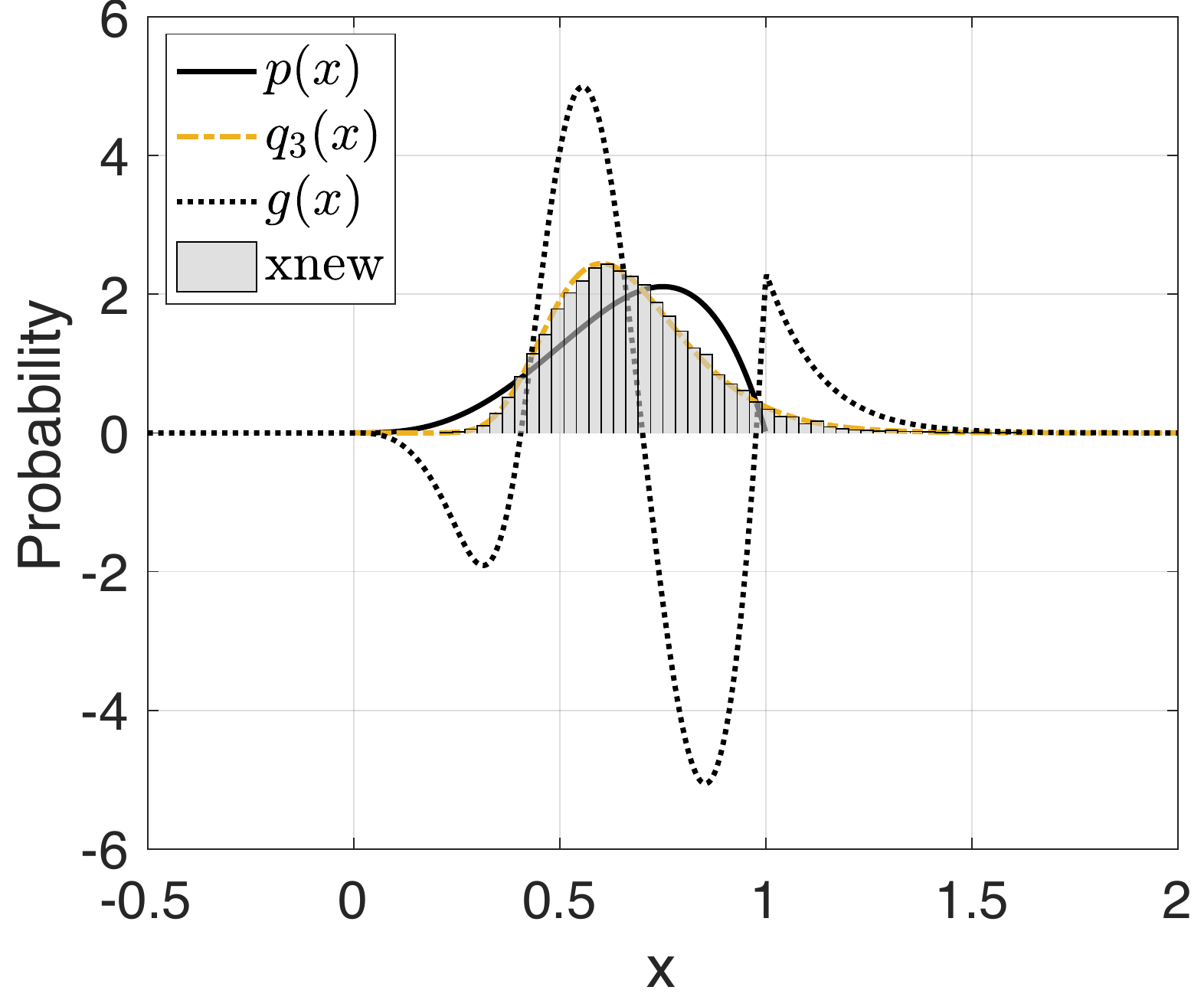}}
%	\subfigure[]{\includegraphics[height=1.6in]{bound_unbound_mix.pdf}}
 	\caption[]{Case 4 - Resampling strategy for original distribution with bounded support and updated distribution with semi-infinite support: (a) original samples, original distribution, and updated distribution and (b) mixed augmenting and filtering.} \label{fig: change_case4}
 \end{figure}

\noindent{\bf Case 4: Bounded to semi-infinite support} -- As in Case 3, importance sampling reweighting and the filtering approach are not applicable here because $\mathcal{S}_p \subseteq  \mathcal{S}_q$. Furthermore, the augmenting strategy theoretically requires infinite added samples because $\lim\limits_{\bm{x}\to 0^+}\dfrac{p(\bm{x})}{q(\bm{x})}=\infty$ as the lognormal, $q(\bm{x})$, decays exponentially and the beta, $p(\bm{x})$, decays polynomially as $\bm{x}\to 0^+$. This is reflected in Table \ref{tab:change_support} by an unrealistically large number of added samples that results from numerical discretization of the pdf. Therefore, the mixed augmenting and filtering strategy is the only viable option. This strategy is efficient requiring only $N_{a+}=1522$ new samples as illustrated in Figure \ref{fig: change_case4} and Table \ref{tab:change_support}.

\noindent{\it Best Option: Mixed Augmenting and Filtering}
 
\begin{figure}[!ht]
 	\centering
 	\subfigure[]{\includegraphics[height=1.6in]{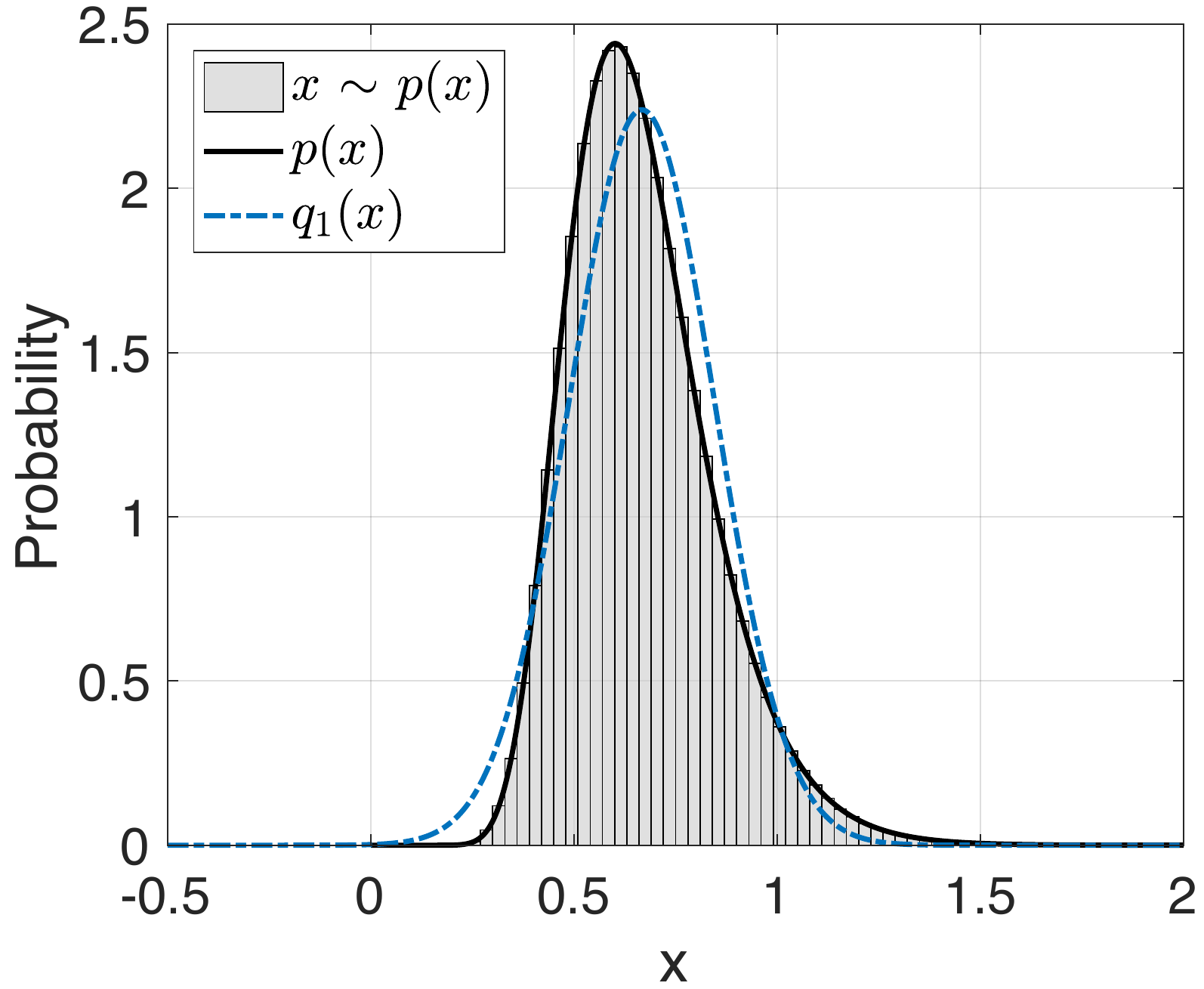}}
 	\subfigure[]{\includegraphics[height=1.6in]{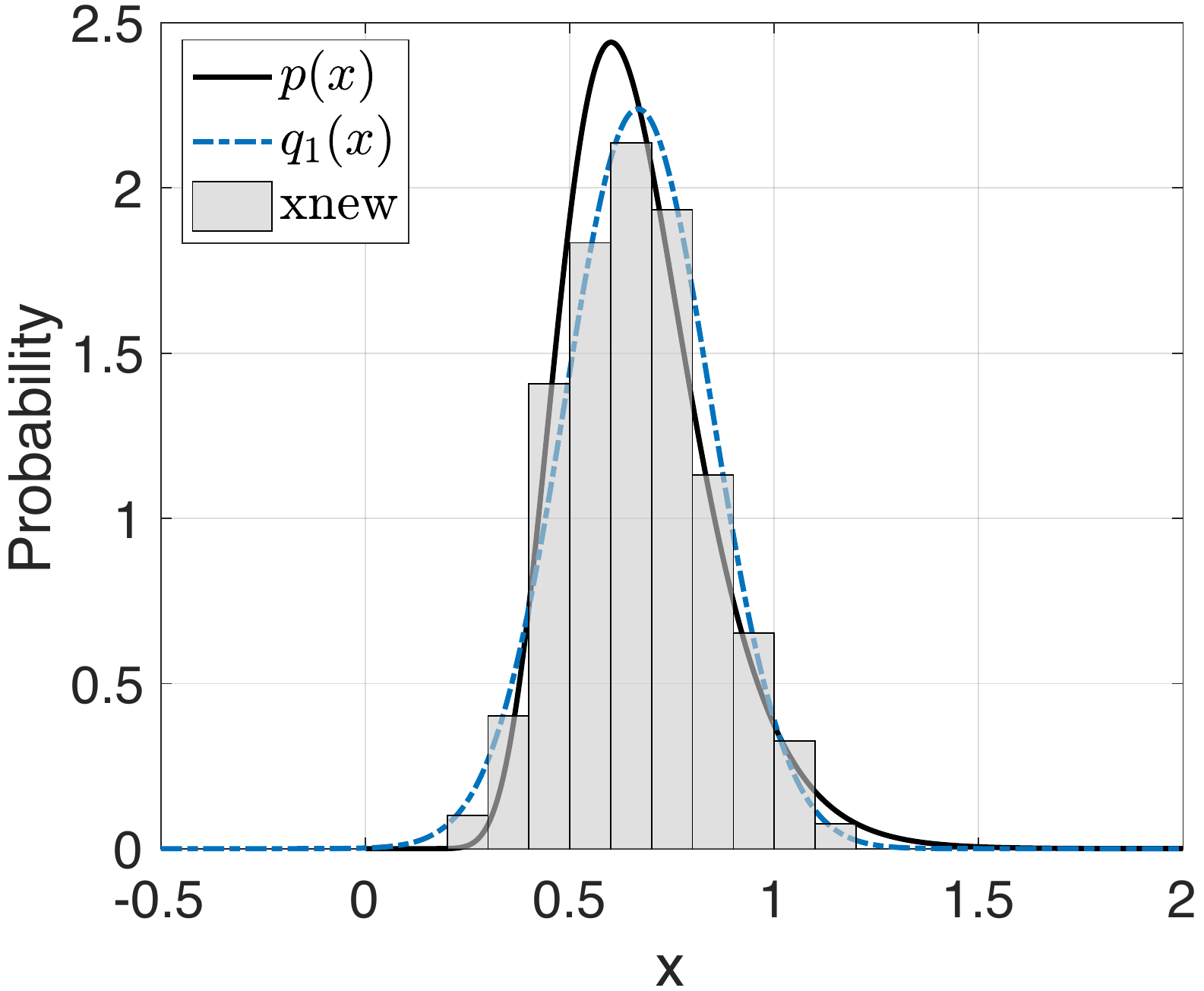}} 
 	\subfigure[]{\includegraphics[height=1.6in]{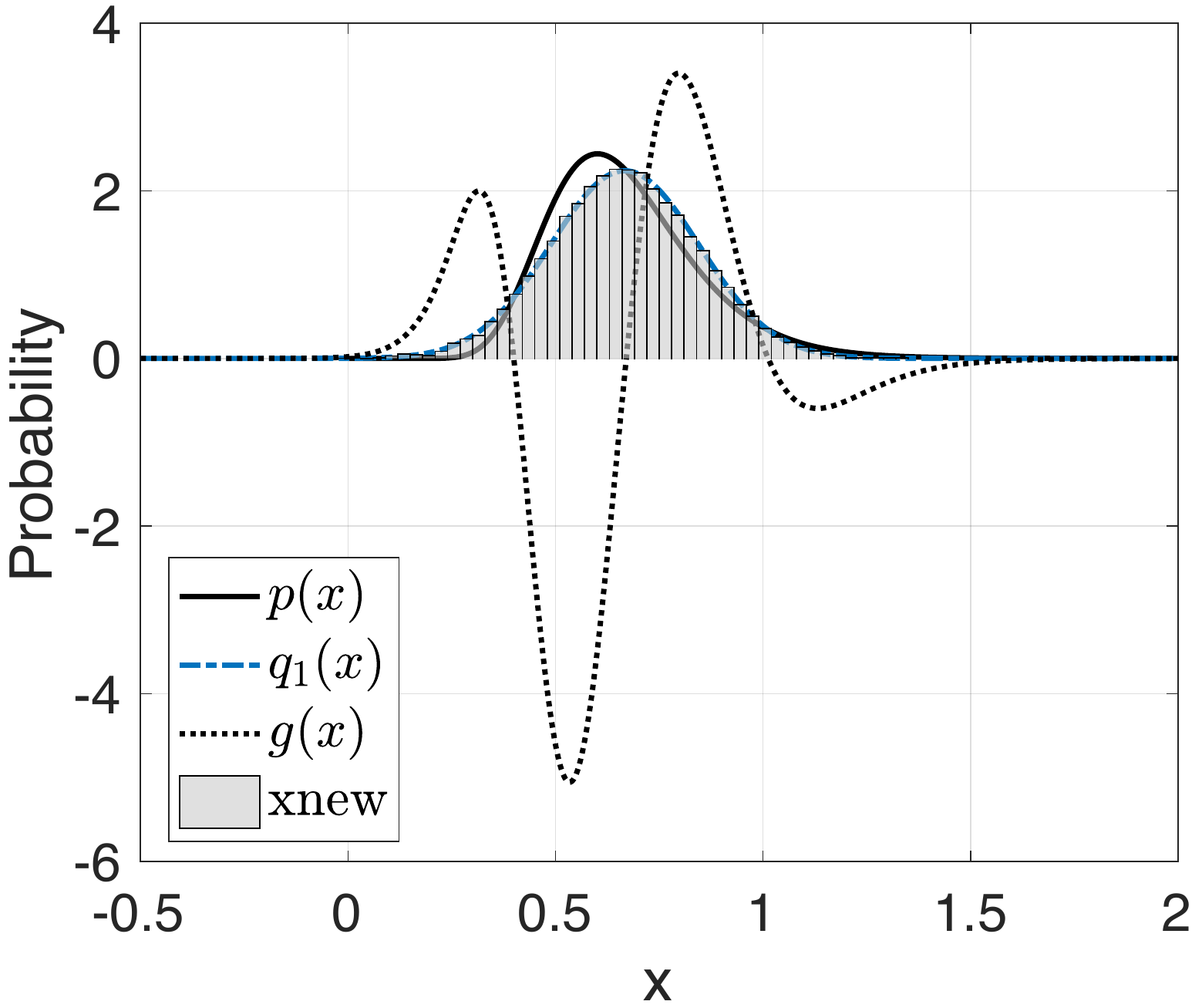}}
%	\subfigure[]{\includegraphics[height=1.6in]{bound_unbound_mix.pdf}}
 	\caption[]{Case 5 - Resampling strategy for original distribution with semi-infinite support and updated distribution with infinite support: (a) original samples, distributions and updated distribution (b) filtering and (c) mixed augmenting and filtering.}  \label{fig: change_case5}
 \end{figure}

\noindent{\bf Case 5: Semi-infinite to infinite support} -- In this case, the support conditions, $\mathcal{S}_p \supseteq  \mathcal{S}_q$, are the same as Cases 1 and 3. But, the effective sample size for importance sampling reweighting, $\hat{N}_{ESS} = 6125$, is considerably smaller than the other two cases. As in those cases, the augmenting strategy cannot be employed because $\dfrac{p(\bm{x})}{q(\bm{x})}=\infty,\forall \bm{x}\le 0$.  Moreover, the number remaining samples in the filtering strategy where $N_{reject}=9718$ is too small for Monte Carlo simulation as illustrated in Figure \ref{fig: change_case5}b. The augmenting and filtering strategy is effective for this problem requiring an additional $N_{a+}=1875$ samples.

\noindent{\it Best Option: Mixed Augmenting and Filtering}\\
 
\begin{figure}[!ht]
 	\centering
 	\subfigure[]{\includegraphics[height=1.6in]{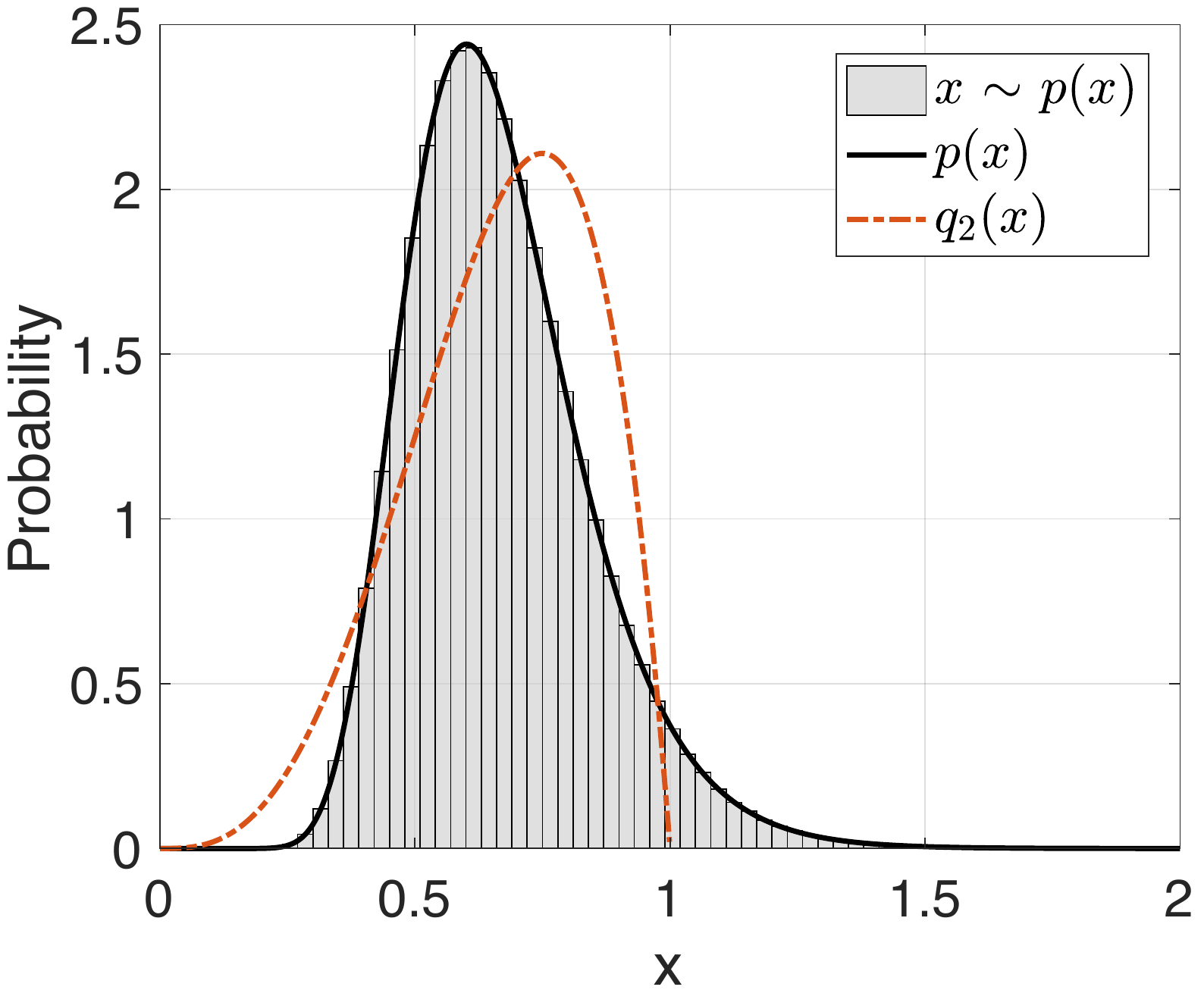}}
%  	\subfigure[]{\includegraphics[height=1.6in]{one_bound_bound_filtering.pdf}} 
 	\subfigure[]{\includegraphics[height=1.6in]{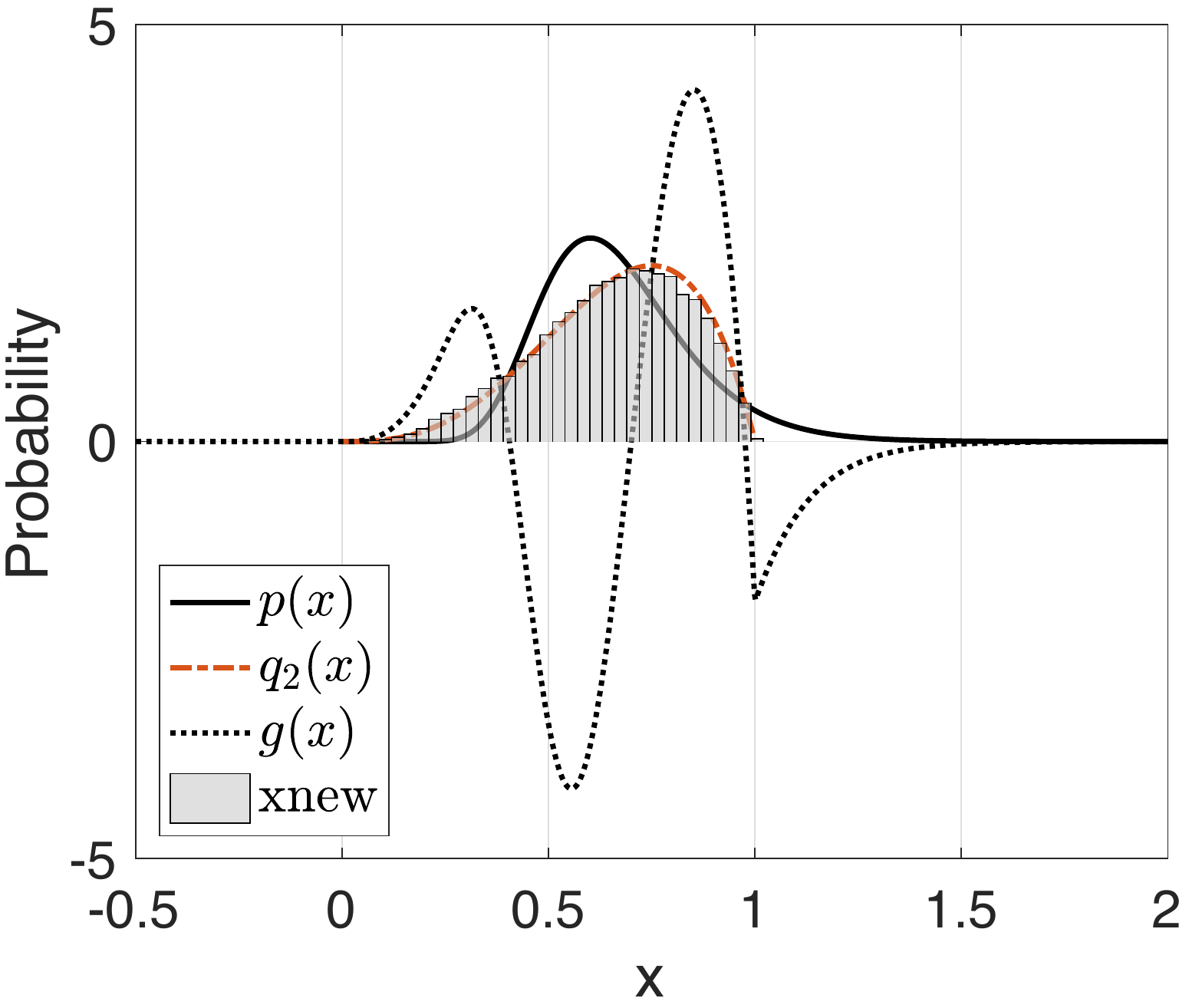}}
 	\caption[]{Case 6 - Resampling strategy for original distribution with semi-infinite support and updated distribution with bounded support: (a) original samples, original distribution, and updated distribution and (b) mixed augmenting and filtering.}  \label{fig: change_case6}
\end{figure}

\noindent{\bf Case 6: Semi-infinite to bounded support} -- With the same support conditions as Cases 3 and 4, $\mathcal{S}_p \subseteq  \mathcal{S}_q$ importance sampling reweighting and filtering cannot be used here. Similarly, the augmenting strategy cannot be employed here because $\dfrac{p(\bm{x})}{q(\bm{x})}=\infty,\forall \bm{x} >1$. Therefore, the mixed augmenting and filtering strategy is the only option and it is quite efficient, $N_{a+}=931$ and robust as illustrated in Figure \ref{fig: change_case6} and Table \ref{tab:change_support}.

\noindent{\it Best Option: Mixed Augmenting and Filtering}\\

These six cases illustrate the performance of each method for differing changes in distribution support. From these results, it is clear that the mixed augmenting and filtering approach is the most robust as it can be applied under any support condition. The purely augmenting and purely filtering approaches, when applicable, are very inefficient and are therefore not recommended under any conditions. Importance sampling reweighting, although not universally applicable, may be advantageous when the change in distribution is small.

\section{Application to Bayesian updating of plate buckling strength}
Uncertainty in the material and geometric properties of ship structural components can significantly impact the performance, reliability and safety of the structural system \cite{ClassNK}. In this work, we study the impact of uncertainty in material properties on buckling strength of a simply supported rectangular plate under uniaxial compression. An analytical formulation for the normalized buckling strength for a pristine plate was first proposed by Faulkner\cite{faulkner1973}
\begin{equation}
\psi =\frac { { \sigma  }_{ u } }{ { \sigma  }_{ 0 } } =\frac { 2 }{ \lambda  } -\frac { 1 }{ \lambda ^{ 2 } }  \label{eq:Faulkner } 
\end{equation}
where $\sigma_u$ is the ultimate stress at failure, $\sigma_0$ is the yield stress, and $\lambda$ is the slenderness of the plate with width \(b\), thickness \(t\), and elastic modulus \(E\) given by 
\begin{equation}
\lambda =\frac { b }{ t } \sqrt { \frac { \sigma _{ 0 } }{ E }  }. \label{eq:slenderness}
\end{equation}
Eq.\ (\ref{eq:slenderness}) was further modified by Carlsen \cite{carlsen1977} to study the effect of residual stresses and non-dimensional initial deflections \(\delta_{0}\)  associated with welding
\begin{equation}
\psi =\left( \frac { 2.1 }{ \lambda  } -\frac { 0.9 }{ \lambda ^{ 2 } }  \right) \left( 1 -\frac { 0.75\delta_{0} }{ \lambda }  \right)\left( 1 -\frac { 2\eta t }{b}  \right) \label{eq:buckling_strenth} 
\end{equation}
where $\eta t$ is the width of the zone of tension residual stress.

The design buckling strength is based on nominal values for the six variables in Eq. (\ref{eq:buckling_strenth}) provided in Table \ref{tab:defination_variables}. However, the actual values of these variables often differ from the design values due to uncertainties in the material properties and ``as built" geometry yielding uncertainty in the buckling strength. Overall, we are therefore interested in investigating the effect of the six uncertain variables shown in Table \ref{tab:defination_variables}, but in this work we focus only on assessing the influence of uncertainty in the yield strength $\sigma_0$ since it is the most sensitive variable identified by Global sensitivity analysis (see Table \ref{tab:defination_variables}) and for clarity of demonstration. 

\begin{table}[H] \footnotesize
\centering
	\caption{ Statistical properties of plate material, geometry and imperfection variables from Hess et al. \cite{hess2002} and Guedes Soares \cite{soares1988}}
	\label{tab:defination_variables}
	\begin{tabular}{cccccc}
		\hline
		Variables & Physical Meaning   & Nominal Value & Mean        & COV    & Global Sensitivity \\ \hline
		\(b\)         & width              &36            & 0.992*36    & 0.028  & 0.017              \\
		\(t\)         & thickness          & 0.75           & 1.05*0.75    & 0.044  & 0.045              \\
		\(\sigma_0\)    & yield strength           & 34            & 1.023*34      & 0.116 & 0.482              \\
		\(E\)         & Young's modulus      & 29000         & 0.987*29000 & 0.076  & 0.194              \\
		\(\delta_0\)    & initial deflection & 0.35          & 1.0*0.35    & 0.05   & 0.043              \\
		\(\eta\)      & residual stress    & 5.25          & 1.0*5.25    & 0.07   & 0.233              \\ \hline
	\end{tabular}
\end{table}

% In this work, we use ABS-B data to illustrate the performance of the proposed resampling algorithm on Bayesian updating. The histogram of total 79 ABS-B data is shown in Fig. \ref{fig:bayes_79data}.  The thick black line is Lognormal distribution with mean $\mu = 34.782$ and standard deviation $\sigma = 34.782*0.116$.  Instead of using multimodel inference in the previous paper (Zhang and Shields MSSP), this study ignores the model-form uncertainty and only selects the single best probability model to represent the limited data. The candidate probability model includes: {'Normal','Lognormal','Gamma','Logistic', 'Weibull','Loglogistic','Nakagami'}. Bayesian model selection is employed to identify the best probability model which has the minimal value among of all candidate models. 

\begin{figure}[!ht]	
	\centering
	\includegraphics[height=2in]{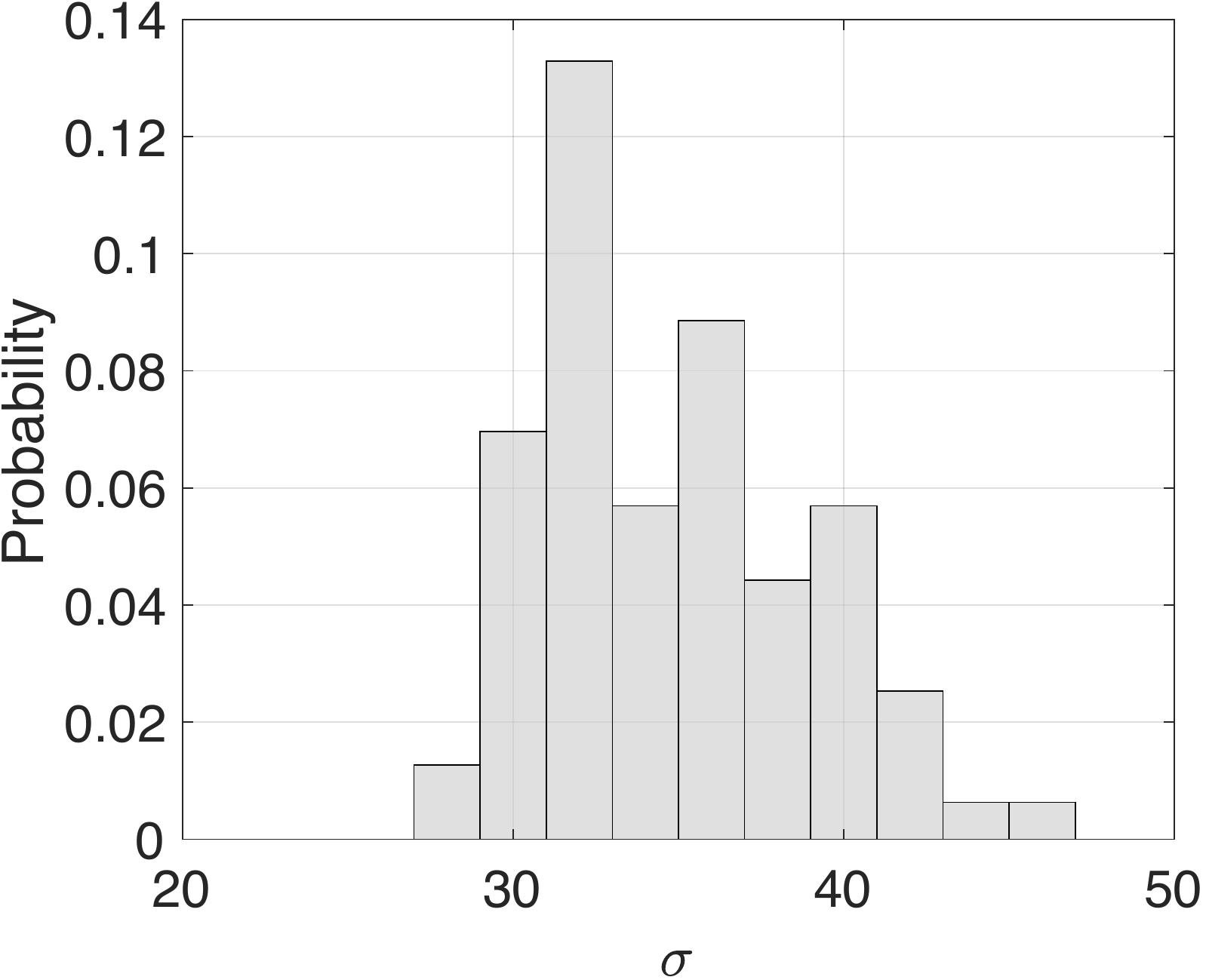}
	\caption[]{Histogram of ABS-B yield stress data.}  \label{fig:bayes_79data}
\end{figure}

The material is considered to be ABS-B type marine grade steel having properties described through historical US Navy testing programs \cite{atua1996, kufman1990, gabriel1962}. A total of 79 yield stress data are available from the historical reports, which are shown in the histogram in Figure \ref{fig:bayes_79data} and described statistically in Table \ref{tab: data_summary} (replicated from \cite{hess2002}).
\begin{table}[!ht] 
\centering
\caption{Statistical summary of ABS-B steel yield stress from historical data.}
\label{tab: data_summary}
\begin{tabular}{ccccccc}
\hline
Steel type & Min  & Max  & Mean   & COV   	& Number of tests & Data Sources                                                                              \\ \hline
% ABS-A      & 31.9 & 39.6 & 36.091 & 0.059 & Lognormal    & 33          & \begin{tabular}[c]{@{}c@{}}Weakly informative \\ but incorrect\end{tabular}  \\
ABS-B      & 27.6 & 46.8 & 34.782 & 0.116 	& 79          &  \cite{atua1996, kufman1990, gabriel1962}                                                                   \\
% ABS-C      & 30.9 & 41.5 & 33.831 & 0.081 & Lognormal    & 13          & \begin{tabular}[c]{@{}c@{}}Weakly informative \\ but incorrect\end{tabular}                                                                \\
% ASTM-A7       & 28.6 & 49.4 & 38.197 & 0.108 & Normal       & 58          & Informative but incorrect \\
 \hline
\end{tabular}
\end{table}

% Let's start from the initial 10 data, as shown in Fig. \ref{fig:bayes_10data}. Gamma distribution, as the most plausible model,  is selected as the best probability model to fit the initial limited 10 data, as shown the thick black curve in Fig. \ref{fig:bayes_10data} (a). Then, we draw 10000 random samples from Gamma distribution $p(x)$, as shown in  Fig. \ref{fig:bayes_10data} (b).  These random samples are taken account the original samples and $p(x)$ is identified as the original distribution. 
For the purposes of this study, we consider that these 79 data are being actively collected and that, at specific intervals of the data collection we will perform Bayesian inference on the data to identify a distribution for Monte Carlo simulation on plate buckling strength. In the Bayesian inference, we consider the following seven parametric probability distributions: Normal, Lognormal, Gamma, Logistic, Weibull, Loglogistic, and Nakagami. Both the distribution form and distribution parameters are selected using the MAP selection criterion described in Section 2.

\begin{figure}[!ht]	
	\centering
	\includegraphics[height=2in]{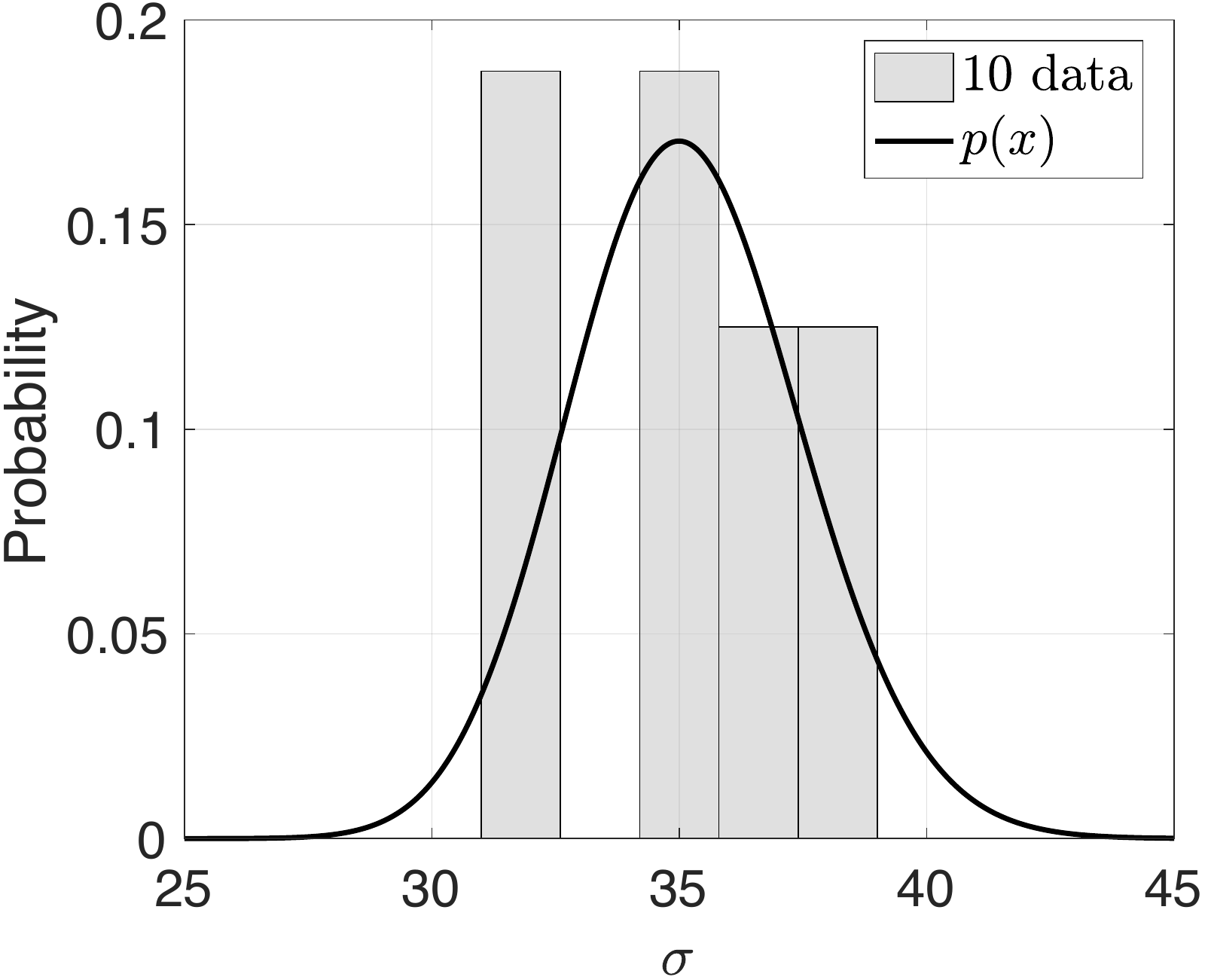}
% 	\subfigure[]{\includegraphics[height=2in]{bayes_3.pdf}}
	\caption[]{Histogram of the initial 10 ABS-B yield stress data and the Gamma distribution fit from Bayesian inference.}  \label{fig:bayes_10data}
\end{figure}

\begin{figure}[!ht]	
	\centering
	\includegraphics[height=2in]{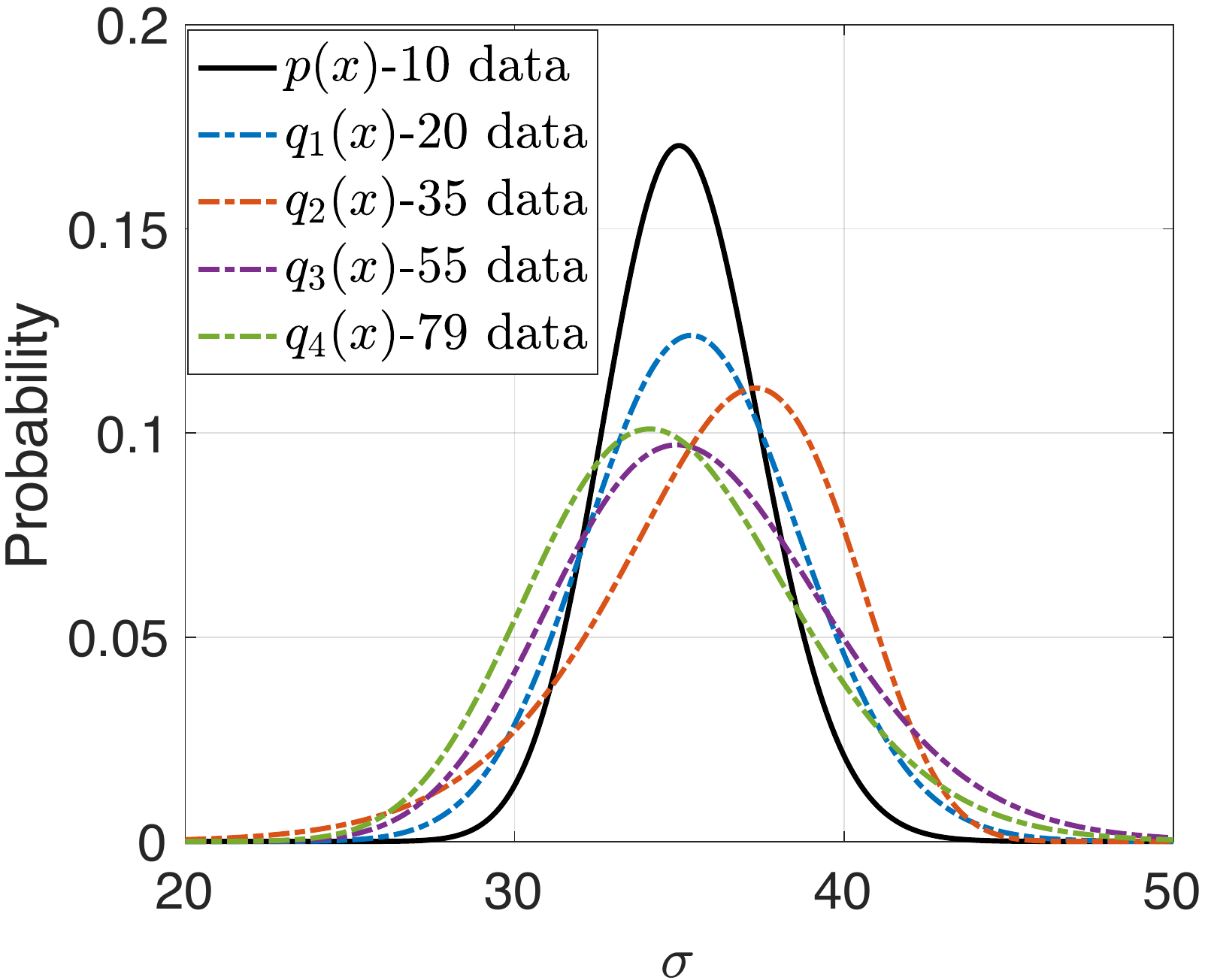}
	\caption[]{Initial yield stress probability density $p(x)$ and updated probability densities for each dataset extension.}  \label{fig:all_dis}
\end{figure}

We start by drawing 10 samples randomly from the 79 yield stress data and consider this as our initial test program. Bayesian inference on these data suggest an initial Gamma distribution fit, $p(\bm{x})$, as illustrated in Figure \ref{fig:bayes_10data}.
We then consider four subsequent extensions of the testing program having 20, 35, 55, and 79 test data. For each extension of the testing program, a new ``updated'' distribution (labeled $q_1(x)-q_4(x)$) is identified for the data using Bayesian inference as shown in Figure \ref{fig:all_dis} and detailed in Table \ref{tab:lastex}.
With each change in distribution, the Monte Carlo sample set (originally drawn from the Gamma distribution, $p(x)$) must be modified to match the new distribution. For each updating, from $q_i(x)\to q_j(x)$, we start by evaluating the effective sampling size for importance sampling reweighting. If $\hat{N}_{ESS}>9000$ (i.e.\ imporance sampling effectively retains more than 90\% of the samples), then importance sampling reweighting is used. Otherwise, the mixed augmenting and filtering strategy is employed to retain a set of 10,000 samples that matches the updated distribution.

\begin{figure}[!ht]	
	\centering
	\includegraphics[height=2in]{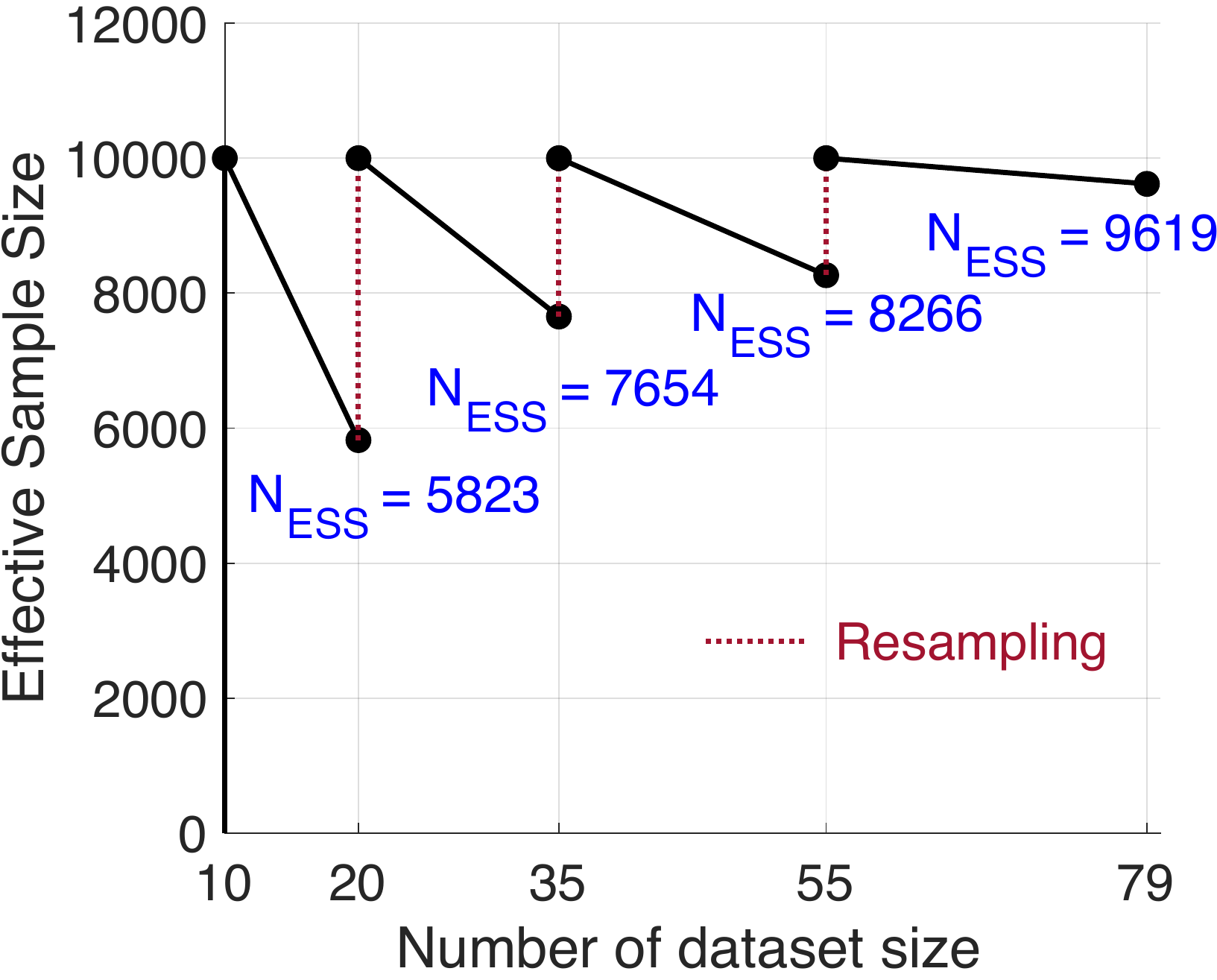}
	\caption[]{Updated sampling densities with common support}  \label{fig:mix_all_dis}
\end{figure}

The results of the sample updating are shown in Table \ref{tab:lastex} where, in the first three updates ($p(x)\to q_1(x)$, $q_1(x)\to q_2(x)$, and $q_2(x)\to q_3(x)$) the distribution changes form and $\sim 1500$ samples must be added/filtered in each case. In the final extension, the best fit distribution does not change and importance sampling reweighting can be used with $\hat{N}_{ESS}=9619$. 
\begin{table}[!ht] 
\centering
\caption{Change of distribution and Monte Carlo samples with increasing dataset size.}
\label{tab:lastex}
\resizebox{\textwidth}{!}{
\begin{tabular}{ccccccc}
\hline
   & Dataset &  &  & Effective  & Total variation & Additional \\
Case & size & Distribution & Parameters & sample size & distance & samples \\ \hline
$p(x)$  & 10              & Gamma       &  (224.6, 0.1565)     & 10000 & 0        & 0                  \\
$q_1(x)$ & 20             & Nakagami    &  (30.6, 1271.3)      & 5823  & 0.314    & 1557               \\
$q_2(x)$ & 35             & Weibull     &  (23.9, 1306.8)      & 7654  & 0.3023   & 1569               \\
$q_3(x)$ & 55             & Lognormal   &  (18.7, 1288.6)      & 8266  & 0.3009   & 1531               \\
$q_4(x)$ & 79             & Lognormal   &  (18.8, 1226.8)      & 9619  & 0.1679   & 0                  \\ \hline
\end{tabular}}
\end{table}
To further illustrate this updating, Figure \ref{fig:mix_all_dis} shows the evolution of the effective sample size as the dataset size increases. 

\begin{figure}[!ht]	
	\centering
	\includegraphics[height=2in]{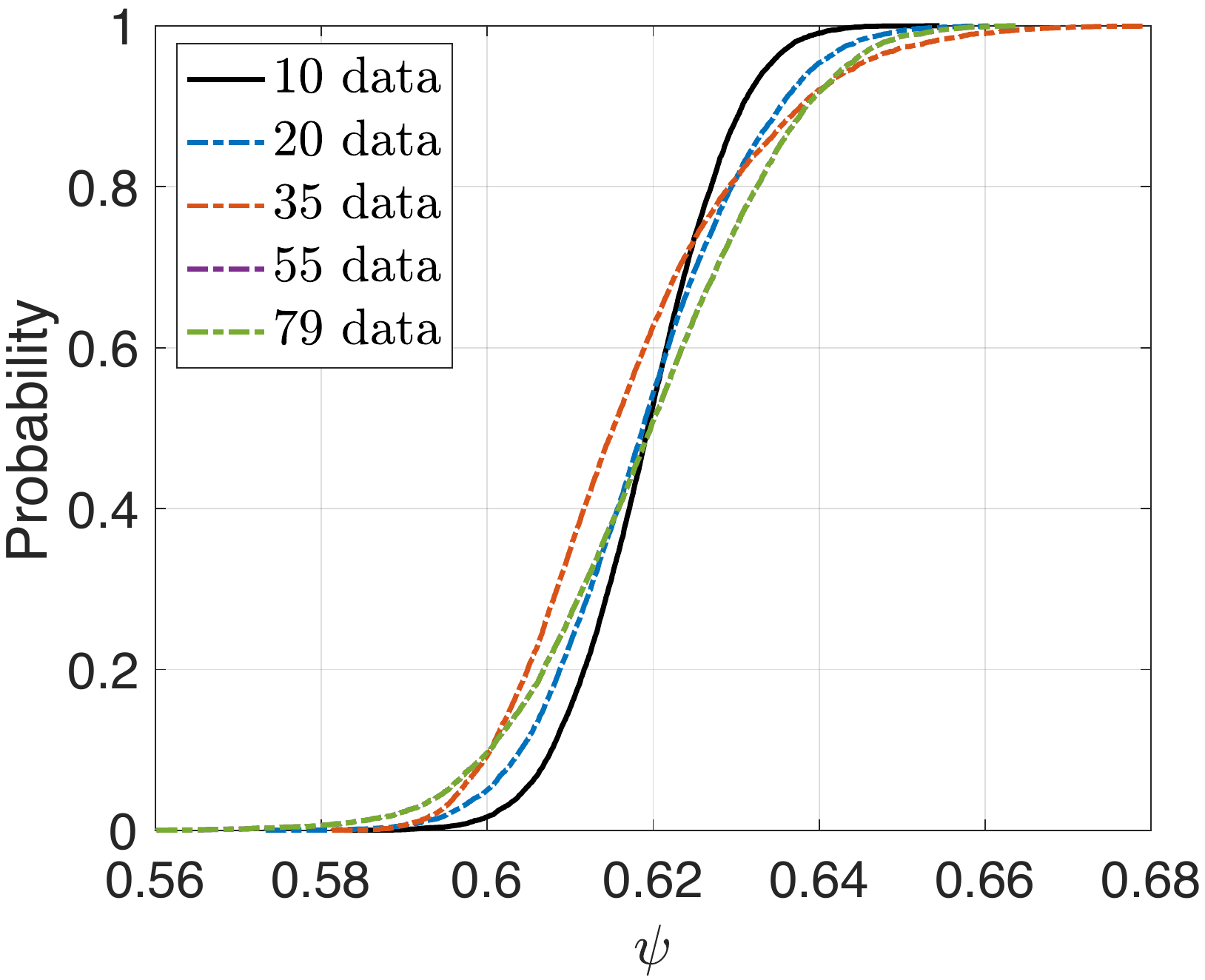}
	\caption[]{Empirical CDFs with the updated sample sets for each dataset size.} \label{fig:bayes_cdf}
\end{figure}

Finally, Figure \ref{fig:bayes_cdf} shows the empirical CDFs from Monte Carlo simulation on Eq.\ \eqref{eq:buckling_strenth} with the updated yield stress distributions for each dataset size. We see that, as the dataset size changes, the buckling strength distribution changes considerably with the yield stress distribution. Previously, to observe these changes with improving data, it would be necessary to either repeat the Monte Carlo simulation or accept diminished accuracy from importance sampling reweighting. To maintain the same accuracy, repeated Monte Carlo simulations for the five cases shown in Figure \ref{fig:bayes_cdf} would require 50,000 simulations. The proposed augmenting and filtering approach, however, requires only 14,657 simulations -- a 70\% savings. While this isn't critical for the simple model considered here, it would be a huge benefit if Eq.\ \eqref{eq:buckling_strenth} were replaced with an expensive finite element model such as that studied in \cite{Nahshon2018}.

\section{Conclusion}
In this work, we overcome the challenge updating Monte Carlo simulations when significant changes are observed in the probability measure of input random variables inferred from Bayesian inference as additional data are collected. Given that a large number of simulations may have been performed using the original distribution, we propose a set of resampling algorithms that serve as alternatives to rerunning new Monte Carlo simulations using the updated probability density. Four methods are discussed. We start from the widely-used importance sampling reweighting algorithm and discuss its limitation to relatively minor probability measure changes. A simple sampling augmenting method and a simple sample filtering strategy are then introduced to modify sample sets to adhere to an updated distribution. But, these approaches are very inefficient. A mixed resampling strategy is finally proposed that combines the benefits of the augmenting and filtering approaches and efficiently updates a sample set according to the updated density.  We illustrate the performance of these four strategies for common support and changing support relationships between the original distribution and updated distribution. It is shown that the mixed augmenting and filtering strategy is the most efficient option when significant change in the distribution is observed. When only small changes occur, importance sampling reweighting is typically effective unless the support of the distribution changes. Finally, we consider an analytical plate buckling example in which data are collected in stages. As new data are collected, the distribution of the yield stress is updated and corresponding Monte Carlo simulation results on the plate buckling strength are efficiently updating using the proposed augmenting and filtering procedure.

Through an example considering the analytical buckling analysis of a simply support plate, we systematically present the mixed strategy in fitting the probability measure changes in Monte Carlo simulation as data are sequentially collected from 10 data to 79 data. It is shown that the mixed strategy only need to draw additional approximated 15$\%$ number of original sample size, meanwhile, the updated samples well match the updated distribution. In other words, approximated 85$\%$ additional computational cost is saved compared with the approach that performs a new Monte Carlo analysis.

\section{Acknowledgement}
The work presented herein has been supported by the Office of Naval Research under Award Number N00014-16-1-2582 with Dr. Paul Hess as program officer.

\appendix
\section{Total variation distance}
The total variation distance between distributions $p(\bm{x}
)$ and $q(\bm{x})$ is defined as
\begin{equation}
d_1=\int_\mathcal{S}|p(\bm{x})-q(\bm{x})|d\bm{x}
\end{equation}
where the total support $\mathcal{S}=\mathcal{S}_p\cup\mathcal{S}_q$. Partitioning the support $\mathcal{S}$ into $\mathcal{S}_+=\{\bm{x}:q(\bm{x})\ge p(\bm{x})\}$ and $\mathcal{S}_-=\{\bm{x}:q(\bm{x})<p(\bm{x})\}$ with $\mathcal{S}=\mathcal{S}_+\cup\mathcal{S}_-$ and $\mathcal{S}_+\cap\mathcal{S}_1=\emptyset$, $d_1$ can be expressed as
\begin{equation}
d_1=\int_{\mathcal{S}_+}q(\bm{x})-p(\bm{x})d\bm{x}+\int_{\mathcal{S}_-}p(\bm{x})-q(\bm{x})d\bm{x}.
\end{equation}
Applying Eqs.\ \eqref{eqn:pi1} - \eqref{eqn:pi4} yields
\begin{equation}
d_1 = \pi_{q+}-\pi_{p+}+\pi_{p-}-\pi_{q-}.
\end{equation}
Recognizing that $\pi_{p+}+\pi_{p-}=\pi_{q+}+\pi_{q-}=1$, we see that $\pi_{q+}-\pi_{p+}=\pi_{p-}-\pi_{q-}$. Therefore,
\begin{equation}
d_1 = 2(\pi_{q+}-\pi_{p+})
\end{equation} 

\newpage
%\textbf{Reference}
%\begin{spacing}{1}
%{\scriptsize
\bibliographystyle{elsarticle-num}
\bibliography{paper}
%}
%\end{spacing}

\end{spacing}
\end{document}